\documentclass[twocolumn,secnumarabic,amssymb, nobibnotes, apa, pra,superscriptaddress]{revtex4-2}
\usepackage{revsymb4-2}

\usepackage{amsmath}
\usepackage{color}
\usepackage{tocloft}
\usepackage[usenames,dvipsnames,svgnames,table]{xcolor}
\usepackage{hyperref}
\usepackage[section]{placeins} 
\usepackage{amssymb}
\usepackage{mathtools}
\usepackage{MnSymbol}
\usepackage{bm}
\usepackage{epstopdf}
\usepackage{braket}
\usepackage{wrapfig}
\usepackage{bbold}
\hypersetup{
     colorlinks   = true,
     citecolor    = blue
}
\usepackage{lipsum}

\newcommand{\At}{{\bm{A}}_\mathrm{MIR}(t)}
\newcommand{\Et}{{\bm{E}}_\mathrm{MIR}(t)}

\newcommand{\kvec}{{\bm{k}}}
\newcommand{\Kvec}{{\bm{K}}}
\newcommand{\avec}{{\bm{a}}}
\newcommand{\bvec}{{\bm{b}}}
\newcommand{\kpar}{{\bm{k}}_{\parallel}}
\newcommand{\dvec}{{\bm{d}}}
\newcommand{\xivec}{{\bm{\xi}}}
\newcommand{\vvec}{{\bm{v}}}

\newcommand{\EvecMIR}{{\bm{E}}_\mathrm{MIR}}
\newcommand{\Fnvec}{{\bm{F}}_n}
\newcommand{\ent}{e^{i\frac{2\pi}{T_0} nt}}
\newcommand{\sigmaM}{\hat{\sigma}_\mathrm{refl}^{(y)}}

\begin{document}

\title{Strong-field physics in three-dimensional topological insulators}

\author{Denitsa Baykusheva}%
\email{denitsab@stanford.edu}
\affiliation{Stanford PULSE Institute, SLAC National Accelerator Laboratory, Menlo Park, California 94025, USA}
\author{Alexis Chac\'{o}n}
\email{achacon@postech.ac.kr }
\affiliation{Center for Nonlinear Studies and Theoretical Division, Los Alamos National Laboratory, Los Alamos, New Mexico 87545, USA}
\affiliation{Department of Physics and Center for Attosecond Science and Technology, POSTECH, 7 Pohang 37673, South Korea}
\affiliation{Max Planck POSTECH/KOREA Research Initiative, Pohang 37673, South Korea}

\author{Dasol Kim}
\affiliation{Max Planck POSTECH/KOREA Research Initiative, Pohang 37673, South Korea}

\author{Dong Eon Kim}
\affiliation{Max Planck POSTECH/KOREA Research Initiative, Pohang 37673, South Korea}


\author{David A. Reis}
\affiliation{Stanford PULSE Institute, SLAC National Accelerator Laboratory, Menlo Park, California 94025, USA}

\author{Shambhu Ghimire}
\affiliation{Stanford PULSE Institute, SLAC National Accelerator Laboratory, Menlo Park, California 94025, USA}

\date{\today}%
\begin{abstract}
We investigate theoretically the strong-field regime of light-matter interactions in the topological-insulator class of quantum materials. In particular, we focus on the process of non-perturbative high-order harmonic generation from the paradigmatic three-dimensional topological insulator bismuth selenide (Bi$_2$Se$_3$) subjected to intense mid-infrared laser fields. We analyze the contributions from the spin-orbit-coupled bulk states  and the topological surface bands separately and reveal a major difference in how their harmonic yields depend on the ellipticity of the laser field. Bulk harmonics show a monotonous decrease in their yield as the ellipticity increases, in a manner reminiscent of high harmonic generation in gaseous media. However, the surface contribution exhibits a highly non-trivial dependence, culminating with a maximum for circularly polarized fields. We attribute the observed anomalous behaviour to: \textit{(i)} the enhanced amplitude and the circular pattern of  the interband dipole and the Berry connections  in the vicinity of the Dirac point; and \textit{(ii)} the influence of the higher-order, ``hexagonal warping'' terms in the Hamiltonian, which are responsible for the hexagonal deformation of the energy surface at higher momenta. The latter are associated directly with spin-orbit-coupling parameters. Our results thus establish the sensitivity of strong-field driven high harmonic emission to the topology of the band structure as well as to the manifestations of spin-orbit interaction. 
\end{abstract} 

\maketitle

\section{Introduction}\label{sec:intro}

Strong-field ionization and subsequent re-scattering processes have been well explored in atoms and molecules in the gas phase. This includes the advanced understanding of the microscopic processes leading to the generation of high-order harmonics (HHG) \cite{Corkum1993,Lewenstein1994}. These insights have laid the foundations of attosecond physics and metrology~\cite{Brabec2000,Corkum2007,Krausz2009,Li2020}, which includes the ability to probe the structure and dynamics of atomic and molecular systems~\cite{Itatani2004, Lepine2014}. HHG has now been extended to condensed matter systems such as bulk crystals~\cite{Ghimire2011,Ghimire2018}, where the underlying microscopic dynamics are rationalized as a combination of the intraband acceleration of carriers~\cite{Ghimire2011, Schubert2014,Luu2015} and the interband dynamics arising from the recollision of electron-hole pairs on a sub-cycle timescale~\cite{Vampa2015}. Representative applications of solid-state HHG include the prospect of an all-optical retrieval of electronic band structures~\cite{Ghimire2011,Vampa2015, Luu2015}, tracking of recollision dynamics of quasi-particles in crystals~\cite{Schubert2014, Hohenleutner2015, McDonald2015}, compact setups for attosecond pulse generation~\cite{You2017b,Li2020}, strong-field dynamics in systems with reduced dimensionality~\cite{Liu2017a}, as well as the reconstruction of the Berry curvature in topologically trivial inversion-symmetry-breaking systems~\cite{Liu2017a, Luu2018}. \\



Here, we theoretically investigate HHG in a new class of matter: three-dimensional topological insulators (3D-TIs)~\cite{Hsieh2008, Xia2009, Hsieh2009b, Chen2009, Moore2010,Hasan2010a,Qi2011}.  In these systems, the cooperative action of strong spin-orbit interaction (SOI) and time-reversal symmetry (TRS) causes  band inversion~\cite{Zhang2009} and leads to the co-existence of insulating bulk bands and conducting gapless surface states with an odd number of Dirac cones in the Brillouin zone (BZ). These gapless surface states are formed near the Fermi level in between the valence and conduction bands of the insulating bulk bands. The topological protection enforced by TRS gives rise to a series of emergent behaviours, in particular   robustness of the surface states against nonmagnetic perturbations,  linear dispersion near the zone center, and a spin texture~\cite{Roushan2009, Hsieh2009} that supports helical, spin-polarized currents~\cite{McIver2012}. Owing to these  properties, TIs represent a potentially unique platform to control and manipulate strong-field-driven dynamics, including those leading to HHG. In this context, 3D-TIs were theoretically shown to support sub-cycle chiral electron dynamics originating from the chirality of Bloch bands near the Gamma point and ``hexagonal warping''~\cite{OliaeiMotlagh2017, OliaeiMotlagh2018}. It has further been predicted that  the topological properties of materials \cite{Hubener2017} can be controlled and manipulated through interactions with strong circularly polarized laser fields. An all-optical, contact-free approach, which can probe the structure and non-equilibrium dynamics of topological materials is therefore highly desired \cite{Morimoto2016a, Shin2019}.\\



The advantage of HHG over conventional spectroscopic methods, such as transport measurements, angle-resolved photoemission spectroscopy, as well as perturbative nonlinear optical methods like photovoltaic effects, Kerr rotation, and second harmonic generation, is the possibility to achieve sub-cycle temporal resolution. Recently, Silva \textit{et al.} considered a Chern insulator as a platform for HHG experiments and predicted that the sub-cycle tunneling dynamics depend strongly on whether the system is in a trivial or a topologically non-trivial phase, and that the topological invariant (in this case the Chern number) can be imprinted on the helicities of the emitted harmonics~\cite{Silva2019}. HHG has  also been scrutinized  as a sensitive probe of topological phase transitions in the Haldane model (through circular dichroism in the harmonic emission)~\cite{Chacon2018} as well as other model systems~\cite{ Bauer2018, Drueeke2019, Juerss2019}. However, the high harmonic response of the topological surface states present in a realistic topological material has not been investigated so far.   \\
 

In this work, we consider the prototypical strong topological insulator Bi$_2$Se$_3$ because of its relatively large band gap ($\sim 0.3$~eV) that makes it particularly suitable for below-band-gap excitation in the mid-infrared (MIR) spectral range. This manuscript is structured as follows. We start in Sec.~\ref{sec:tbm} by presenting the crystal symmetries and introducing the tight-binding model (TBM) Hamiltonian adopted from Ref.~\cite{Mao2011}. After discussing the spectrum of the bulk states (BSs), we proceed with the derivation of an effective 2D Hamiltonian for the topological surface states (TSSs) (Sec.~\ref{sec:tbm_surf}) and discuss the incorporation of the TBM results into the framework of the semiconductor Bloch equations in Sec.~\ref{sec:el_dyn}.  Section~\ref{sec:results} includes HHG results. Our calculations show distinctly different ellipticity responses of bulk versus surface states. Whereas the response of the bulk states is shown to strongly resemble the case of monoatomic gases, characterized by a fast, monotonic decay of the HHG yield as a function of ellipticity, the surface states showcase a non-trivial behavior, culminating in an enhanced yield for circularly polarized fields. We attribute this behaviour to the presence of a chiral, ``vortex''-like pattern in the interband transition matrix elements and the Berry connections in the vicinity of the Dirac cone~\cite{Liu2018a}, and to the influence of the higher-order (hexagonal ``warping'') terms~\cite{Fu2009a}. Importantly, the latter mechanism directly relates the ellipticity sensitivity of the HHG response to the  spin-orbit coupling (SOC) terms in the Hamiltonian. We conclude with a short summary in Sec.~\ref{sec:conclusion}.\\



\section{Electronic structure calculations}\label{sec:tbm}

\subsection{Crystal structure}\label{sec:crystal}

\begin{figure}
\centering
\includegraphics[scale=0.325]{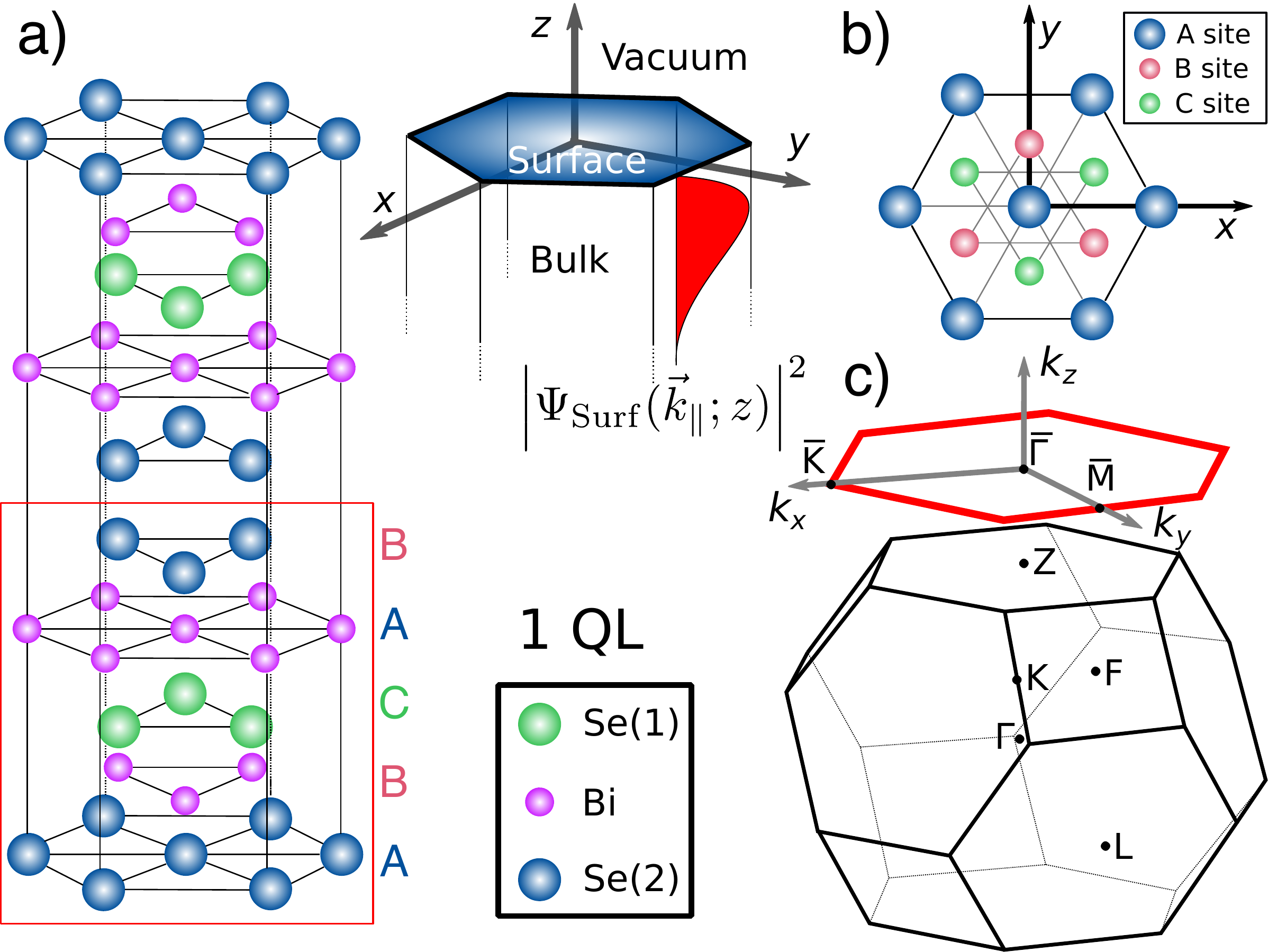}
\caption{a) Crystal structure of Bi$_2$Se$_3$, comprising alternating Bi and Se layers, stacked along the $z$-direction. Five consecutive layers form a quintuple layer (QL, cp. red rectangle), the building block of the lattice. Each QL comprises five atoms: 2 equivalent Bi sites, 2 equivalent Se sites (Se(1)), and a third Se atom, Se(2), which assumes the role of an inversion center. The hexagonal lattice constants are  $a=4.14$~\AA\ and $c=28.70$~\AA.\ The space-dependent wavefunction of the surface states (squared magnitude), $\Psi_\mathrm{Surf}(\kpar;z)$, is sketched as a red-shaded surface and illustrates the employed boundary conditions. b) Schematic representation of the  $C_{3v}$-symmetric Bi$_2$Se$_3$ (111)-surface (rhombohedral convention), exposing a top Se-layer and underlying Bi and Se'-layers. c) Sketch of the 3D Brillouin zone of bulk Bi$_2$Se$_3$ (black) with the four time-reversal-invariant points indicated ($\Gamma, L, F, Z$). The projected 2D BZ of the (111) surface is shown as a red hexagon, with labelled high-symmetry points $\overline{\Gamma}, \overline{K}, \overline{M}$.  }\label{fig:crystalBZ}
\end{figure}

We employ the generic TBM Hamiltonian put forward in Ref.~\cite{Mao2011} for materials of the Bi$_2$Se$_3$-family sharing the rhombohedral crystal lattice belonging to the $D_{3d}^5$ ($R\overline{3}m$, $\sharp 166$) space group. The crystal structure of Bi$_2$Se$_3$ is depicted schematically in Fig.~\ref{fig:crystalBZ}~a, together with the corresponding 3D Brillouin zone and the projected 2D surface BZ. Bi$_2$Se$_3$ is a layered materials with five atoms in the unit cell. The five atoms constitute a ``quintuple layer'', and each layer is organized into a triangular lattice. These are stacked along the $z$-direction and held together by weak van-der-Waals interactions. The spatial symmetries of the rhombohedral point group include: \textit{(i)} inversion symmetry $\hat{i}$ (IS), \textit{(ii)} two-fold rotation along the $x$-direction $\hat{\mathcal{R}}_2^{(x)}$, \textit{(iii)} three-fold rotation around the $z$-axis, $\hat{\mathcal{R}}_3^{(z)}$. Although formally not a \textit{spatial } symmetry of the $D_{3d}^5$-group, the electronic wavefunctions of Bi$_2$Se$_3$ are also characterized by time-reversal symmetry $\hat{\mathcal{T}}$ (TRS). We also briefly discuss the symmetry properties of the (111)-surface (depicted in panel b of Fig.~\ref{fig:crystalBZ}), which can be formally classified as belonging to the $C_{3v}$-group. Inversion symmetry is necessarily lost at the boundary, whereas the three-fold rotation $\hat{\mathcal{R}}_3^{(z)}$ is preserved as well as a mirror plane $\sigmaM$ coincident with the $y$-axis. There are in total three equivalent mirror planes parallel to the $\overline{\Gamma M}$-high-symmetry lines in the projected 2D BZ, and one of them is chosen as the $k_y$-direction in the coordinate system employed here. The TBM is constructed from the four levels closest to the Fermi level which form the basis for each site:
\begin{equation}\label{eq:Pz_basis}
\left\{ \ket{P^+_z,\uparrow}, \ket{P^-_z,\uparrow}, \ket{P^+_z,\downarrow}, \ket{P^-_z,\downarrow} \right\}.
\end{equation}
The superscripts $\pm$ denote the parity of the state, and $\ket{P^+_z\uparrow (\downarrow)}$ and $\ket{P^-_z\uparrow (\downarrow)}$ are derived from atomic $p_z$-orbitals of the Bi and Se atoms, respectively. The $\uparrow (\downarrow)$-symbols denote the spin state.

In the basis defined above, the tight-binding Hamiltonian in momentum space has the generic form:
\begin{equation}\label{eq:Htbm_generic}
\hat{\tilde{H}}(\kvec) = \hat{\epsilon}(\kvec) + \sum_{i=1}^3\left( \hat{t}_{\avec_i}e^{i\kvec\cdot\avec_i} + \hat{t}_{\bvec_i}e^{i\kvec\cdot\bvec_i} + \mathrm{h. c.}\right),
\end{equation}
where $\hat{\epsilon}(\kvec)$ is a diagonal (on-site) energy term. The sets of vectors $\{\pm\avec_i\}$ and $\{\pm\bvec_i\}$ in Eq.~(\ref{eq:Htbm_generic}) indicate the positions of the six intra- and inter-layer neighbours on each lattice site and are listed explicitly in  Appendix~\ref{app:tbm}. Correspondingly, $\hat{t}_{\avec_i}$ and $\hat{t}_{\bvec_i}$ denote the intralayer and the interlayer hopping parameters. The Hamiltonian in Eq.~(\ref{eq:Htbm_generic}) can also be recast in the form:
\begin{equation}\label{eq:Htbm_h}
\hat{\tilde{H}}(\kvec) = h_0(\kvec) + \sum_{i=1}^5 h_i(\kvec)\Gamma_i
\end{equation}
where $\Gamma_i$ are the Dirac matrices defined in terms of the Pauli matrices $\hat{\sigma}_i$ and $\hat{\tau}_i$ in Eq.~(\ref{eq:Gamma}). The auxiliary functions $h_i(\kvec)$ in Eq.~\ref{eq:Htbm_h} are given in the Appendix. In Sections~\ref{sec:spinpol} and~\ref{sec:asymptotic}, we study their low-energy behaviour $\kvec\sim \bm{0}$ in the context of the surface-state spin polarization and the surface Bloch band topology. 

\subsection{Bulk states}\label{sec:tbm_bulk}

We next apply the unitary transformation ($\hat{U}_1$, s. Eq.~(\ref{eq:U1})) introduced by Liu \textit{et al.}~\cite{Liu2010}:
\begin{equation}\label{eq:Hk3}
\hat{H}(\kvec) = \hat{U}_1\hat{\tilde{H}}(\kvec) \hat{U}_1^T.
\end{equation}
Diagonalizing the resulting $\hat{H}(\kvec)$ yields the eigenspectrum and the eigenfunctions of the bulk states. The spectrum is doubly degenerate as a consequence of the combined action of IS and TRS. The energies of the bulk valence $(-)$ and conduction $(+)$ bands can be expressed as:
\begin{equation}
\mathcal{E}_\mathrm{B}^\pm(\kvec) = h_0(\kvec) \pm \sqrt{\sum_{i=1}^5h_i^2(\kvec)}.
\end{equation}

\begin{figure}
\centering
\includegraphics[scale=0.135]{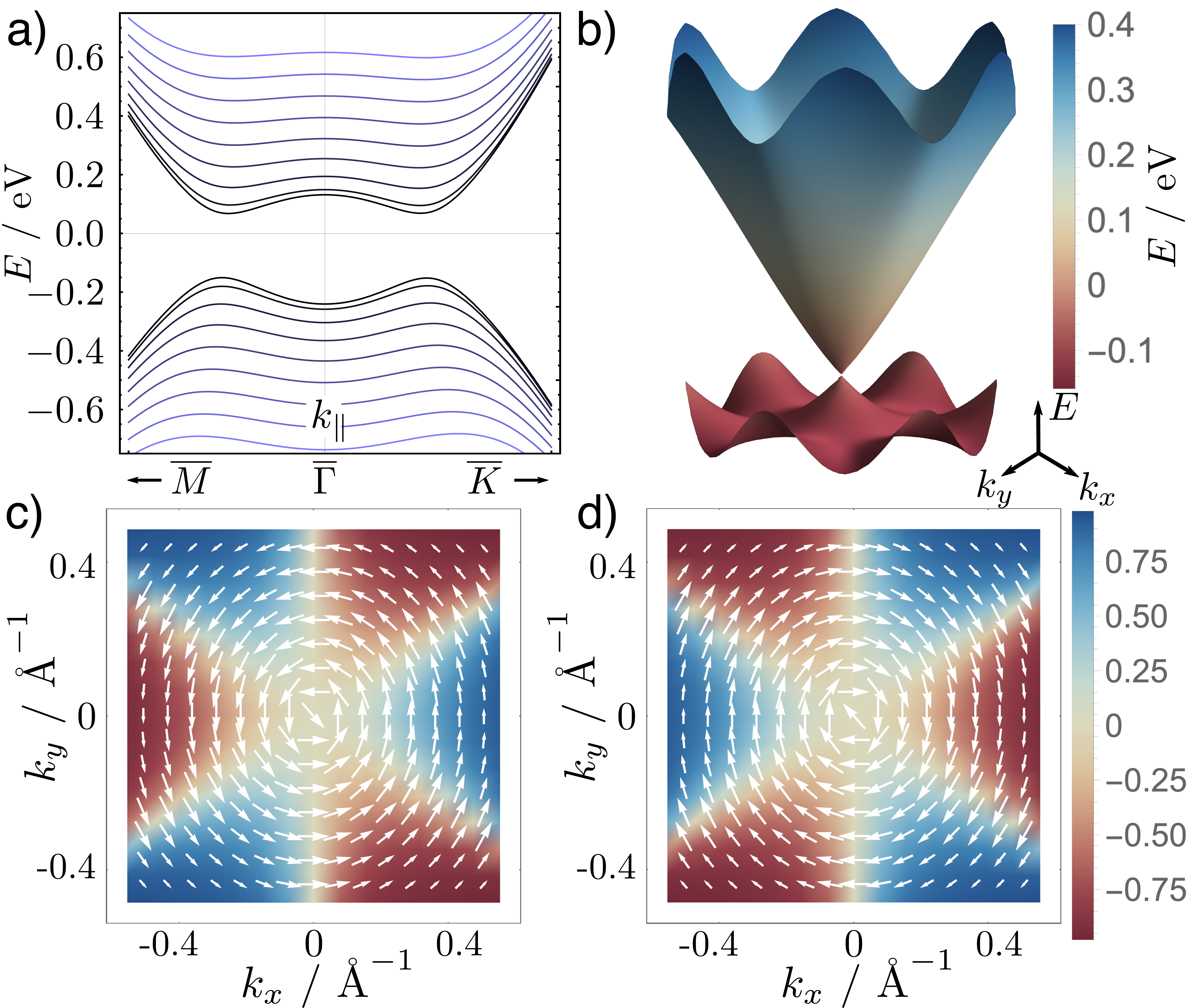}
\caption{a) Energy dispersion of the bulk states $E_\mathrm{B}^\pm(k_x,k_y,k_z)$ along the high-symmatry directions $\overline{\Gamma M}$ and $\overline{\Gamma K}$ in inverse space, shown for different $k_z$ values. The black lines correspond to the plane $k_z=0$, projected band dispersion curves pertaining to increasing $k_z$ are given in progressively lighter blue colors, whereby the $k_z$ is varied by $\Delta k_z=1.5\cdot10^{-3}$~\AA$^{-1}$. The abscissa covers the range from $k_x=0$ to $k_x = 0.36$~$\mathrm{\AA}^{-1}$ ($\Gamma K$-direction) and  from $k_y=0$ to $k_y = 0.31$~$\mathrm{\AA}^{-1}$ ($\Gamma M$-direction).
b) Energy dispersion of the surface modes $E_{2D}^\mathrm{S; (\pm)}(k_x,k_y)$ resulting from the employed TBM model, given in the disk defined by $k_\parallel \le 0.4$~\AA$^{-1}$. Near the $\overline{\Gamma}$ point, the dispersion is nearly linear. At higher momenta, the effect ``hexagonal warping'' is seen as a consequence of the higher-order contributions. c) and d) Spin polarization of the surface lower (c) and upper (d) Dirac cones over a selected portion of the BZ. The white arrows indicate the magnitude and the direction of the \textit{in-plane} polarization, whereas the color coding corresponds to the magnitude of the spin polarization in $\hat{z}$-direction (\textit{out-of-plane}). We note that the Berry curvature in momentum space follows a similar pattern as the \textit{out-of-plane} spin polarization. }\label{fig:bandstr}
\end{figure}

The wavefunctions are doubly degenerate as well, and the two eigenvectors (spinors), labelled by $\nu=\{1,2\}$, have the form:
\begin{equation}\label{eq:psiB1}
\psi_{\mathrm{B},\nu=1}^\pm(\kvec) =\mathcal{N}^{\pm}_\mathrm{B}(\kvec) f_{1\kvec} \begin{pmatrix}
-i\left(h_5(\kvec)\pm\sqrt{\sum_{i=1}^5 h_i^2(\kvec)}\right)\\
-\left(h_3(\kvec)+ih_4(\kvec)\right) \\
0\\
h_1(\kvec)+ih_2(\kvec)
\end{pmatrix}
\end{equation}

and:

\begin{equation}\label{eq:psiB2}
\psi_{\mathrm{B},\nu=2}^\pm(\kvec) =\mathcal{N}^{\pm}_\mathrm{B}(\kvec)f_{2\kvec} \begin{pmatrix}
0 \\
h_1(\kvec)-ih_2(\kvec) \\
-i\left(h_5(\kvec)\pm\sqrt{\sum_{i=1}^5 h_i^2(\kvec)}\right)\\
h_3(\kvec)-ih_4(\kvec) 
\end{pmatrix}.
\end{equation}

In the above, $\mathcal{N}^{\pm}_\mathrm{B}(\kvec)$ is a normalization constant given in Eq.~(\ref{eq:Nb}), and $f_{\nu\kvec} = \left(h_1(\kvec)+i(-1)^{\nu-1}h_2(\kvec)\right)^{-1}$. By construction, the spinors $\psi_{\mathrm{B},\nu=1,2}^{\pm}(\kvec)$ form a Kramers doublet, i.e. they are related by a time reversal operation ($\hat{\mathcal{T}} = i(\hat{\sigma}_y\otimes\hat{\mathbb{1}}_2)\hat{\mathcal{K}}$ with $\hat{\mathcal{K}}$ being the complex-conjugation operator): $\psi_{\mathrm{B},\nu=1}(\kvec) = -i(\hat{\sigma}_y\otimes\hat{\mathbb{1}}_2)\psi^\ast_{\mathrm{B},\nu=2}(-\kvec)$. TRS has profound implications for the physics of topological insulators. By virtue of the Kramers theorem, no time-reversal-invariant perturbation can induce gap-opening at the surface Dirac cone~\cite{Zhang2009}.\\

The TBM parameters used in the subsequent calculations are listed in Tab.~\ref{tab:tbm}, whereas Fig.~\ref{fig:bandstr} depicts the resulting band dispersions $\mathcal{E}_\mathrm{B}^\pm(\kvec)$ along the parallel momentum $\kpar$ for selected values of $k_z$. Note that for simplicity, we neglect the interlayer spin-flip hopping, i.e. we set $B_{14}=0$.\\

\subsection{Surface states}\label{sec:tbm_surf}

For the purposes of describing the topological surface states, we first derive an effective two-band Hamiltonian ($\hat{H}_{2D}^\mathrm{S}(\kpar)$), based on the generic $4\times4$-TBM-Hamiltonian $\hat{H}(\kvec)$ in Eq.~\ref{eq:Hk3}. The detailed procedure, adapted from Ref.~\cite{Liu2010}, is outlined in the Appendix~\ref{app:surf_deriv}. At this place, we restrict the discussion to a brief recapitulation of the main steps. The point of departure in our \textit{ansatz} is to impose open boundary conditions onto the Hamiltonian in Eq.~\ref{eq:Hk3} by postulating that the surface state wavefunction $\Psi_\mathrm{Surf}(\kpar, z)$ vanishes at the crystal-vacuum interface (defined as $z=0$ in Fig.~\ref{fig:crystalBZ}) and decays exponentially into the bulk for $z\rightarrow -\infty$, as illustrated by the red-shaded surface in Fig.~\ref{fig:crystalBZ}~a. With the aid of this procedure, we construct a general Hamiltonian ($\hat{H}^{(2)}(\kvec)$, cp.\ Eq.~(\ref{eq:Hk3Taylor})) describing both bulk and surface states and use its low-energy limit to obtain expressions for the surface state wavefunctions at the zero-energy Dirac point ($\overline{\Gamma}$-point in the 2D BZ). Due to the presence of spin and orbital degrees of freedom, the latter are degenerate at this special point of the BZ. We then split the general Hamiltonian $\hat{H}^{(2)}(\kvec)$ into one part dependent on the in-plane momentum coordinates ($\kpar = (k_x, k_y)^T$) and another term $\hat{H}^{(2)}(\kpar=\bm{0},\bm{k}_\perp)$ independent on $\kpar$. This approach can be regarded as doing degenerate perturbation theory in terms of the \textit{in-plane} momentum $\kpar$, whereby the parallel perturbation Hamiltonian is then projected onto the basis of the degenerate ground states, yielding the effective Hamiltonian for the states localized near the surface. In the end, following the steps presented in Appendix~\ref{app:surf_deriv}, the full expression for the effective Hamiltonian, corrected for the energy of the unperturbed states, and its spectrum, are given by:

\begin{widetext}
\begin{equation}\label{eq:HS2D}
\hat{H}_{2D}^{S}(\kpar) = \left(h_0^{z_0}(\kpar)+\frac{B_0\left(-h_5^{z_0}(\kpar)+h_5^\Gamma\right)}{B_{11}} \right) 
\hat{\mathbb{1}}_2 + \sqrt{1-\frac{B_0^2}{B_{11}^2}} \left[ h_1^{z_0}(\kpar) \hat{\sigma}_x + h_2^{z_0}(\kpar) \hat{\sigma}_y  + h_3^{z_0}(\kpar) \hat{\sigma}_z 
\right]
\end{equation}
and:
\begin{equation}\label{eq:E2D}
\mathcal{E}_{2D}^\mathrm{S;(\pm)}(\kpar) = 6A_0+h_0^{z_0}(\kpar)-h_0^\Gamma + \frac{B_0}{B_{11}}\left(6B_{11}-h_5^{z_0}+h_5^\Gamma\right)
\pm \frac{\sqrt{(-B_0^2+B_{11}^2) \sum_{i=1}^3\left(h_i^{z_0}(\kpar)\right)^2}}{B_{11}},
\end{equation}
\end{widetext}
where the sign $\pm$ corresponds to the lower ($-$) or the upper ($+$) Dirac cones, respectively. The quantities $h_i^{z_0}(\kpar)$ and $h_i^\Gamma$ are defined in Appendix~\ref{app:surf_deriv}~\ref{sec:H2D}, whereas $A_{ij}$ and $B_{ij}$ are the TBM parameters with values listed in Tab.~\ref{tab:tbm} in Appendix~\ref{app:tbm}.  $\hat{\sigma}_{x,y,z}$ are the conventional Pauli matrices operating in real spin space. The surface mode eigenstates (defined over the entire 2D surface Brillouin zone) read:
\begin{equation}\label{eq:psiS}
\psi_\mathrm{S}^\pm (\kpar) = \mathcal{N}^\pm_\mathrm{S}(\kpar)f_{\kpar}\begin{pmatrix}
h_3^{z_0}(\kpar)\pm\sqrt{\sum_{i=1}^3\left(h_i^{z_0}(\kpar)\right)^2} \\
h_1^{z_0}(\kpar) + ih_2^{z_0}(\kpar) 
\end{pmatrix}
\end{equation}
with $f_{\kpar} = \left(h_1^{z_0}(\kpar)+ih_2^{z_0}(\kpar)\right)^{-1}$ and a normalization constant $\mathcal{N}_\mathrm{S}^\pm(\kpar)$ defined in Eq.~\ref{eq:Ns}.\\

\subsection{Spin polarization}\label{sec:spinpol}

In the following, we briefly examine the spin structure of the surface modes derived in Sec.~\ref{sec:tbm_surf}. The spin polarization of the TSSs is calculated by evaluating the expectation values of the Pauli matrices $\{\hat{\sigma}_i\}$ (with $i=\{1,2,3\}$ corresponding to the axes $\{x,y,z\}$) over the eigenmodes $\psi_\mathrm{S}^\pm (\kpar)$:
\begin{eqnarray}\label{eq:spin_exp}
\left\langle\hat{\sigma}_i\right\rangle_{\pm} & \equiv & \left\langle\psi_\mathrm{S}^\pm\left|\hat{\sigma}_i\right|\psi_\mathrm{S}^\pm\right\rangle = \pm\frac{h_i^{(z_0)}(\kpar)}{\sqrt{\sum_{j=1}^3 \left(h_j^{(z_0)}(\kpar)\right)^2}}.
\end{eqnarray}

The spin polarization $\left(\left\langle\hat{\sigma}_x\right\rangle_{\pm}, \left\langle\hat{\sigma}_y\right\rangle_{\pm}, \left\langle\hat{\sigma}_z\right\rangle_{\pm}\right)^T$ pertaining to the lower and upper Dirac cones is displayed as a vector density plot in panels~c and~d of Fig.~\ref{fig:bandstr}. On the basis of these results, one can deduce that the TBM model employed here recovers the theoretically~\cite{Liu2010} and experimentally~\cite{Wang2011} established characteristic that for low momenta, the spin polarization of the surface states is predominantly \textit{in-plane}, whereby spin and momentum are ``locked'' such that the spin is always perpendicular to the in-plane momentum $\kpar$. At high momenta, a significant \textit{out-of-plane} spin component (i.e. in the $\hat{z}$-direction) develops as a result of the hexagonal warping. 

Within the framework of the TBM presented in Sec.~\ref{sec:tbm_bulk}, this observation can be quantitatively accounted for by examining the low-momentum limit of Eq.~(\ref{eq:spin_exp}). The spin polarization in momentum space and the resulting spin-momentum locking at low momenta are tied to the following terms in the 2D Hamiltonian:
\begin{equation}
\hat{H}_{2D}^\mathrm{S} \propto h_1^{z_0}(\kpar) \hat{\sigma}_x + h_2^{z_0}(\kpar)\hat{\sigma}_y + h_3^{z_0}(\kpar)\hat{\sigma}_z.
\end{equation} 
For low momenta, the functions $h_i^{z_0}(\kpar)$ can be expanded up to the third order ($\mathcal{O}(k_\parallel^3)$).
Taking into account the fact that in the adopted TBM parametrization we have neglected the inter-plane spin-flip transfer, i.e. $B_{14}=0$ (cp. Tab.~\ref{tab:tbm}), the asymptotic expressions simplify to:
\begin{eqnarray}
h_1^{z_0}(\kpar\rightarrow\bm{0}) & \sim &   3A_{14}a k_y - \frac{3}{8}A_{14}a^3k_yk_\parallel^2  \label{eq:h1TaylorFull} \\
h_2^{z_0}(\kpar\rightarrow\bm{0}) & \sim &  -3A_{14}a k_x + \frac{3}{8}A_{14}a^3k_x k_\parallel^2 \label{eq:h2TaylorFull} \\
h_3^{z_0}(\kpar\rightarrow\bm{0}) & \sim &  -\frac{1}{4}A_{12} a^3 k_x (k_x^2-3k_y^2), \label{eq:h3TaylorFull}
\end{eqnarray}
where $k_\parallel=\sqrt{k_x^2+k_y^2}$.\\

From the above, it follows that $h_1^{z_0}(\kpar) \hat{\sigma}_x + h_2^{z_0}(\kpar)\hat{\sigma}_y \propto A_{14} a (k_y\hat{\sigma}_x-k_x\hat{\sigma}_y)$ for very small $\kpar$. In contrast, the \textit{out-of-plane} component  gains in importance only at higher momenta as $h_3^{z_0}(\kpar)$ is of third order in $\kpar$ according to Eq.~(\ref{eq:h3TaylorFull}). Further, the magnitude of the \textit{in-plane} spin polarization is controlled by the intralayer spin-flip parameter $A_{14}$, whereas the polarization in $\hat{z}$-direction is proportional to the intralayer hopping $A_{12}$. In Sec.~\ref{sec:asymptotic}, the interconnection between the hopping constants $A_{12}$ and $A_{14}$ and the strength of the spin-orbit interaction will be revisited again in the context of the optical response of the surface states to intense circularly polarized laser fields.\\


\section{Semiconductor Bloch equations}\label{sec:el_dyn}

The microscopic interaction of the intense MIR laser fields with the bulk (Sec.~\ref{sec:tbm_bulk}) and the surface (Sec.~\ref{sec:tbm_surf}) states is solved within the framework of the semiconductor Bloch equations (SBEs) in the basis of ``accelerated'' Bloch functions, closely following previous works~\cite{Kira2011,Schubert2014,Hohenleutner2015, Luu2016,Li2019, Yue2020}.  We solve the SBEs for the time-dependent populations $\rho^{\Kvec}_{mm}(t)$ and  coherences ($\rho^{\Kvec}_{mm'}(t), m\neq m'$), which are explicitly propagated according to:
\begin{eqnarray}
\dot{\rho}^{\Kvec}_{m'm}(t) & = & i\Big[\Delta \mathcal{E}_{m'm}(\Kvec+\At) \label{eq:rho_mpm}  \\
& +&  \Et \cdot\Delta\xivec_{m'm}(\Kvec+\At) + \frac{i}{T_2}\Big]\rho^{\Kvec}_{m'm}(t) \nonumber \\
&+& i\sum_{m''\neq m'}
\Et\cdot \dvec_{m'm''}^\ast(\Kvec+\At) \rho^{\Kvec}_{m''m}(t)\nonumber \\
& - &  i\sum_{m''\neq m}
\Et\cdot \dvec_{mm''}(\Kvec+\At) \rho^{\Kvec}_{m'm''}(t) \nonumber \\
\dot{\rho}^{\Kvec}_{mm}(t) & = & \label{eq:rho_mm} \\
  -  2 \mathcal{I}m & \Big\{ & \sum_{m''\neq m}  \Et \cdot \dvec_{mm''}^\ast(\Kvec+\At)\rho^{\Kvec}_{m''m}(t)\Big\}.\nonumber
\end{eqnarray}
Decoherence  due to scattering effects has been taken into account via the phenomenological dephasing constant $T_2$. The index $m$ runs over the number of bands, $\Delta \mathcal{E}_{m'm}(\kvec)$ is the difference between the energies of the bands $m'$ and $m$: $\Delta \mathcal{E}_{m'm}(\kvec) = \mathcal{E}_{m'}(\kvec)-\mathcal{E}_m(\kvec)$, where $\mathcal{E}_m$ is either $\mathcal{E}_\mathrm{B}^\pm$ or $\mathcal{E}_{2D}^\mathrm{S; \pm}$. $\Delta\xivec_{m'm}(\kvec)$ denotes the difference between the corresponding Berry connections $\xivec_{mm}(\kvec)$. The latter are defined as: $\xivec_{mm}(\kvec) = i\bra{u_{\kvec,m}}{\bm{\nabla}}_{\kvec}\ket{u_{\kvec,m}}$ with $\ket{u_{\kvec,m}}$ being the periodic part of the Bloch wavefunction. $\dvec_{m'm}(\kvec)$ denotes the interband transition matrix element $i\bra{u_{\kvec,m'}}{\bm{\nabla}}_{\kvec}\ket{u_{\kvec,m}}, m'\neq m$, also referred to as ``non-Abelian Berry connection'' in earlier works~\cite{OliaeiMotlagh2018} on strong-field dynamics in TIs. $\kvec$ is the crystal momentum, whereas $\Kvec = \kvec-\At$ is the quasi-canonical crystal momentum in the presence of the vector potential $\At$ associated with the external laser electric field, defined as $\Et = -\partial_t \At$.  The laser electric field excites both intraband (${\bm{J}}_\mathrm{ra}(t)$) as well as interband (${\bm{J}}_\mathrm{er}(t)$) electron dynamics, which can be calculated from the time-dependent populations $\rho^{\Kvec}_{mm}(t)$ and coherences $\rho^{\Kvec}_{m'm}(t)$ in the following manner:
\begin{eqnarray}
{\bm{J}}_\mathrm{ra}(t) & = &\sum_m\int_{\overline{\mathrm{BZ}}}\vvec_m(\Kvec+\At)\rho^{\Kvec}_{mm}(t)\mathrm{d} \Kvec^2\label{eq:Jra}
\end{eqnarray}
and
\begin{eqnarray}
{\bm{J}}_\mathrm{er}(t) & = & \frac{\mathrm{d}}{\mathrm{d}t} \int_{\overline{\mathrm{BZ}}} \sum_{m'\neq m} \dvec_{m'm}(\Kvec+\At)\rho^{\Kvec}_{m'm}(t)\mathrm{d} \Kvec^2 \nonumber \\
& + &\mathrm{c.\ c.}\label{eq:Jer}
\end{eqnarray}

The band velocity $\vvec_m(\kvec) = \vvec_{\mathrm{gr},m}(\kvec) + \vvec_{\mathrm{an},m}(\kvec)$ in Eq.~(\ref{eq:Jra}) comprises both  the group velocity in absence of Berry curvature $\vvec_{\mathrm{gr},m}(\kvec) = {\bm{\nabla}}_{\kvec}\mathcal{E}_m(\kvec)$ as well as the contribution from the anomalous velocity $\vvec_{\mathrm{an},m}(\kvec) = -\Et\times{\bm{\Omega}}_m(\kvec)$, where ${\bm{\Omega}}_m(\kvec)$ is the Berry curvature of the band $m$. The Bloch functions $\ket{u_{m}(\kvec)}$ required for the evaluation of all matrix elements in Eqs.~(\ref{eq:rho_mpm}-\ref{eq:rho_mm}) are evaluated with the aid of the  eigenspinors derived from the TBM Hamiltonian, i.e. Eqs.~(\ref{eq:psiB1}-\ref{eq:psiB2}) and~(\ref{eq:psiS}). \\

Although all calculation results reported in the next Sections have been obtained by numerically propagating Eqs.~\eqref{eq:rho_mpm} and~\eqref{eq:rho_mm}, a physical insight can also be gained by writing the inter and intra-band currents in closed form using the approximation $\rho_{vv}^\Kvec - \rho^\Kvec_{cc} \approx 1$~\cite{Vampa2014, Chacon2018}. Thereby, the subscripts $c$ and $v$ pertain to either the conduction and valence bands (BSs), or  the upper and lower Dirac cones (TSSs). In this way, we can decouple Eqs.~\eqref{eq:rho_mpm} and~\eqref{eq:rho_mm}, and the $i$\textsuperscript{th} vectorial-component ($i=x,\,y$) of intraband current contribution can be evaluated as: 
\vspace{-.15cm}
\begin{align}
J^{(i)}_\mathrm{ra}(t) &=&  \sum_{m} \int_{\rm \overline{BZ}} \mathrm{d} {\Kvec}^2 \,v^{(i)}_m\left({{\Kvec}}+ \At\right) \, \rho_{mm}^{\Kvec}(t),
\label{eqn:intra2}
\end{align}

\noindent where the occupation of the $m^{\rm th}$ state, $\rho_{mm}^{\Kvec}(t)$, is given by:
\begin{eqnarray}
\rho_{mm}^{\Kvec}(t)&=&(-1)^m\sum_{j,k} \int_{t_0}^{t'} \mathrm
{d}t' \,E^{(k)}(t')\left\rvert d^{(k)}_{cv}({\Kvec} + {\bm{A}_\mathrm{MIR}}(t')) \right\rvert \nonumber \\ &
\times & \int_{t_0}^t \mathrm{d}t'' \,E^{(j)}(t'')\left\rvert d^{(j)}_{cv}({\Kvec} + {\bm{A}_\mathrm{MIR}}(t''))\right\rvert 
\label{eqn:occup}\\ 
\hspace{-1.25cm}&\times &\,e^{-i S^{(j)}({\Kvec},t',t'') - \frac{t'-t''}{T_2}+ i\left(\varphi_{cv}^{(j)}(\Kvec, t')-\varphi_{cv}^{(k)}(\Kvec, t')\right)} 
+ \mathrm{c.c.}
\nonumber
\end{eqnarray}
For the interband current, we have:
\begin{eqnarray}
J^{(i)}_\mathrm{er}(t) 
 & = & 
-i\sum_{j}{\frac{\mathrm{d}}{\mathrm{d}t}} 
\int_{t_0}^{t} \mathrm{d}t' 
\int_{\rm  \overline{BZ}} \mathrm{d}  {\Kvec}^2\, 
\left\lvert {d}^{(i)}_{cv}\left({\Kvec}+ {\bm{A}_\mathrm{MIR}}(t)\right)\right\rvert \nonumber\\
& \times & \left\lvert d^{(j)}_{cv}\left({\Kvec}+ {\bm{A}_\mathrm{MIR}}(t')\right)\right\rvert E^{(j)}(t') 
\qquad 
\label{eqn:inter2} 
\\ \nonumber  
& \times & \,e^{-iS^{(j)}({\Kvec},t,t')-(t-t')/T_2 + i\left(\varphi_{cv}^{(j)}({\Kvec},t)-\varphi_{cv}^{(i)}({\Kvec},t)\right) }+\mathrm{c.c.},
\end{eqnarray}

\noindent where  $\scriptstyle S^{(j)}({\Kvec},t,t') = \int_{t'}^{t}\left[\Delta \mathcal{E}_{cv}({\Kvec}, (t'')) +  {{\EvecMIR}}(t'')\cdot{{\bm{ \mathcal D}}}^{(j)}_{cv}({\Kvec},t'') \right]\mathrm{d}t''$ is the electron-hole pair accumulation phase between the birth event $t'$ and the emission event $t$. We define ${\bm{ \mathcal D}}^{(j)}_{cv}({\kvec}) = \Delta{\bm{ \xi}}_{cv}({\kvec}) + \bm{\nabla}_{\kvec} \varphi_{cv}^{(j)}({\kvec})$ as the covariant Berry connection, whereas  $\bm{\nabla}_{\kvec} \varphi_{cv}^{(j)}({\kvec})$ denotes the dipole phase derivative.
These terms  appear naturally in the acquired electron-hole pair phase and, together with the dipole amplitude ($\left\lvert d^{(j)}_{cv}\left({\kvec}\right)\right\rvert $), are coupled to the MIR driving field and hence dictate the ways in which electronic structure features are encoded in the HHG spectra. In Sec.~\ref{sec:mechanisms}, we elaborate on the details of this coupling in the case of the TSSs. The interband dipole matrix element between upper and lower Dirac cones is displayed in terms of its real and imaginary parts in panels a and b of Fig.~\ref{fig:dipole_vortex}. In Fig.~\ref{fig:app_vortex} of Appendix~\ref{app:vortex}, we present the Berry connection difference $\Delta{\bm{ \xi}}_{cv}({\kpar})$ (panel a) as well the gradients of the dipole phase (panels b and c). The fact that the magnitude of the phase gradients $\bm{\nabla}_{\kpar} \varphi_{cv}^{(j)}({\kpar})$ exceeds $\Delta{\bm{ \xi}}_{cv}({\kpar})$ (cp. Fig.~\ref{fig:app_vortex}), together with the strongly enhanced magnitude of the dipole around the $\overline{\Gamma}$-point due to the singularity, implies that the transition dipole has the predominant influence on the HHG spectra.   Finally, we stress that the quantity ${\bm{ \mathcal D}}^{(j)}_{cv}({\kvec})$, as well as the total (intra and inter-band) currents, are Bloch-wavefunction-gauge-invariant~\cite{Chacon2018}. 
 \\

In the above, the integration is performed over the shifted Brilloiun zone $\overline{\mathrm{BZ}} = \mathrm{BZ}-\At$.  We consider a laser field normally incident on the (111)-surface of the TI-system (rhombohedral convention) $x-y$-plane, cp. Fig.~\ref{fig:crystalBZ}). Under the assumption that the initiated electron dynamics is confined to the incident plane, we restrict the momentum space integration in Eqs.~(\ref{eq:Jra}-\ref{eq:Jer}) to two dimensions ($k_x, k_y$). Whereas the motion of the surface-state electrons is confined to \textit{in-plane} momenta by construction, for the bulk states,  where the  band structure is inherently three-dimensional, this approximation implies that the analysis is restricted to the $(k_x,k_y, k_z=0)$-time-reversal-invariant plane. Extending the BZ integration to include the $k_z$-direction would require extensive computational resources that are beyond the capacity currently at our disposal.\\

In the remaining Sections, we study the strong-field dynamics of bulk and surface states separately. The 2D Hamiltonian for the TSSs in Eq.~(\ref{eq:HS2D}) yields two bands corresponding to the lower ($-$) and the upper ($+$) Dirac cones. As inversion symmetry is broken at the surface, the Berry curvature of the surface bands possesses a non-vanishing component in the $z$-direction:
\begin{equation}
\Omega_{\pm}^\mathrm{(S)}(\kpar) = i\bra{ {\bm{\nabla}}_{\kpar}\psi_\pm^\mathrm{(S)}(\kpar)}\times
\ket{{\bm{\nabla}}_{\kpar}\psi_\pm^\mathrm{(S)}(\kpar)}.
\end{equation}
The $4\times4$-Hamiltonian describing the bulk band structure in Eq.~(\ref{eq:Htbm_h}) yields two pairs of degenerate bands corresponding to the valence ($-$) and conduction ($+$) bands, separated by a band gap of $0.37$~eV at the $\Gamma$-point (for comparison, the experimental band gap is reported as $\approx0.3$~eV). Hence, we propagate the full $4\times4$ density matrix resulting from Eqs.~(\ref{eq:rho_mpm})-(\ref{eq:rho_mm}). The degenerate nature of the bands arising from the combination of TRS and IS has profound consequences for the Berry curvature and the anomalous velocity. In the presences of degeneracies, the definition of the Berry curvature has to be extended to a tensor definition~\cite{Xiao2010, Yang2014,Shindou2005, Gradhand2012} by the covariant derivatives:
\begin{eqnarray}
& & \left({\bm{\Omega}}_m^{\mathrm{(B)}}(\kvec)\right)_{ij}=i\left\langle{\bm{\nabla}}_{\kvec} u_m^{(i)}(\kvec)\right|\times \left|{\bm{\nabla}}_{\kvec}u_m^{(j)}(\kvec)\right\rangle\label{eq:berry_nonAbel}\\
& &  - i \sum_{l=1}^2\left\langle{\bm{\nabla}}_{\kvec}u_m^{(i)}(\kvec) \right|\left. u_m^{(l)}({\kvec})\right\rangle \times \left\langle
u_m^{(l)}(\kvec) \right|\left. {\bm{\nabla}}_{\kvec}u_m^{(j)}(\kvec)\right\rangle, \nonumber
\end{eqnarray}
where the indices $i,j$ run over the degenerate components. The anomalous current is proportional to the trace of this tensor, i.e. $\vvec_{\mathrm{an},m}(\kvec) \propto \mathrm{Tr}\left\{ \left({\bm{\Omega}}_m^{\mathrm{(B)}}(\kvec)\right)_{ij} \right\}$, which forms a gauge-invariant quantity. For the bulk bands, the trace evaluates to zero as $\left({\bm{\Omega}}_m^{\mathrm{(B)}}(\kvec)\right)_{ii} = -\left({\bm{\Omega}}_m^{\mathrm{(B)}}(\kvec)\right)_{jj}$ owing to IS and TRS.\\

\section{HHG results}\label{sec:results}




\begin{figure}
\begin{center}
\includegraphics[scale=0.4]{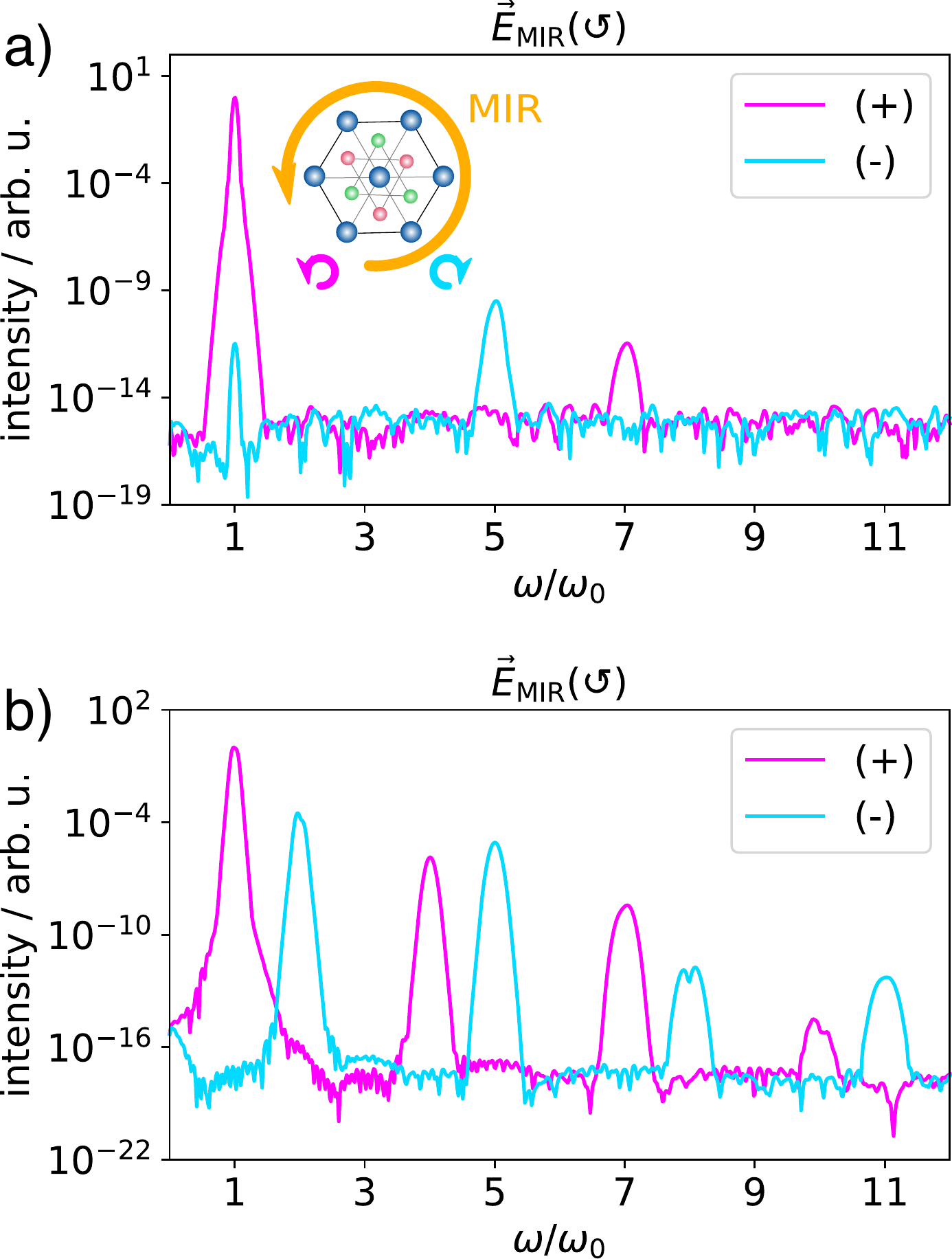}
\caption{HHG spectra of Bi$_2$Se$_3$ driven by circularly polarized fields, for both bulk (a) and surface states (b). The relative orientation of the (111)-surface (real space) and the MIR polarization vector is sketched in the inset, whereby the propagation direction points towards the reader.  The MIR pulse is left-circularly polarized with  $I_0 = 0.0025$~$\mathrm{TW/cm^2}$ and a FWHM duration of 12 cycles. The helicity of the emitted harmonics is encoded in the color: \textit{left, co-rotating} orders are shown in magenta, \textit{right, counter-rotating}  in cyan. Bulk and surface states obey different selection rules: $\omega=(6n\pm1)\omega_0$ for the bulk (a) vs.  $\omega=(3n\pm1)\omega_0$ for the surface (b). In both calculations, the dephasing time is set at $T_2=1.25$~fs.}\label{fig:lin_circ_spectrum}
\end{center}
\end{figure}


	
\subsection{Recovery of selection rules for circular polarization}\label{sec:res_cpl}

High-order harmonic spectra from the bulk and surface states driven by a left circularly polarized MIR laser field with a center wavelength of $\lambda_\mathrm{MIR}=7.5$~$\mathrm{\mu m}$ are shown in panels a and b of Fig.~\ref{fig:lin_circ_spectrum}), respectively. In order to rationalize the observed  spectral features, we discuss in detail the selection rules based on dynamical symmetry analysis in Appendix~\ref{app:dyn_sym}. Essentially,  the three-fold crystal symmetry ($\hat{\mathcal{R}}_3^{(z)}$) precludes emission of every third harmonic multiple of the fundamental frequency. Indeed, for both bulk and surface states, harmonic orders (HOs) 3, 6, and 9 are missing, as evident from panels a and b. For the surface states, the the selesction rule for ``allowed'' harmonic orders reads $\omega = \left(3n\pm1\right)\omega_0$, with $n\in \mathcal{N}$ and $\omega_0$ being the driving frequency. In addition, the $(3n+1)$-th orders  are co-rotating, whereas the $(3n-1)$-th orders are counter-rotating with respect to the helicity of the laser field. In our results, magenta-color (HOs 4, 7, and 10) represents co-rotating and cyan-color (HOs 2, 4, 8, and 11) represents counter-rotating harmonics, respectively.  The presence of inversion symmetry in the bulk, as discussed in the preceding Sec.~\ref{sec:crystal}, precludes additional even-order harmonics, leading to a more restrictive selection rule: $\omega = \left(6n\pm1\right)\omega_0$. This is also consistent to our observation in panel a. We note that selection rules  for harmonic generation in circularly polarized fields were first derived  within a perturbative analysis in Ref.~\cite{Tang1971} and verified experimentally for the non-perturbative regime in Ref.~\cite{Saito2017}. Having established the  essential selection rules for fully circular MIR fields, we move to the more general case of elliptical polarization and focus on how the harmonic yield changes as we vary the laser ellipticity in small steps.

\subsection{Non-trivial ellipticity dependence}\label{sec:anom_ell}

\begin{figure}[htp]
\centering
\includegraphics[scale=0.275]{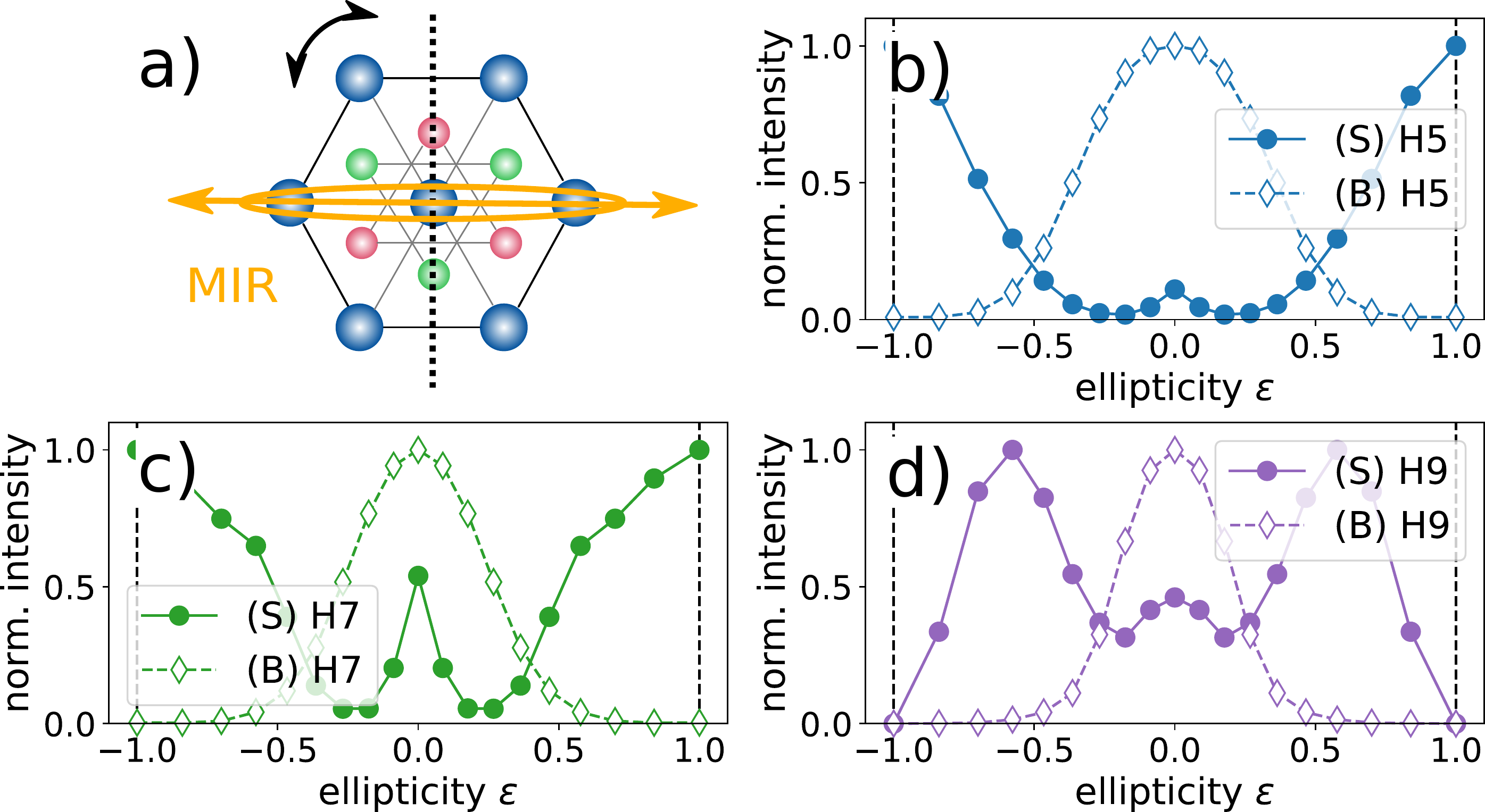}
\caption{a) Sketch of the excitation geometry, where the major axis of the MIR  ellipse (orange) remains  perpendicular to the mirror axis (dashed line) of the crystal throughout the measurement. Panels b) - d) show the calculated ellipticity response for harmonic orders $5$, $7$, and $9$. The calculations (12-cycle Gaussian pulse, $I_0=0.0085$~TW/cm$^2$, $\lambda_\mathrm{MIR}$=7.5~$\mathrm{\mu m}$, $T_2=1.25$~fs) pertain to the contributions from the bulk  ((B), dashed lines, diamonds) and from the surface states ((S), solid lines, circles/squares). }
\label{fig:ell75micron}
\end{figure}

We select  few representative harmonics and plot their total yields as a function of laser ellipticity, both for the bulk and surface states. During the ellipticity scan, the major axis of the ellipse is kept fixed along the $x$-axis ($\EvecMIR\parallel\overline{\Gamma K}$ in momentum space), as shown in the panel a of Figure~\ref{fig:ell75micron}. The intensity of the MIR driver ($\lambda_\mathrm{MIR}=7.5$~$\mathrm{\mu m}$) is $I_0=0.0085$~TW/cm$^2$. As evident from panels~b, c, and d, HOs 5, 7 and 9 from the bulk (represented by dashed lines with diamond symbols) exhibit a monotonic decrease as the ellipticity increases. We note that the ellipticity profiles are normalized with respect to their  maxima. A closer look to these profiles shows a decrease in FWHM as  the harmonic order increases from 5 to 9. These behaviors are similar to the ellipticity dependence in atomic HHG, as well as in a number of solid-state materials studied by HHG in recent years~\cite{Ghimire2011, Liu2017a}.\\

However, the ellipticity dependence  of HHG from the surface states is profoundly different. As it can be discerned from panels b, d, and f, HOs 5, 7, and 9 (represented by solid lines with filled circle symbols)  from the surface exhibit a substantial enhancement as the laser ellipticity increases. In particular,  HO~5 reaches a maximum at circular polarization, with a factor of $\sim$10 higher intensity with respect to the linear polarization case. Because this observation is in contrast with the manifestations of re-collision physics observed in atomic and molecular HHG, it calls for a detailed investigation. In order to track the origin of the non-trivial behaviour, we perform a detailed analysis of the characteristic quantities that govern the population dynamics in the lower and upper Dirac cones by virtue of Eqs.~(\ref{eqn:intra2}) and~\eqref{eqn:inter2}, i.e. the complex interband transition moments and the Berry connections~\cite{Chacon2018,Yue2020,Yue2020a}. 

\subsection{Mechanisms for  HHG in elliptical fields}\label{sec:mechanisms}

\subsubsection{Low-momentum limit: band topology}\label{sec:asymptotic}

\begin{figure}
\vspace{1cm}
\centering
\includegraphics[scale=0.275]{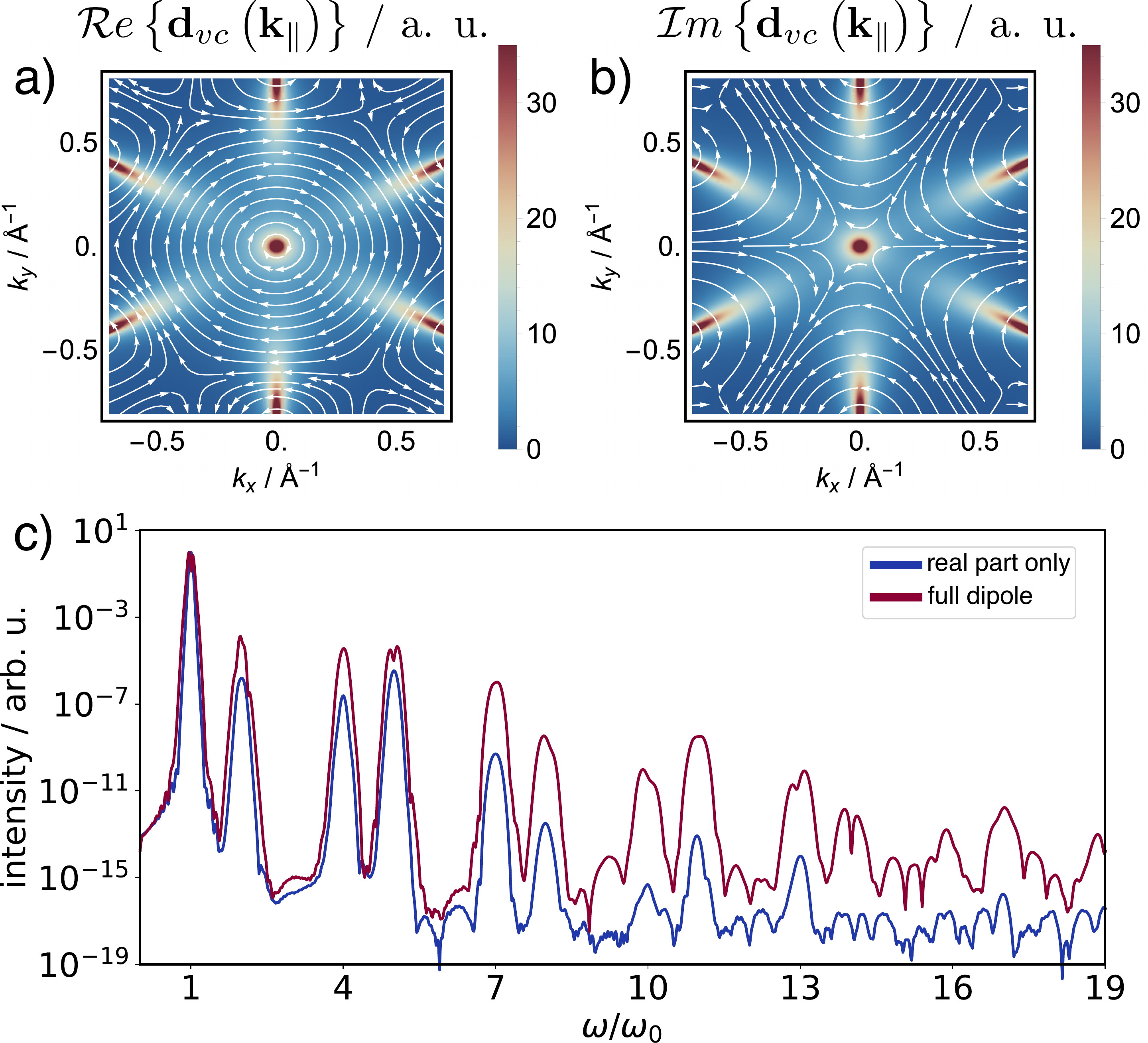}
\caption{ Panels a) and b): Real (a) and imaginary (b) parts of the interband transition matrix element $\dvec_{vc}(\kpar)$ between the surface valence and conduction bands. The streamlines (white) indicate the local direction of the vector fields in momentum space. Panel c): HHG spectra of the TSSs illuminated by a LCP MIR pulse with $I_0=0.0075$~$\mathrm{TW/cm^2}$, $\lambda_\mathrm{MIR}\sim7.5$~$\mathrm{\mu m}$, $T_2=1.25$~fs, and a duration of 12 cycles. The blue curve corresponds to a calculation which considers only the real part of the interband dipole $\dvec_{cv}(\kvec)$, i.e. leading term in Eqs.~(\ref{eq:dcvTaylorx}-\ref{eq:dcvTaylory}). The red curve pertains to the full expression for the dipole.}
\label{fig:dipole_vortex}
\end{figure}

In this section, we examine the low-momentum behavior of the interband transition dipole moment $\dvec_{cv}(\kpar)$. All characteristic quantities are associated to the surface states, unless otherwise noted explicitly. With a series expansion of  $\dvec_{cv}(\kpar)$ around the $\overline{\Gamma}$-point,  the elements of the interband transition dipole vector can be approximated as:
\begin{eqnarray}\label{eq:dcvTaylorx} 
d_{cv}^{(x)}(\kvec_\parallel) & \sim & \frac{{k_y} \left(a^2 k_{\parallel}^2-8\right)^2}{128 k_{\parallel}^2} \nonumber \\
& - & i \frac{A_{12}}{A_{14}}a^2 \frac{1}{1536 \left| k_{\parallel}\right| ^3} 
\Big[192 k_{\parallel}^2 ({k_x}-{k_y}) ({k_x}+{k_y})  \nonumber \\
& - &  {k_x}^2 \left(a^2 k_{\parallel}^2-8\right) \left(3 a^2 k_{\parallel}^2-8\right) \left({k_x}^2-3 {k_y}^2\right)\Big]
\end{eqnarray}
and:
\begin{eqnarray}
d_{cv}^{(y)}(\kvec_\parallel) & \sim & -\frac{{k_x} \left(a^2 k_{\parallel}^2-8\right)^2}{128 k_{\parallel}^2} \nonumber \\ 
& + & i \frac{A_{12}}{A_{14}}a^2 {k_x} {k_y} \frac{1}{1536 \left| k_{\parallel}\right| ^3} 
 \Big[384 k_{\parallel}^2   \nonumber \\
 & + & \left(a^2 k_{\parallel}^2-8\right) \left(3 a^2 k_{\parallel}^2-8\right) \left({k_x}^2-3 {k_y}^2\right)\Big]. \label{eq:dcvTaylory}
\end{eqnarray}
For very low momenta $\kpar$, the dominant terms are given by $d_{cv}^{(x)}(\kvec_\parallel) \propto \tfrac{1}{2k_\parallel^2}k_y$ and
$d_{cv}^{(y)}(\kvec_\parallel) \propto -\tfrac{1}{2k_\parallel^2}k_x$. This implies that the direction of the transition dipole is perpendicular to the electron crystal momentum $\kpar$, in a manner reminiscent of the ``spin-momentum'' locking, i.e. the orthogonal mutual orientation of the \textit{in-plane} spin polarization and $\kpar$ on the TI surface, discussed in Sec.~\ref{sec:spinpol} (cp. Eqs.~(\ref{eq:h1TaylorFull}-\ref{eq:h3TaylorFull}) therein) and visualized in Fig.~\ref{fig:bandstr}. The orientation of real part $\dvec_{cv}(\kpar\sim\overline{\Gamma})$ forms a chiral ``vortex'' feature, as evident from the plot in Fig.~\ref{fig:dipole_vortex}~a.\\

\begin{figure}
\centering
\includegraphics[scale=0.35]{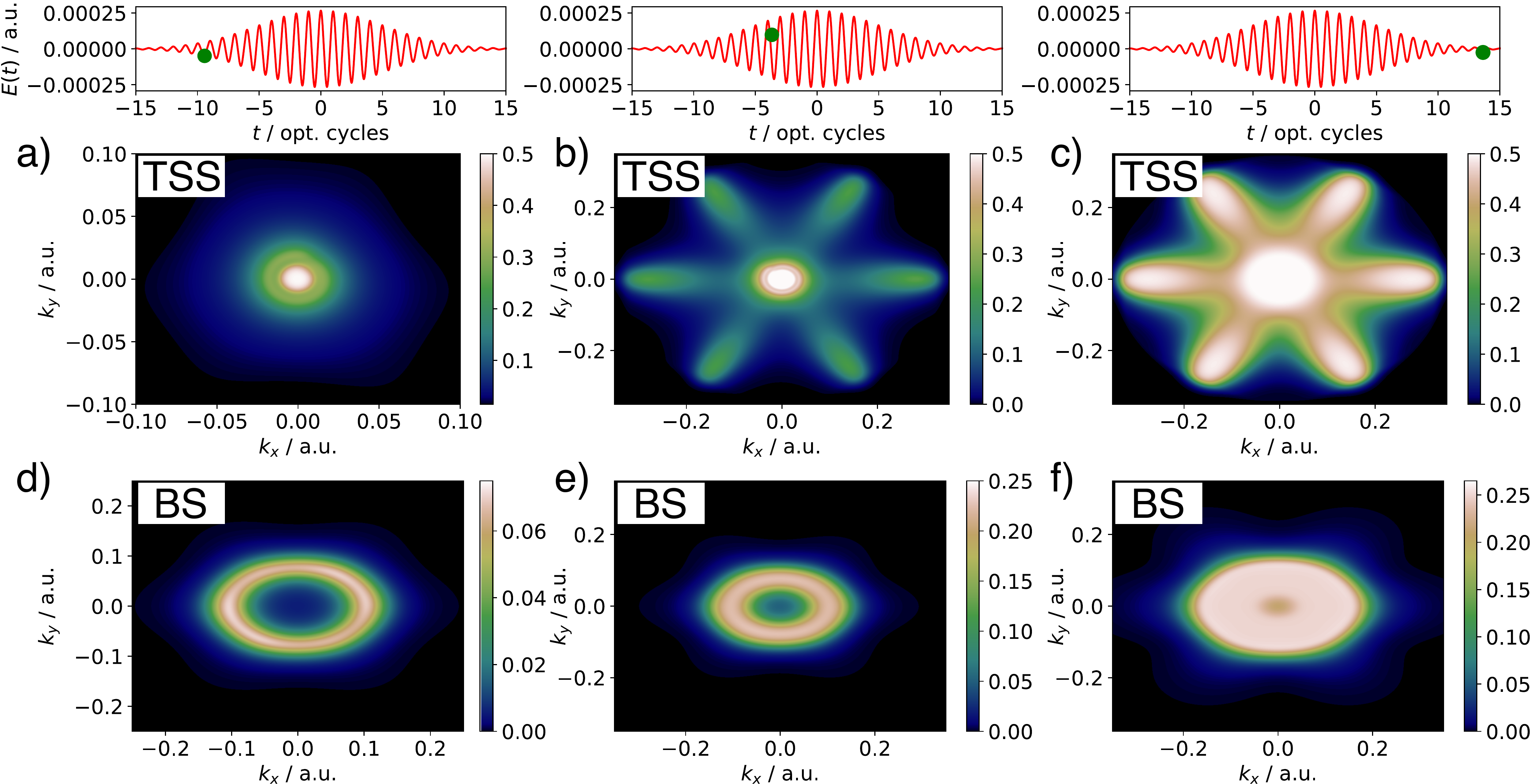}
\caption{\textit{Upper row}: Momentum-resolved temporal snapshot of the population distribution in the upper Dirac cone at different time points during the interaction with a LCP MIR driving field ($I_0=0.0025$~$\mathrm{TW/cm^2}$, 12 cycles), shown at three different time points (a, b, and c) of the pulse envelope. The top panels show the $x$-component of the electric field amplitude of the MIR driving pulse (in atomic units). \textit{Upper row}: Panels d), e), and f) show the corresponding population distribution for one of the degenerate components of the bulk conduction states ($\psi^+_{\mathrm{B},\nu =1}$) for the same conditions as for the TSSs. Note that $\rho_{cc}^{\kpar}\in [0,1]$ for the TSSs and  $\rho_{c(\nu=1)c(\nu=1)}^{\kpar} = \rho_{c(\nu=2)c(\nu=2)}^{\kpar}\in [0,1/2]$ for the BSs. }
\label{fig:momentum_snapshot}
\end{figure}

This last feature of the surface band topology, together with the strong localization  of the transition dipole magnitude in the vicinity of the $\overline{\Gamma}$-point, implies a pronounced sensitivity to the vectorial nature of the coupling to the external oscillating electromagnetic field. In particular, CPL driving fields couple more efficiently due to the non-vanishing $x$- and $y$-components of the instantaneous polarization vector.  This enhancement mechanism for low-momentum range is reminiscent to the one invoked to explain the non-trivial ellipticity response in graphene~\cite{Liu2018a}, where the HHG yield was found to maximize at finite ellipticities ($|\epsilon|\sim 0.32$~\cite{Yoshikawa2017}). In the case of 3D-TIs such as Bi$_2$Se$_3$, this mechanism precipitates the efficient generation of low-order harmonics (HO~$\le 5$) in highly elliptical fields. As the interband dipole-momentum-locking is mediated by the real part of $\dvec_{cv}(\kpar\sim\overline{\Gamma})$, this last statement can be verified by studying the effect of the imaginary part of the dipole vector on the emitted HHG. In panel~c of Fig.~\ref{fig:dipole_vortex}, we present HHG spectra of Bi$_2$Se$_3$ for a left-circularly polarized MIR field calculated including only the real part of $\dvec_{cv}(\kpar\sim\overline{\Gamma})$, i.e. the leading terms in Eqs.~(\ref{eq:dcvTaylorx}-\ref{eq:dcvTaylory}) (blue), and compare them to the results of a full calculation (red curve). The intensity of the low-order harmonics such as HO~5 remains only slightly affected compared to higher orders in the range from HO~11 to HO~19, implying that the chiral vorticity of the dipole vector is the dominant mechanism for HHG in this spectral range. In addition, the vortex feature in the case of the TSSs of Bi$_2$Se$_3$ leaves an imprint on the population dynamics in the upper band. The vorticity of the interband dipole $\dvec_{cv}(\kvec_\parallel)$ leads to the formation of a chiral vortex pattern in the electron population distribution, as evident from the momentum-resolved occupations of the upper Dirac cone depicted in Figs.~\ref{fig:momentum_snapshot}~a-c.  For comparison, in Fig.~\ref{fig:momentum_snapshot} (bottom row) we also show the corresponding population evolutions for the bulk states, which do not exhibit these chiral features. Consequently, we conclude that the ellipticity response is highly sensitive to the details of the topology of the Bloch bands, particularly near the Dirac cone. \\

\onecolumngrid

\twocolumngrid

\begin{widetext}

\end{widetext}
\newpage

\subsubsection{High momentum limit: hexagonal warping}\label{sec:hex_warp}

We now turn to the dynamics in the high-momentum regions of the BZ, which are governed predominantly by the imaginary part of $\dvec_{cv}(\kpar)$, as implied by the results presented in Fig.~\ref{fig:dipole_vortex}~c. In fact, the presence of imaginary components in $\dvec_{cv}(\kpar)$ is a characteristic feature of TSSs in 3D-TIs that is distinctly different than other gapless systems with linear dispersion, such as  graphene, as elaborated in Refs.~\cite{OliaeiMotlagh2017,OliaeiMotlagh2018}. We now show that this feature is mediated by the strong spin-orbit coupling in the 3D-TI system, and that it gives rise to the pronounced anomalous ellipticity behavior of the higher orders.\\

\begin{figure}[htp!]
\centering
\includegraphics[scale=0.34]{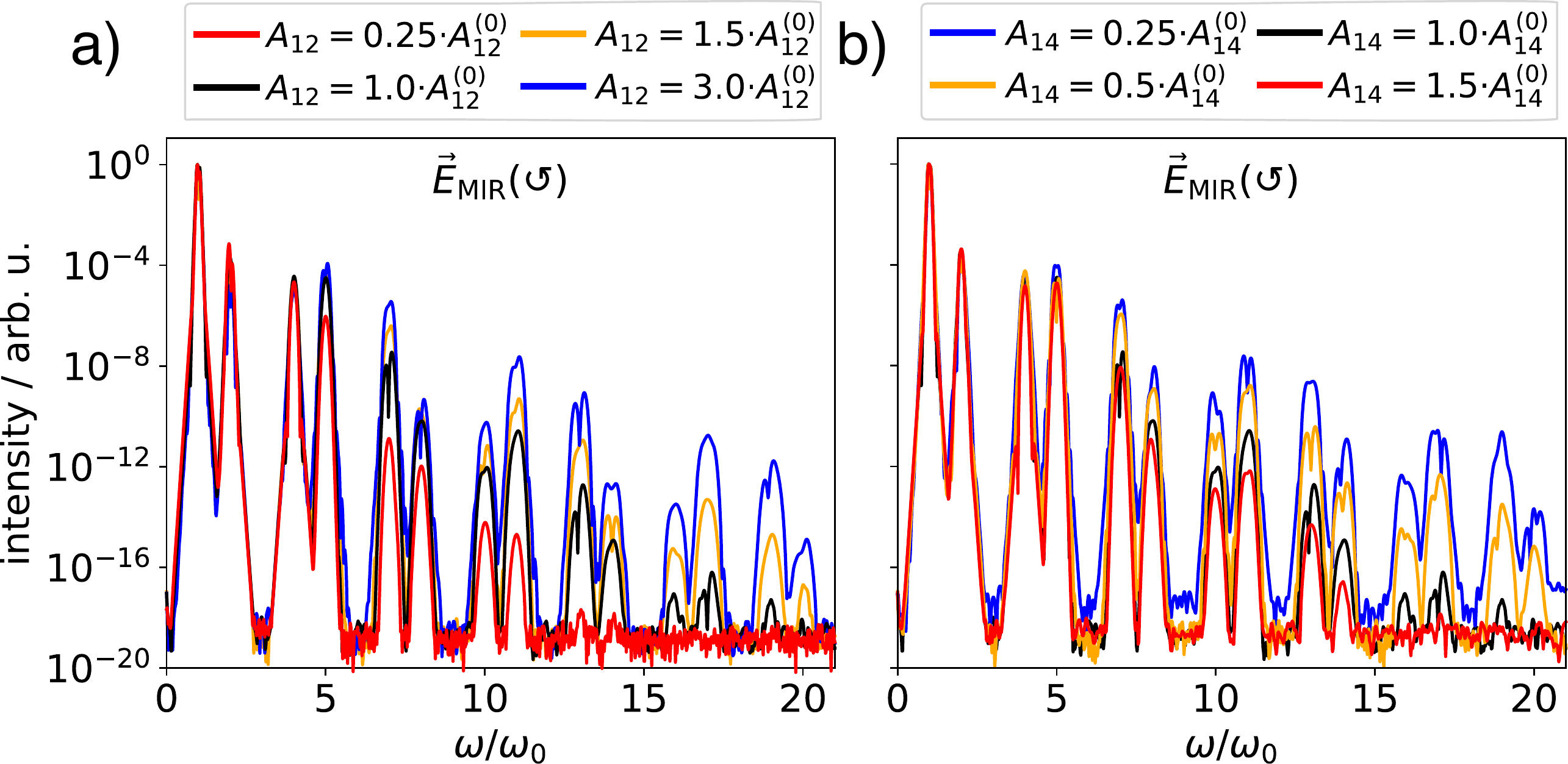}
\caption{ HHG spectra emitted from the surface states driven by a 12-cycle left-circularly polarized pulse with a peak intensity of $I_0=0.004\ \mathrm{TW\: cm}^{-2}$, whereby one of the TBM parameters $A_{12}$ (panel a) or $A_{14}$ (panel b) is varied (s. legend). The spectra corresponding to the parameters listed in Tab.~\ref{tab:tbm} are shown in black.  }
\label{fig:CPL_ratio_dep}
\end{figure}

Our analysis starts by noting that the higher-order (imaginary) component of $\dvec_{cv}(\kpar)$ is directly proportional to the ratio $\tfrac{A_{12}}{A_{14}}$, i.e. the TBM coefficients linked to the \textit{in-plane} spin polarization ($h_1^{z_0}(\kpar)\hat{\sigma}_x + h_2^{z_0}(\kpar)\sigmaM \sim 3A_{14}a\left(k_y\hat{\sigma}_x-k_x\sigmaM\right)$) and its \textit{out-of-plane} ($A_{12}$) component ($h_3^{z_0}(\kpar)\hat{\sigma}_z\sim -\tfrac{1}{8}a^3A_{12}(k_+^3+k_-^3)\hat{\sigma}_z$). The latter term coupled to $\hat{\sigma}_z$ is the analogon of the cubic Dresselhaus spin-orbit term in bulk rhombohedral structures, as noted in Ref.~\cite{Fu2009a}. This relationship reveals the sensitivity of the yield of higher-order harmonics to the details of the SOC parameters in a system with a strong SOI. Similar considerations apply to the Berry connections ${\bm{\xi}}_{mm}$ ($m=c,v$) as well. We consider directly the difference between the Berry connections of upper and lower Dirac cones $\Delta \xivec_{cv}(\kpar) = \xivec_{cc}(\kpar)-\xivec_{vv}(\kpar)$ that enters the SBEs in Eq.~(\ref{eq:rho_mpm}):
\begin{eqnarray}
\Delta\xi_{cv}^{(x)} & \sim & \frac{A_{12}}{A_{14}} a^2 \frac{ {k_x} {k_y} \left(a^2 k_\parallel^2-8\right)^2 \left({k_x}^2-3 {k_y}^2\right)}{768  \left| k_\parallel\right| ^3}	 \label{xiTaylorx} \\
\Delta\xi_{cv}^{(y)} & \sim & -\frac{A_{12}}{A_{14}} a^2 \frac{ {k_x}^2 \left(a^2 k_\parallel^2-8\right)^2 \left({k_x}^2-3 {k_y}^2\right)}{768  \left| k_\parallel\right| ^3} \label{xiTaylory}.
\end{eqnarray}
As in the case of interband dipole $\dvec_{cv}(\kpar)$, the magnitude of $\Delta\xivec_{cv}(\kpar)$ near $\overline{\Gamma}$ is controlled by the ratio $\tfrac{A_{12}}{A_{14}}$. The corresponding vector field plot revealing the vorticity of the Berry connection difference in the BZ are shown in Fig.~\ref{fig:app_vortex}~a. \\

To put the above considerations on a more quantitative basis, we next investigate the sensitivity of the HHG efficiency of Bi$_2$Se$_3$ in CPL fields to the variations of the two TBM parameters $A_{12}$ and $A_{14}$. In panels a and b of Fig.~\ref{fig:CPL_ratio_dep}, we present HHG spectra obtained for different $A_{12}$ and $A_{14}$ values, respectively. Thereby, we have assured that the parameter range spanned by the selected $\{A_{12},A_{14}\}$ values does not alter the underlying band structure of the model Bi$_2$Se$_3$ appreciably. The results in Fig.~\ref{fig:CPL_ratio_dep} imply that increasing $A_{12}$ resp. decreasing $A_{14}$, i.e. maximizing the $\tfrac{A_{12}}{A_{14}}$-ratio, leads to a  pronounced enhancement of the HHG yield in CPL fields. In particular, increasing the $A_{12}$-value by a factor of 3 results in an enhancement of the yield of higher-order harmonics (HO~$>15$) by several orders of magnitude. The same tendency is observed when $A_{14}$ is decreased by a factor of $2-4$, s. panel b of Fig.~\ref{fig:CPL_ratio_dep}. Although similar tendencies are also present in the case of bulk states (cp. Fig.~\ref{fig:CPL_ratio_dep_bulk} in Appendix~\ref{app:bulk_ell}), the overall efficiency of the HHG driven by CPL fields in this case is much weaker.\\

\onecolumngrid
 
\begin{figure}[htp!]
\includegraphics[scale=0.325]{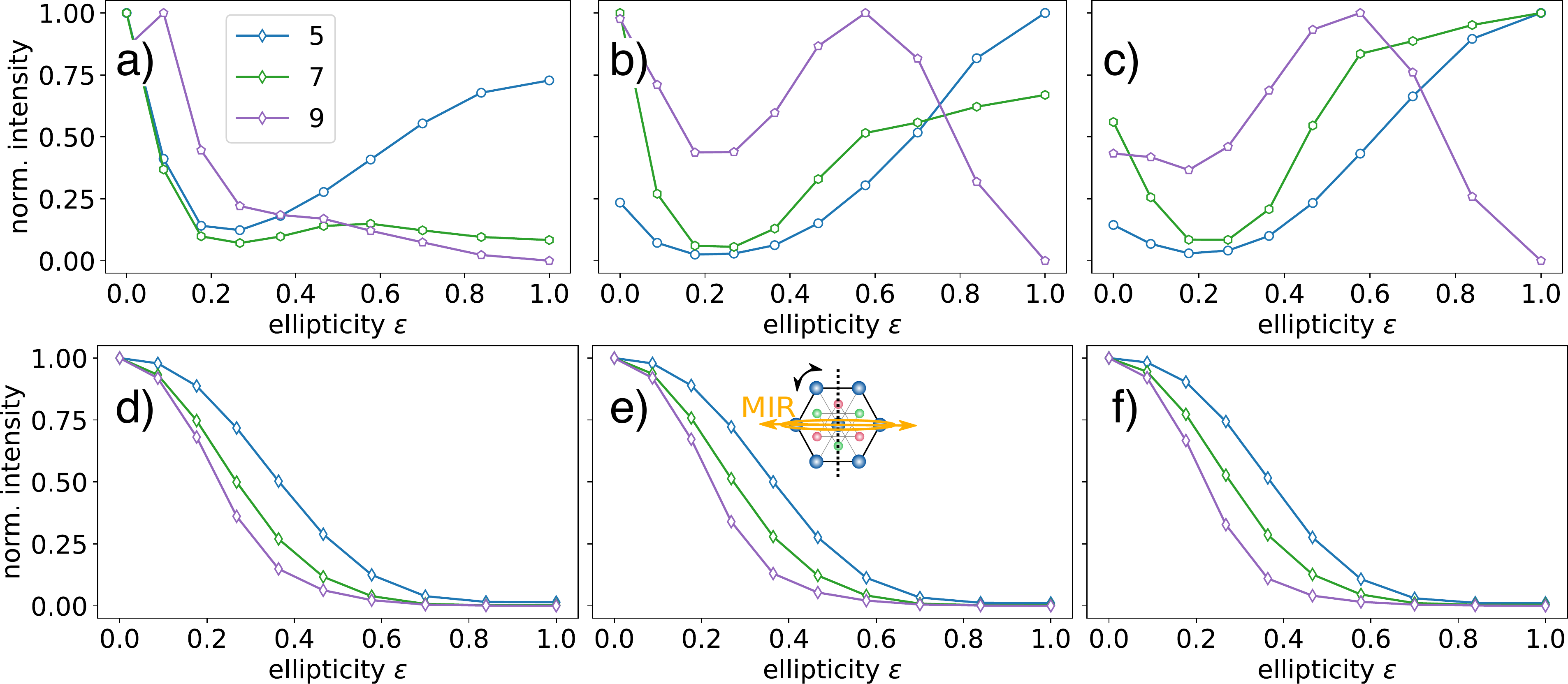}
\caption{ Ellipticity dependence of the HHG yields of HOs 5,~7, and~9 of a MIR pulse with $\lambda_\mathrm{MIR}=7.5$~$\mathrm{\mu m}$ and different driving peak intensities: $I_0=0.0045$~$\mathrm{TW/cm^2}$ in panels a) and d), $I_0=0.006$~$\mathrm{TW/cm^2}$ in panels b) and e), and $I_0=0.01$~$\mathrm{TW/cm^2}$ in panels c) and f). The upper row corresponds to the surface states, bulk states are plotted in the bottom row. As illustrated in the inset of panel e), the main axis of the polarization ellipse $\mathbf{\hat{e}}$ (orange ellipse) is set perpendicular to the mirror plane ($\sigmaM$, dashed line). The pulse has a Gaussian profile with a FWHM duration of 12 cycles. }
\label{fig:CPL_ellscans}
\end{figure}

\twocolumngrid

\begin{widetext}

\end{widetext}

Finally, we show that the above analysis implies that the high-momentum-limit mechanism is intensity-dependent, as it is mediated by the higher-order ($\mathcal{O}(k_\parallel^n), n\ge 3$) terms in the expansions of $\dvec_{cv}(\kpar)$ and $\Delta\xivec_{cv}(\kpar)$ around $\overline{\Gamma}$. For this aim, we calculate the ellipticity dependence for three different peak laser intensities, and compare results with their bulk counterparts. The results for the surface and bulk bands are presented in the upper and bottom row of Fig.~\ref{fig:CPL_ellscans} respectively. 
From these results, it can be inferred that for surface states, increasing the peak intensity brings about an anomalous ellipticity dependence, manifested in an increased HHG yields for highly-elliptical and CPL fields. This result is consistent with the notion that for higher peak electric field amplitudes, the strong-field dynamics is predominantly governed by the higher-momentum regions of the BZ, populated by coupling to the higher-order terms in Eqs.~(\ref{eq:dcvTaylorx}-\ref{eq:dcvTaylory}). Notably, HO~5 exhibits an anomalous dependence that remains robust for the entire intensity range considered, as the corresponding dynamics for this order originated from low-momentum range. An analogous intensity-dependent behavior is clearly not present in bulk harmonics shown in the bottom row.\\

\section{Conclusion and Outlook}\label{sec:conclusion}

Summarizing, we have investigated a strong-field driven phenomenon on three-dimensional topological insulator Bi$_2$Se$_3$ crystal lattice subjected to intense ultrashort fields in the mid-infrared spectral domain. Specifically, we have studied the high-harmonic response of the bulk states and the surface modes. To this end, we have integrated a simple tight-binding model (Ref.~\cite{Mao2011}) into the framework of the semiconductor Bloch equations formulated in the length gauge. Starting with a TBM comprising the four electronic states closest to the Fermi energy, we have outlined the derivation of the bulk eigenstates as well as the construction of an effective 2D surface Hamiltonian that allows us to treat  the topological surface states. Our analysis  accounts for geometrical effects in the strong field dynamics by  incorporating the complex dipole elements, Berry connections, and Berry curvature into the SBE treatment.  We have studied the general characteristics of the high-harmonic emission from bulk and surface states driven by circularly as well as linearly (see Appendix~\ref{sec:res_lin}) polarized MIR fields, and have elucidated the different dynamical symmetries that govern the non-linear response. This symmetry analysis establishes a potential approach to disentangle the contributions from bulk and surface in an all-optical experimental setting, for example with generation of even-order harmonics from the surface. \\

We have conducted a detailed analysis of the ellipticity dependence of the harmonic yield, and found a profound difference in the ellipticity profiles of the bulk and the surface states. Specifically, our results indicate that the topological  surface states of Bi$_2$Se$_3$ exhibit an anomalous ellipticity behavior, manifested in a pronounced enhancement of the harmonic yield for circularly polarized fields. With the aid of detailed analytical analysis as well as numerical calculations, we have attributed this behavior to two mechanisms operating predominantly in  the low- and the high-momentum regions of the BZ. The low-momentum range mechanism relies on the characteristic topological features of the Bloch bands that give rise to a vortex structure in the interband dipole moments and Berry connections in momentum space, manifested in a perpendicular  ``locking'' between the transition dipole and momentum vectors. The high-momentum range mechanism is  relevant for high peak amplitudes of the incident field, and is mediated by the ``warping'' terms in the surface Hamiltonian that cause the hexagonal deformation of the energy surface. Representing the counterpart of the cubic Dresselhaus spin-orbit terms in rhombohedral structures, the sensitivity of the emitted HHG spectra to these components of the Hamiltonian underlines its potential to serve as an all-optical probe of spin-orbit interaction features. \\


While the detailed results presented in this manuscript are specific for Bi$_2$Se$_3$, they are equally generalizable to any member of the tetradymite family by adopting appropriate tight-binding parameters. Moreover, the theoretical framework developed in this work allows the investigation of questions of fundamental importance such as topological phase transitions or the influence of the band inversion of the strong field dynamics by modifying the phase diagram of the tight-binding model employed in the SBE framework. We believe that these detailed theoretical results will serve as a guide for future experiments. \\

Finally, it is worth commenting on the limitations of our model, in particular the adopted electronic structure calculation strategy. As a consequence of the decoupling of the surface states from the bulk, our treatment cannot account for laser-induced transitions between surface and bulk bands. Further, using the solutions at the $\overline{\Gamma}$-point as a basis for deriving the effective surface model, as explained in Sec.~\ref{sec:tbm_surf}, implies that the TSS dispersions and wave functions are quantitatively accurate only in the low energy limit. Finally, our model does not incorporate couplings to higher-lying bands.  Nevertheless, the intuition gained by examining this simplified model can provide useful insights into the complex physics of 3D-TIs in strong laser fields. Even for the highest intensities considered in this work ($I_0=0.01$~TW/cm$^2$ for $\lambda_\mathrm{MIR}=7.5$~$\mathrm{\mu m}$), electron excursion trajectories are expected to cover $\Delta k\sim e E_\mathrm{MIR,0}/(\hbar \omega_0 )\approx 0.17$~$\mathrm{\AA}$, i.e. less than $20$~$\%$ of the BZ. The above-enumerated effects are anticipated to gain importance at intensities even higher than the ones considered in this work, in which case the trajectory of the driven electron covers a large portion of the BZ.

\begin{acknowledgements}
At Stanford/SLAC this work is supported by the US Department of Energy, Office of Science, Basic Energy Sciences, Chemical Sciences, Geosciences, and Biosciences Division through the AMOS program. D.B. gratefully acknowledges support from the Swiss National Science Foundation (SNSF) through project No: P2EZP2\_184255. A.C., D.K., and D.E.K acknowledge financial support by the Max Planck POSTECH/KOREA Research Initiative Program (Grant No.~2016K1A4A4A0192202) through the National Research Foundation of Korea (NRF) funded by the Ministry of Science, ICT and Future Planning, Korea Institute for Advancement of Technology (KIAT) grant funded by the Korea  Government (MOTIE) (P0008763, The Competency Development Program for Industry Specialists) and LANL LDRD project. 
\end{acknowledgements}



\section*{Appendix}

\appendix

\section{Additional details on the TBM Hamiltonian}\label{app:tbm}

The tight-binding model considered in this work accounts for nearest-neighbor (NN) intra-layer interactions ($\hat{t}_{\avec_i}$) as well as inter-layer hoppings ($\hat{t}_{\bvec_i}$). The NN vectors $\pm\avec_i$ and $\pm\bvec_i$ in Cartesian coordinates are explicitly given by:

\begin{eqnarray}
\avec_1 = (a,0,0)^T &\qquad & \bvec_1 = \left(0,\frac{\sqrt{3}a}{3}, c\right)^T \\
\avec_2 = \left(-\frac{a}{2},\frac{\sqrt{3}a}{2},0\right)^T & \qquad & \bvec_2 = \left(-\frac{a}{2},-\frac{\sqrt{3}a}{6},c\right)^T\\
\avec_3 =  \left(-\frac{a}{2},-\frac{\sqrt{3}a}{2},0\right)^T 
 &\qquad & \bvec_3 =  \left(\frac{a}{2},-\frac{\sqrt{3}a}{6}, c\right)^T .
\end{eqnarray}
The set of vectors $\pm\bvec$ are also the lattice vectors.\\

\begin{figure}[htb!]
\centering
\includegraphics[scale=0.375]{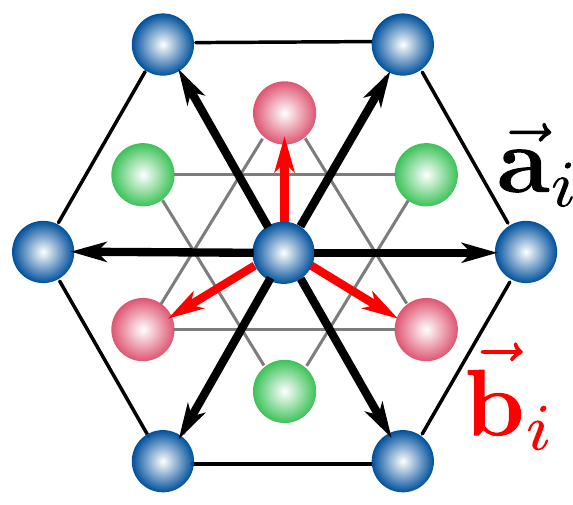}
\caption{Simplified representation of the lattice structure illustrating the nearest-neighbour vectors $\pm\avec$ and $\pm\bvec$. }
\label{fig:spinHHG}
\end{figure}

The $\Gamma$-matrices employed in Eq.~(\ref{eq:Htbm_h}) are defined as:
\begin{eqnarray}
\Gamma_1 & = & \hat{\sigma}_1\otimes\hat{\tau}_1, \Gamma_2 = \hat{\sigma}_2\otimes\hat{\tau}_1, \Gamma_3 = \hat{\sigma}_3\otimes\hat{\tau}_1\nonumber \\ 
\Gamma_4 & = & \hat{\mathbb{I}}_2\otimes\hat{\tau}_2, \Gamma_5 = \hat{\mathbb{I}}_2\otimes\hat{\tau}_3.\label{eq:Gamma}
\end{eqnarray}
In the above, the two sets of Pauli matrices $\{\hat{\tau}_i\}$ and $\{\hat{\sigma}_i\}$ can be interpreted as operating on the orbital ($\hat{\tau}$) and the spin ($\hat{\sigma}$) degrees of freedom, respectively.\\

In the following, we define the auxiliary functions $h_i(\kvec)$ used in the TBM Hamiltonian of Eq.~(\ref{eq:Htbm_h}):
\begin{eqnarray}
h_0 (\kvec) & = & 2A_0\sum_{i=1}^{3}\cos(\kvec\cdot\avec_i) + 2B_0\sum_{i=1}^3\cos(\kvec\cdot\bvec_i) \\
h_1 (\kvec) & = & -2A_{14}\sin \omega\left[\sin(\kvec\cdot\avec_2)-\sin(\kvec\cdot\avec_3)\right]  \nonumber \\
& + & 2B_{14}\left[ \sin(\kvec\cdot\bvec_1) + \cos\omega(\sin(\kvec\cdot\bvec_2)\right. \nonumber \\
& + & \left. \sin(\kvec\cdot\bvec_3))\right] \\
h_2 (\kvec) & = & -2B_{14}\sin \omega\left[\sin(\kvec\cdot\bvec_2)-\sin(\kvec\cdot\bvec_3)\right] \nonumber \\
& - & 2A_{14}\left[ \sin(\kvec\cdot\avec_1) + \cos\omega(\sin(\kvec\cdot\avec_2) \right. \nonumber \\
& + & \left. \sin(\kvec\cdot\avec_3))\right] \\
h_3 (\kvec) & = & 2A_{12}\sum_{i=1}^3\sin(\kvec\cdot\avec_i) \\
h_4 (\kvec) & = & -2 B_{12}\sum_{i=1}^3\sin(\kvec\cdot\bvec_i) \\
h_5 (\kvec) & = & 2A_{11}\sum_{i=1}^3\cos(\kvec\cdot\avec_i)\nonumber \\
& + & 2B_{11}\sum_{i=1}^3\cos(\kvec\cdot\bvec_i) + m_{11},
\end{eqnarray}
where $\omega=-2\pi/3$. $m_{11}$ in $h_5(\kvec)$ controls the band inversion and, in this work, is chosen such that the resulting band structure describes a strong topological insulator, s. discussion in Ref.~\cite{Mao2011}.\\

The unitary transformation matrix $\hat{U}_1$ employed in Eq.~(\ref{eq:Hk3}) is given by~\cite{Liu2010}:

\begin{equation}\label{eq:U1}
\hat{U}_1 = \begin{pmatrix}
1 &  0 & 0 & 0 \\
0 & -i & 0 & 0 \\
0 &  0 & 1 & 0 \\
0 &  0 & 0 & i
\end{pmatrix}. 
\end{equation}

The normalization constant of the bulk spinors in Eq.~(\ref{eq:psiB1}-\ref{eq:psiB2}) reads:
\begin{eqnarray}
\mathcal{N}_\mathrm{B}^\pm &=& \frac{1}{\sqrt{2}}\Bigg\{ \left({h_1(\kvec)}^2+{h_2(\kvec)}^2\right) \label{eq:Nb}\\
&\times &  \Bigg(\pm {h_5(\kvec)}  \sqrt{\sum_{i=1}^5\left(h_i(\kvec)\right)^2}  + \sum_{i=1}^5\left(h_i(\kvec)\right)^2 \Bigg)^{-1}\Bigg\}^{1/2}.\nonumber
\end{eqnarray}

In contrast to the $\kvec\cdot\bm{p}$-perturbative model~\cite{Liu2010} frequently employed to study the low-energy physics of 3D-TIs (s. Refs.~\cite{OliaeiMotlagh2017,OliaeiMotlagh2018}), the TBM Hamiltonian defined in Eq.~(\ref{eq:Htbm_h}) retains its validity over the entire Brillouin zone and is periodic.

\begin{table}[h]
\begin{tabular}{| p{1.5cm} p{1.5cm} p{1.5cm} p{1.5cm}|}
\hline
\hline
 & & &\\
$\hat{t}_{\avec_i}$ / \mbox{eV} & & $\hat{t}_{\bvec_i}$ / \mbox{eV} &  \\
 & & &\\
\hline
$A_0$     & $-0.0255$  & $B_0$    & $0.0164$ \\
$A_{11}$  & $ 0.1937$  & $B_{11}$ & $0.1203$ \\
$A_{12}$  & $ 0.2240$  & $B_{12}$ & $0.3263$  \\
$A_{14}$  & $ 0.0551$  & $B_{14}$ & $0$      \\
\hline 
$m_{11}$  & $-1.6978$  & & \\
\hline
\hline
\end{tabular}
\caption{Parameters for the TBM Hamiltonian used in this work.}\label{tab:tbm}
\end{table}

\section{Derivation of the effective 2D surface Hamiltonian}\label{app:surf_deriv}

\subsubsection{Surface-state spinors at the $\Gamma$-point}\label{sec:surfGamma}

In this Section, we briefly outline the derivation of the effective (2D) Hamiltonian for describing the surface electrons. The approach follows closely the procedures outlined in Refs.~\cite{Liu2010,Shan2010}. We start by obtaining approximate expressions for the TSS Hamiltonian and the wavefunctions at the $\Gamma$-point ($\kvec=0$); the latter will be subsequently used as a basis for constructing the 2D model. For this aim, we start by expanding the TBM Hamiltonian in Eq.~\ref{eq:Htbm_h} up to the second order:

\begin{widetext}
\begin{equation}
\hat{H}^{(2)}(\kvec)\equiv
\begin{pmatrix}
h_0^{(2)}(\kvec) + h_5^{(2)}(\kvec) & -6B_{12}c k_z & 0 & a(3A_{14}+\sqrt{3}B_{14})k_- \\
-6 B_{12}c k_z & h_0^{(2)}(\kvec) - h_5^{(2)}(\kvec) & a(3A_{14}+\sqrt{3}B_{14})k_- & 0 \\
0 & a(3A_{14}+\sqrt{3}B_{14})k_+ & h_0^{(2)}(\kvec) + h_5^{(2)}(\kvec) &  6 B_{12}c k_z \\
a(3A_{14}+\sqrt{3}B_{14})k_+ & 0 & 6 B_{12}c k_z & h_0^{(2)}(\kvec) - h_5^{(2)}(\kvec)
\end{pmatrix}.\label{eq:Hk3Taylor}
\end{equation} 
\end{widetext}
\newpage
In the above, $k_\pm = k_x\pm i k_y$. The terms $h_i^{(2)}(\kvec)$ denote the second-order Taylor expansions of the functions $h_i(\kvec)$ around $\overline{\Gamma}$. In particular:
\begin{eqnarray}
h_0^{(2)}(\kvec) & = & 6(A_0+B_0) - \frac{1}{2}a^2(3A_0+B_0)k_\parallel^2\nonumber\\
&-& 3B_0c^2k_z^2
\end{eqnarray}
\begin{eqnarray}
h_5^{(2)}(\kvec) & = & 6(A_{11} + B_{11})-\frac{1}{2}a^2(3A_{11}+B_{11})k_\parallel^2\nonumber\\
&-& 3B_{11}c^2k_z^2+m_{11}.
\end{eqnarray}
The low-momentum Hamiltonian in Eq.~(\ref{eq:Hk3Taylor}) is of equivalent form as the $\kvec\cdot{\bm{p}}$-Hamiltonian derived by Liu~\textit{et al.} in Ref.~\cite{Liu2010}. 
To obtain the general surface-state Hamiltonian, open boundary conditions are applied, i.e. we restrict the surface mode to the half-space defined by $z<0$ and let the corresponding TSS wavefunction vanish at $z=0$ and $z\rightarrow-\infty$. The resulting breaking of the translational symmetry can be accommodated via the substitution $k_z\rightarrow -i\partial_z$ in Eq.~(\ref{eq:Hk3Taylor}):
\begin{equation}
\hat{\mathcal{H}}^\mathrm{S}(\kpar;-i\partial_z) \equiv \hat{H}^{(2)}(k_x,k_y,k_z\rightarrow -i\partial_z).\label{eq:Hlam}
\end{equation} 
The following \textit{ansatz} is used for the wavefunction:
\begin{equation}\label{eq:psi_totS}
\Psi_\mathrm{Surf}(\kpar;z)\propto \psi_\lambda e^{\lambda z},
\end{equation}
where $\psi_\lambda$ is a 4-component spinor of the form:
 
\begin{equation}
\psi_\lambda=\begin{pmatrix}
\psi_\uparrow\\ 
\psi_\downarrow 
\end{pmatrix} = 
\begin{pmatrix}
\psi_{1\uparrow}\\
\psi_{1\uparrow}\\
\psi_{2\downarrow}\\ 
\psi_{2\downarrow}
\end{pmatrix},
\end{equation}
and $\lambda$ denotes a parameter which is chosen to satisfy the boundary conditions.
Substituting  Eq.~\ref{eq:Hlam} into $\hat{\mathcal{H}}^\mathrm{S}(\kpar;-i\partial_z)\Psi_\mathrm{Surf}(\kpar;z)=E\Psi_\mathrm{Surf}(\kpar;z)$ leads to the time-independent Schr\"odinger equation for the spinors $\psi_\lambda$:
\begin{equation}
\hat{\mathcal{H}}^\mathrm{S}(\kpar;\lambda) \psi_\lambda = E\psi_\lambda,\label{eq:TISE_Surf}
\end{equation}
which constitutes a second-order differential equation, consequently, the superposition defined by Eq.~\ref{eq:psi_totS} features at most 8 components, with coefficients constrained by the imposed boundary conditions. The general solution for the low-energy (second-order) Hamiltonian has been derived in multiple works  (see, e.g., Refs.~\cite{Shan2010, Shun-QingShen2012}), therefore, we refrain from presenting it here. At the center of the BZ ($\overline{\Gamma}$-point), the Hamiltonian $\hat{\mathcal{H}}^\mathrm{S}(\kpar;\lambda)$ becomes block-diagonal, i.e.:
\begin{widetext}
\begin{equation}
\hat{H}^{\Gamma} = \begin{pmatrix}
h_0^{\Gamma}+h_5^{\Gamma}+3(B_0+B_{11})c^2\lambda^2 & 6iB_{12}c\lambda &0 & 0\\
6iB_{12}c\lambda & h_0^{\Gamma}-h_5^{\Gamma}+3(B_0-B_{11})c^2\lambda^2 &0 &0 \\
0 & 0 & h_0^{\Gamma}+h_5^{\Gamma}+3(B_0+B_{11})c^2\lambda^2 & -6iB_{12}c\lambda \\
0 & 0 & -6iB_{12}c\lambda & h_0^{\Gamma}-h_5^{\Gamma}+3(B_0-B_{11})c^2\lambda^2
\end{pmatrix}
\end{equation}
\end{widetext}
where $\hat{H}^{\Gamma} \equiv \hat{\mathcal{H}}^\mathrm{S}(\kpar=\bm{0};\lambda)$ and $h_i^{\Gamma}$ corresponds to the value of the function $h_i(\kvec)$ at the $\overline{\Gamma}$ point. The eigenvectors of the Hamiltonian $\hat{H}^\Gamma$ are doubly degenerate (time-reversal symmetry), and the two spinors have the structure:
\begin{eqnarray}
\psi_\mathrm{S}^\uparrow & = &  \begin{pmatrix}
\phi \\
\bm{0}
\end{pmatrix} \\
\psi_\mathrm{S}^\downarrow & = & \begin{pmatrix}
\bm{0}\\
\hat{\tau}_z\phi
\end{pmatrix},
\end{eqnarray}
where $\phi$ is a 2-vector. 
With the basis defined in Eq.~\ref{eq:Pz_basis}, the two spinors at the $\overline{\Gamma}$-point thus correspond to pure ``spin-up'' and ``spin-down'' components. 
Recasting the results in Ref.~\cite{Shan2010} in terms of the TBM parameters, our \textit{ansatz} for the low-order solutions at $\kvec=\bm{0}$ becomes:

\begin{eqnarray}
\psi_\mathrm{S}^\uparrow & = & \frac{1}{\sqrt{2}} \begin{pmatrix}
i\sqrt{\frac{B_{11}-B_0}{B_{11}}} \\
\sqrt{\frac{B_{11}+B_0}{B_{11}}} \\
0 \\
0
\end{pmatrix}\left( e^{\lambda_1 z} - e^{\lambda_2 z} \right) 
\end{eqnarray}

and:

\begin{eqnarray}
\psi_\mathrm{S}^\downarrow & = & \frac{1}{\sqrt{2}} \begin{pmatrix}
0 \\
0 \\
 i\sqrt{\frac{B_{11}-B_0}{B_{11}}} \\
 -\sqrt{\frac{B_{11}+B_0}{B_{11}}} 
\end{pmatrix}\left( e^{\lambda_1 z} - e^{\lambda_2 z} \right).
\end{eqnarray}

In this work, we are primarily interested in the electron dynamics on the surface, and we drop the spatial part ($\propto e^{\lambda z}$). \\
\subsubsection{Effective 2D surface Hamiltonian in the TBM formalism}\label{sec:H2D}

In the following, we illustrate the construction of the approximate 2D Hamiltonian for the surface states. We start with the full Hamiltonian given by Eq.~(\ref{eq:Hk3}) and perform a Taylor expansion only in $k_z$. We then split the resulting Hamiltonian $\hat{H}^{(z_2)}(\kvec)$ into two parts: one term independent of $\kpar$ and another term depending on $\kpar$:
\begin{equation}\label{eq:Hsplit}
\hat{H}^{(z_2)}(\kvec;k_z\rightarrow -i\partial_z) = \hat{H}_0(\kvec=\bm{0}; -i\partial_z) + \hat{H}_\parallel(\kpar)
\end{equation}
with $\kpar = (k_x,k_y)^T$. In the language of degenerate perturbation theory, the second term in Eq.~(\ref{eq:Hsplit}) can be understood as a ``perturbation'' term and, in the TBM formalism employed here, is given by:\\

\begin{widetext}
\begin{eqnarray}
&\hat{H}_\parallel(\kpar) & =  \hat{U}_1\left( h_0^{z_0}(\kpar)\hat{\mathbb{I}}_4 + \sum_{i=1}^5 h_i^{z_0}(\kpar)\Gamma_i  - 
h_0^\Gamma \hat{\mathbb{I}}_4  - h_5^\Gamma \Gamma_5 \right) \hat{U}_1^T \\
& = & \begin{pmatrix}
h_0^{z_0}(\kpar)-h_0^\Gamma + h_5^{z_0}(\kpar)-h_5^\Gamma & i\left(h_3^{z_0}(\kpar)-i h_4^{z_0}(\kpar)\right) & 0 & -i\left(h_1^{z_0}(\kpar)-i h_2^{z_0}(\kpar)\right) \\
-i\left(h_3^{z_0}(\kpar)+i h_4^{z_0}(\kpar)\right) & h_0^{z_0}(\kpar)-h_0^\Gamma - h_5^{z_0}(\kpar)+h_5^\Gamma & 
-i\left(h_1^{z_0}(\kpar)-i h_2^{z_0}(\kpar)\right) & 0 \\
0 & i\left(h_1^{z_0}(\kpar)+i h_2^{z_0}(\kpar)\right) & h_0^{z_0}(\kpar)-h_0^\Gamma + h_5^{z_0}(\kpar)-h_5^\Gamma &
i\left(h_3^{z_0}(\kpar)+i h_4^{z_0}(\kpar)\right) \\
i\left(h_1^{z_0}(\kpar)+i h_2^{z_0}(\kpar)\right) & 0 & i\left(-h_3^{z_0}(\kpar)+i h_4^{z_0}(\kpar)\right) & 
h_0^{z_0}(\kpar)-h_0^\Gamma - h_5^{z_0}(\kpar)+h_5^\Gamma
\end{pmatrix} .\nonumber
\end{eqnarray}
\end{widetext}

In the above, $h_i^{z_0}(\kpar)\equiv h_i(k_x,k_y,k_z=0)$, $h_i^\Gamma \equiv h_i(\kvec=\bm{0})$, and $h_i(\kvec)$ are the auxiliary functions defined in the Appendix.
Afterwards, the effective 2D Hamiltonian $\Delta\hat{H}_{2D}^\mathrm{S}$ is constructed by taking the matrix elements of $\hat{H}_\parallel(\kpar)$ with the spinor part of the basis states $\psi_\mathrm{S}^{\uparrow,\downarrow} \equiv \ket{\psi_\mathrm{S}^\sigma}$:
\begin{equation}
\left(\Delta\hat{H}_{2D}^\mathrm{S}\right)_{\sigma,\sigma'} = \left\langle\psi_\mathrm{S}^{\sigma} \left| \hat{H}_\parallel  \right| \psi_\mathrm{S}^{\sigma'}\right\rangle.
\end{equation}
The explicit expression individual matrix elements are given in the next Section. 


\subsubsection{Matrix elements of the effective surface Hamiltonian $\Delta H_{2D}^{\mathrm{S})}$}\label{app:H2D}

The  matrix elements of $\Delta H_{2D}^{\mathrm{S}} (\kpar)$ are given by:
\begin{eqnarray}
\left\langle\psi_\mathrm{S}^{\uparrow} \left| \hat{H}_\parallel  \right| \psi_\mathrm{S}^{\uparrow}\right\rangle & = & 
h_0^{z_0}(\kpar)-h_0^\Gamma + \sqrt{1-\frac{B_0^2}{B_{11}^2}}h_3^{z_0}(\kpar)\nonumber \\
&+ &\frac{B_0(-h_5^{z_0}(\kpar)+h_5^\Gamma)}{B_{11}} 
\end{eqnarray}
\begin{eqnarray}
\left\langle\psi_\mathrm{S}^{\downarrow} \left| \hat{H}_\parallel  \right| \psi_\mathrm{S}^{\downarrow}\right\rangle & = & 
h_0^{z_0}(\kpar)-h_0^\Gamma - \sqrt{1-\frac{B_0^2}{B_{11}^2}}h_3^{z_0}(\kpar)\nonumber \\
& +&\frac{B_0(-h_5^{z_0}(\kpar)+h_5^\Gamma)}{B_{11}} 
\end{eqnarray}
\begin{eqnarray}
\left\langle\psi_\mathrm{S}^{\uparrow} \left| \hat{H}_\parallel  \right| \psi_\mathrm{S}^{\downarrow}\right\rangle & = & 
\sqrt{1-\frac{B_0^2}{B_{11}^2}}(h_1^{z_0}(\kpar)-ih_2^{z_0}(\kpar)) 
\end{eqnarray}
\begin{eqnarray}
\left\langle\psi_\mathrm{S}^{\downarrow} \left| \hat{H}_\parallel  \right| \psi_\mathrm{S}^{\uparrow}\right\rangle & = & 
\sqrt{1-\frac{B_0^2}{B_{11}^2}}(h_1^{z_0}(\kpar)+ih_2^{z_0}(\kpar)).
\end{eqnarray}
Adding the energies of the unperturbed states to the diagonal elements in the above expressions gives the Hamiltonian $H_{2D}^\mathrm{(S)}(\kpar)$ reported in Eq.~(\ref{eq:HS2D}).\\

The normalization constant $\mathcal{N}_\mathrm{S}^\pm(\kpar)$ of the surface wavefunctions in Eq.~(\ref{eq:psiS}) is given by:
\begin{equation}
\label{eq:Ns}
\mathcal{N}_\mathrm{S}^\pm(\kpar) = \frac{1}{\sqrt{2}}\sqrt{1\mp \frac{h_3^{z_0}(\kpar)}{\sum_{i=1}^3\left(h^{z_0}_i(\kpar)\right)^2}}.
\end{equation}

\section{Calculation details}\label{app:calc}

We employ the following expression for defining the temporal profile of the vector potential:
\begin{equation}
\At = - A_0 g_\mathrm{env}(t) 
\begin{pmatrix}
\sin(\omega_0 t) & + & \cos(2\alpha_\mathrm{QWP})\cos(\omega_0 t ) \\
& - & \sin(2\alpha_\mathrm{QWP})\cos(\omega_0 t )\\
\end{pmatrix},\label{eq:Adef}
\end{equation}
where $\omega_0$ denotes the angular frequency of the driving laser field, $A_0=\tfrac{E_0}{\omega_0}$ is the peak amplitude of the vector potential (corresponding to peak electric field $E_0$), and $\alpha_\mathrm{QWP}$ is the angle with respect to the fast axis of a quarter wave plate in cases where an elliptically or circularly polarized (CPL) field is considered. $\alpha_\mathrm{QWP}=0$ corresponds to horizontally-polarized (P-polarization) laser field, whereas $\alpha_\mathrm{QWP}= +/- \tfrac{\pi}{4}$ yields left/right CPL field (LCP/RCP), respectively. The function $g_\mathrm{env}(t)$ in Eq.~(\ref{eq:Adef}) represents a Gaussian envelope. In the text, we report the full-width-half-maximum (FWHM) pulse duration in number of cycles. Interaction with the magnetic component of the laser field is neglected.

We consider ultrashort ($10-12$ optical cycles in FWHM duration with a Gaussian profile) driving pulses with a photon energy lying below the bulk-band-gap, i.e. in the far midinfrared (MIR) range ($\hbar \omega_0\sim$ $0.165$~eV) and a peak electric field amplitude of $1\--3$~\mbox{MV/cm}. In accordance to previous theoretical studies employing the SBE formalism in this spectral domain~\cite{Hohenleutner2015}, we set the dephasing time to $T_2=1.25$~fs. We solve the SBEs for the bulk and the surface in the length gauge on a two-dimensional momentum grid typically $640\times640$ points (BS) or $960\times960$ points (TSS). Momentum-space integration is performed over $\lesssim 85$~$\%$ of the first BZ by applying a circular ``mask'' in the BZ. Prior to Fourier transformation, the time-dependent currents are filtered by a Hanning window. The HHG spectra are normalized with respect the linear response (i.e. the maximum). \\


\section{HHG driven by linear polarization}\label{sec:res_lin}

Here we cover the case of a Bi$_2$Se$_3$ crystal excited by a linearly polarized 15-cylce MIR pulse ($\lambda_\mathrm{MIR} = 7.5$~$\mathrm{\mu m}$, $I_0 = 0.002$~TW/cm$^{-2}$) with polarization vector aligned either along  $k_x$ ($\EvecMIR\parallel\overline{\Gamma K}$ in the 2D BZ, cp. red hexagon in Fig.~\ref{fig:crystalBZ}~c) or along $k_y$ ($\EvecMIR\parallel\overline{\Gamma M}$ in the 2D BZ). For the surface states, the latter choice corresponds to a situation where the MIR electric field direction coincides with the mirror plane ($\EvecMIR\parallel\sigmaM$), whereas it is orthogonal to it ($\EvecMIR\perp\sigmaM$) in the former case. In Fig.~\ref{fig:lin_circ_spectrum}, we depict the resulting harmonic spectra of the combined (intra- + interband) emission for the two different orientations of the MIR polarization with respect to the high-symmetry directions in the 2D BZ (${\bm{E}}_\mathrm{MIR}\parallel\overline{\Gamma M}$ in panels a and d and ${\bm{E}}_\mathrm{MIR}\parallel\overline{\Gamma K}$ in b and e). \\

\begin{figure}
\begin{center}
\includegraphics[scale=0.28]{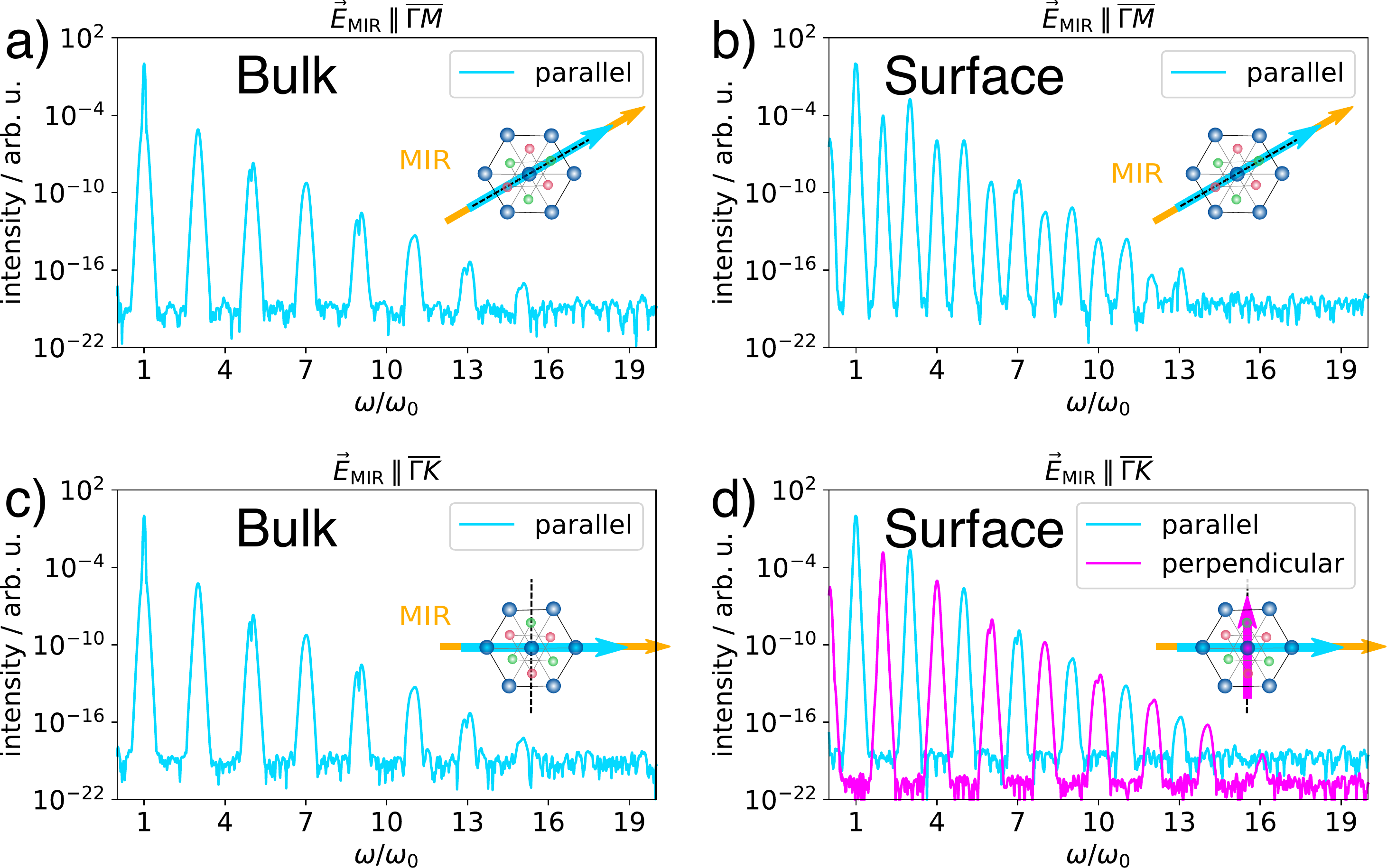}
\caption{HHG spectra of Bi$_2$Se$_3$ driven by linearly polarized fields, for both bulk and surface states. The relative orientation of the (111)-surface (real space) and the MIR polarization vector is sketched in the insets. Panels a) and c) show spectra of the bulk states for a linearly polarized MIR laser pulse oriented along the $\overline{\Gamma M}$ and $\overline{\Gamma K}$ directions, respectively. All emitted harmonics are parallel with respect to $\Et$ (cyan). The laser field has a peak intensity of $I_0 = 0.002$~$\mathrm{TW/cm^2}$ and a FWHM duration of 15 cycles. Panels b) and d) show the corresponding spectra for the surface states. The polarization of the even harmonics flips from parallel (cyan) when $\Et\parallel\overline{\Gamma M}$ (i.e. $\Et\parallel\sigmaM$) to perpendicular (magenta) with respect to the driving field when $\Et\parallel\overline{\Gamma K}$ (i.e. $\Et\perp\sigmaM$). In all calculations, the dephasing time is set at $T_2=1.25$~fs.}\label{fig:lin_spectrum}
\end{center}
\end{figure}

From the results in Fig.~\ref{fig:lin_spectrum}, the following tendencies can be discerned. First, the inversion-symmetric bulk bands support only odd harmonics, linearly polarized along the polarization direction of the MIR field, whereas all orthogonally polarized (with respect to $\EvecMIR$) contributions to the total current vanish (s. panels a and c). This  follows from dynamical symmetry analysis~\cite{Neufeld2019} after taking into account the fact that the reciprocal $\hat{k}_x$- and $\hat{k}_y$-directions correspond to the $\hat{\mathcal{R}}_2^{(x)}$- and $\sigmaM$-symmetry operations in real space. A more rigorous treatment is provided in the Appendix~\ref{app:dyn_sym}. In addition, the absence of orthogonal current component also reflects the zero trace of the non-Abelian Berry curvature associated with the BSs (s. Sec.~\ref{sec:el_dyn}). This results into a null anomalous velocity contribution of the BSs. \\

As a direct consequence of the breaking of IS at the TI surface, even harmonics appear in the spectra from the TSSs (cp. panel~b and d of Fig.~\ref{fig:lin_circ_spectrum}). As in the case of the BSs, the polarization of the odd harmonics follows the polarization of the driving MIR field $\EvecMIR$. For the even harmonics, this holds only when the laser field is parallel to the mirror plane $\sigmaM$ ($\EvecMIR\parallel \overline{\Gamma M}$, panel b), in which case the dynamical symmetry conservation requires that current component orthogonal to $\sigmaM$ must cancel out (s. Appendix~\ref{app:dyn_sym}). On the contrary, when $\EvecMIR\perp \sigmaM$, i.e. when the MIR is aligned along $\overline{\Gamma K}$, the only even harmonic contributions are generated perpendicular to the driving field polarization (cp. magenta line in Fig.~\ref{fig:lin_circ_spectrum}~d). These results are consistent with experimental findings in inversion-symmetry-breaking systems such as ZnO~\cite{Jiang2019}, GaSe~\cite{Langer2016}, $\alpha$-SiO$_2$ ($\alpha$-quartz~\cite{Luu2018}), or 2D monolayers (MoS$_2$)~\cite{Liu2017a}, as well as with a number of previous theoretical results~\cite{Yue2020}.\\

\section{Dynamical symmetries of the $D_{3d}^5$ spatial group}\label{app:dyn_sym}

We outline the derivation of the dynamical symmetry (DS) selection rules for the three cases considered in Sec.~\ref{sec:results} of the main text as well as Appendix~\ref{sec:res_lin}: a linearly polarized MIR field $\EvecMIR$ polarized along $x$, along $y$, and circularly polarized. Thereby, we follow closely the procedure derived in Ref.~\cite{Neufeld2019}. We consider the adjoints of the spatial symmetry operators outlined in Sec.~\ref{sec:crystal} ($\hat{i}$, $\hat{\mathcal{R}}_2^{(x)}$, $\hat{\mathcal{R}}_3^{(z)}$, $\sigmaM$) with the temporal transformations $\hat{\tau}_n$, where $\hat{\tau}_n$ denotes the temporal translation by $T_0/n$ with $T_0$ being the fundamental optical cycle: $\hat{\tau}_n\EvecMIR(t)=\EvecMIR(t+T_0/n)$. Selection rules are derived by studying the effect of each DS adjoint on a time-dependent observable $\bm{o}(t)$, expanded as a Fourier series with coefficients $\Fnvec$: $\bm{o}(t)=\sum_n\Fnvec\ent$.

\subsection{Laser field linearly polarized along the $x$-direction}

For the bulk states, the inversion symmetry $\hat{i}$ and the two-fold rotational axis along $x$ ($\hat{\mathcal{R}}_2^{(x)}$) lead to the following two dynamical symmetry restrictions when $\EvecMIR\parallel x$:
\begin{eqnarray}
\sum_n \hat{i}\cdot\Fnvec \hat{\tau}_2 \ent = \sum_n \Fnvec \ent \nonumber \\ \Leftrightarrow \begin{pmatrix}
-F_{n,x} \\
-F_{n,y} \\
\end{pmatrix}e^{in\pi}=\begin{pmatrix}
F_{n,x} \\
F_{n,y} \\
\end{pmatrix}\label{eq:linx_i}
\end{eqnarray}
and
\begin{eqnarray}
\sum_n \hat{\mathcal{R}}_2^{(x)}\cdot\Fnvec \hat{\tau}_2 \ent = \sum_n \Fnvec \ent \nonumber \\ \Leftrightarrow \begin{pmatrix}
F_{n,x} \\
-F_{n,y} \\
\end{pmatrix}=\begin{pmatrix}
F_{n,x} \\
F_{n,y} \\
\end{pmatrix}.\label{eq:linx_Rx}
\end{eqnarray}
Condition~(\ref{eq:linx_Rx}) implies that all polarization components of the emitted HHG perpendicular to the driving field vanish. For the HHG emission parallel to the field, the inversion symmetry (Eq.~(\ref{eq:linx_i})) implies that $e^{i\pi n}=-1$, which is fulfilled for odd values of $n$ only. Summarizing, only odd-order harmonics, linearly polarized along the driver field are emitted.\\

For the surface states, the absence of inversion symmetry and the presence of a mirror axis $\sigmaM$ along $y$ result in the following DS:
\begin{eqnarray}
\sum_n \sigmaM\cdot\Fnvec \hat{\tau}_2 \ent = \sum_n \Fnvec \ent \nonumber \\ \Leftrightarrow \begin{pmatrix}
-F_{n,x} \\
F_{n,y} \\
\end{pmatrix}e^{in\pi}=\begin{pmatrix}
F_{n,x} \\
F_{n,y} \\
\end{pmatrix}\label{eq:linx_sy}.
\end{eqnarray}
Emission along the polarization axis, i.e. $\EvecMIR\parallel x$, is subject to the condition $e^{i\pi n}=-1$ and thus restricted to odd harmonics only. The orthogonal emission has to comply to the restriction $e^{i\pi n}=1$ and supports only even harmonic orders of the driving field.\\

\subsection{Laser field linearly polarized along the $y$-direction}
In a manner analogous to the above, we obtain the following DSs for the bulk states in the case $\EvecMIR\parallel y$:
\begin{eqnarray}
\sum_n \hat{i}\cdot\Fnvec \hat{\tau}_2 \ent = \sum_n \Fnvec \ent \nonumber \\ \Leftrightarrow \begin{pmatrix}
-F_{n,x} \\
-F_{n,y} \\
\end{pmatrix}e^{in\pi}=\begin{pmatrix}
F_{n,x} \\
F_{n,y} \\
\end{pmatrix}\label{eq:liny_i}
\end{eqnarray}
and
\begin{eqnarray}
\sum_n \hat{\mathcal{R}}_2^{(x)}\cdot\Fnvec \hat{\tau}_2 \ent = \sum_n \Fnvec \ent \nonumber \\ \Leftrightarrow \begin{pmatrix}
-F_{n,x} \\
F_{n,y} \\
\end{pmatrix}=\begin{pmatrix}
F_{n,x} \\
F_{n,y} \\
\end{pmatrix}.\label{eq:liny_Rx}
\end{eqnarray}
The last condition implies that harmonics along the $x$-direction, or, perpendicular to the driving field, are symmetry-forbidden. Harmonic emission is directed along $y$ and thus follows the laser polarization, whereby $n$ is restricted to odd numbers only (due to $e^{i\pi n}=-1$).\\

For the surface states, when $\EvecMIR\parallel \sigmaM$, the DS rules reduce to:
\begin{eqnarray}
\sum_n \sigmaM\cdot\Fnvec  \ent = \sum_n \Fnvec \ent \nonumber \\ \Leftrightarrow \begin{pmatrix}
-F_{n,x} \\
F_{n,y} \\
\end{pmatrix}=\begin{pmatrix}
F_{n,x} \\
F_{n,y} \\
\end{pmatrix}\label{eq:liny_sy}.
\end{eqnarray}
This condition implies that whereas all perpendicular components along $x$ ($F_{n,x}$) vanish, the parallel component comprises both even and odd harmonics.

\subsection{Circularly polarized laser fields}
The DS pertaining to the case of  CPL MIR drivers is most easily tackled by adopting the spherical basis for the vectors $\Fnvec$, i.e. $\Fnvec = (F_{n,+},F_{n,-})^T$ with $F_{n,\pm} = F_{n,x}\pm iF_{n,y}$. In the presence of discrete three-fold rotational symmetry $\hat{\mathcal{R}}_3^{(z)}$, the DS reads:
\begin{eqnarray}
\sum_n\hat{\mathcal{R}}_3^{(z)}\cdot\Fnvec \hat{\tau}_3\ent = \sum_n \Fnvec \ent \nonumber \\
\Leftrightarrow 
\begin{pmatrix}
e^{-i2\pi/3}F_{n,+}\\
e^{i2\pi/3}F_{n,-}
\end{pmatrix}e^{i n 2\pi/3}
= 
\begin{pmatrix}
F_{n,+}\\
F_{n,-}
\end{pmatrix},
\end{eqnarray}
which implies $e^{i(n\mp1)2\pi/3}=1$, or $n=3N\pm1$, where $N$ is an integer. In other words, each third harmonic multiple is precluded by symmetry. This consideration holds for surface and bulk states alike. For the bulk states, inversion symmetry still holds and further restricts the emitted harmonics to odd multiples only, implying an effective selection rule of $n=6N\pm1$. Further, the individual members of the pairs $n=3N\pm1$ or $n=6N\pm1$ have alternating helicities.\\
\clearpage

%
%
%
%
%


\section{Additional vector field plots}\label{app:vortex}

\begin{figure}[htp!]
\centering
\includegraphics[scale=0.2]{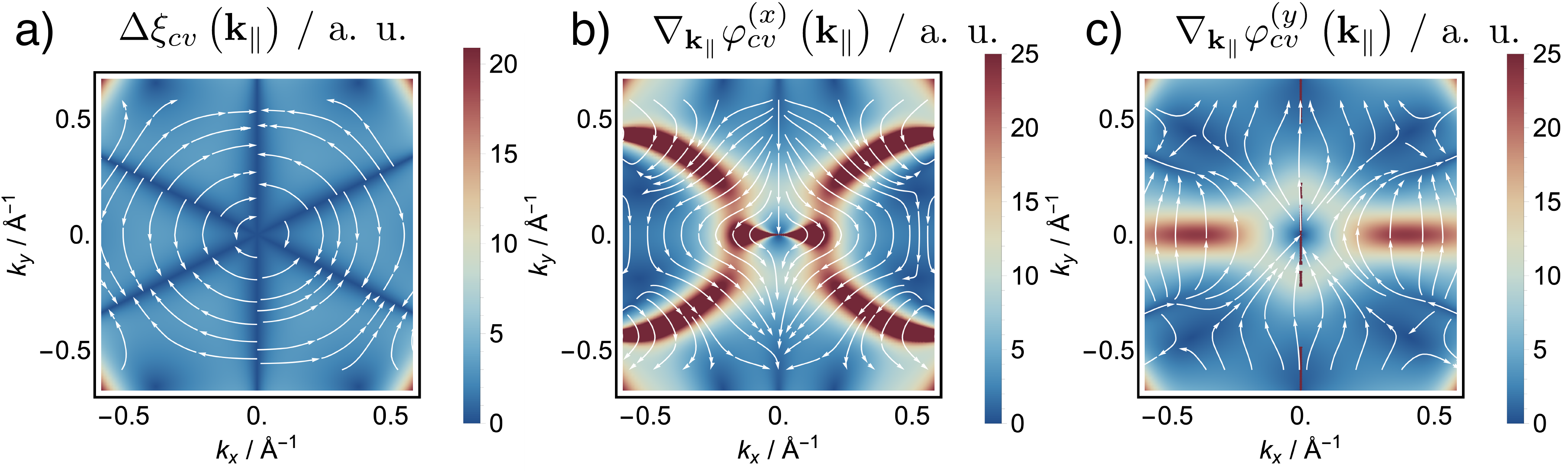}
\caption{ Panel a): Vector field plot of the difference of the Berry connections of upper and lower Dirac cones of the TSS. The color quantifies the absolute magnitude of $\Delta\xivec_{cv}(\kpar)$. The streamlines (white) indicate the local direction of the vector fields in momentum space. Panels b) and c): Stream plots of the derivatives of the $x$- (b) and the $y$-components (c) of the phase of the interband dipole matrix element $\dvec_{cv}(\kpar)$.  }
\label{fig:app_vortex}
\end{figure}
\section{Additional calculations of the bulk ellipticity response}\label{app:bulk_ell}

\begin{figure}[h!]
\centering
\includegraphics[scale=0.34]{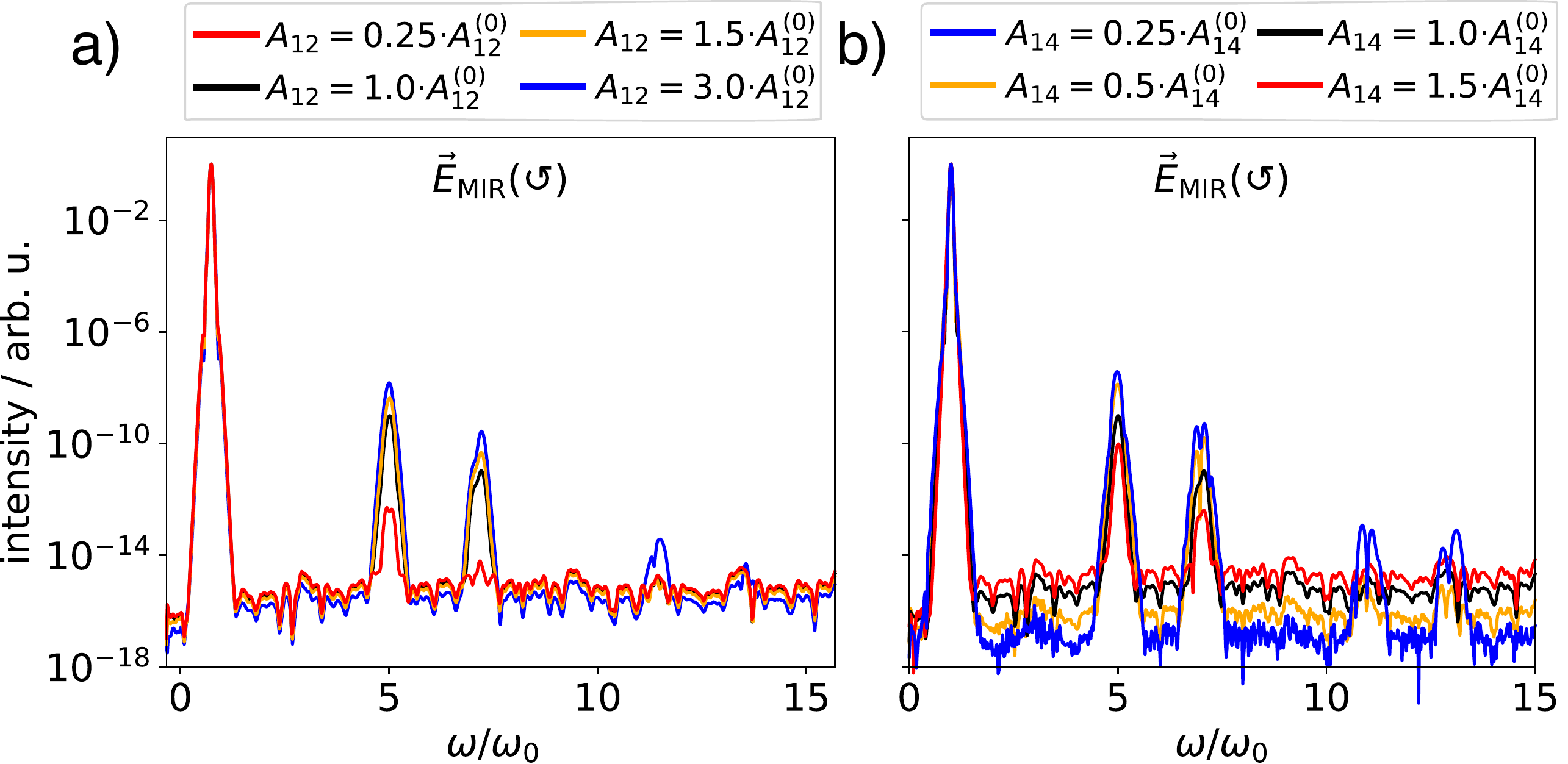}
\caption{ HHG spectra emitted from the bulk states driven by a 12-cycle left-circularly polarized pulse with a peak intensity of $I_0=0.004\ \mathrm{TW\: cm}^{-2}$, whereby one of the TBM parameters $A_{12}$ (panel a) or $A_{14}$ (panel b) is varied (s. legend). The spectra corresponding to the parameters listed in Tab.~\ref{tab:tbm} are shown in black.  }
\label{fig:CPL_ratio_dep_bulk}

\end{figure}

This section contains complementary calculations related to the ellipticity dependence of the bulk states. Figure~\ref{fig:CPL_ratio_dep_bulk} shows the effect of the variation of the TBM parameters $A_{12}$ and $A_{14}$ on the emitted HHG under illumination with MIR CPL fields. 


%
%

\mbox{}\newline
\clearpage

\bibliographystyle{apsrev4-1}
\bibliography{Bi2Se3_HHG}

\begin{thebibliography}{62}%
\makeatletter
\providecommand \@ifxundefined [1]{%
 \@ifx{#1\undefined}
}%
\providecommand \@ifnum [1]{%
 \ifnum #1\expandafter \@firstoftwo
 \else \expandafter \@secondoftwo
 \fi
}%
\providecommand \@ifx [1]{%
 \ifx #1\expandafter \@firstoftwo
 \else \expandafter \@secondoftwo
 \fi
}%
\providecommand \natexlab [1]{#1}%
\providecommand \enquote  [1]{``#1''}%
\providecommand \bibnamefont  [1]{#1}%
\providecommand \bibfnamefont [1]{#1}%
\providecommand \citenamefont [1]{#1}%
\providecommand \href@noop [0]{\@secondoftwo}%
\providecommand \href [0]{\begingroup \@sanitize@url \@href}%
\providecommand \@href[1]{\@@startlink{#1}\@@href}%
\providecommand \@@href[1]{\endgroup#1\@@endlink}%
\providecommand \@sanitize@url [0]{\catcode `\\12\catcode `\$12\catcode
  `\&12\catcode `\#12\catcode `\^12\catcode `\_12\catcode `\%12\relax}%
\providecommand \@@startlink[1]{}%
\providecommand \@@endlink[0]{}%
\providecommand \url  [0]{\begingroup\@sanitize@url \@url }%
\providecommand \@url [1]{\endgroup\@href {#1}{\urlprefix }}%
\providecommand \urlprefix  [0]{URL }%
\providecommand \Eprint [0]{\href }%
\providecommand \doibase [0]{http://dx.doi.org/}%
\providecommand \selectlanguage [0]{\@gobble}%
\providecommand \bibinfo  [0]{\@secondoftwo}%
\providecommand \bibfield  [0]{\@secondoftwo}%
\providecommand \translation [1]{[#1]}%
\providecommand \BibitemOpen [0]{}%
\providecommand \bibitemStop [0]{}%
\providecommand \bibitemNoStop [0]{.\EOS\space}%
\providecommand \EOS [0]{\spacefactor3000\relax}%
\providecommand \BibitemShut  [1]{\csname bibitem#1\endcsname}%
\let\auto@bib@innerbib\@empty
\bibitem [{\citenamefont {Corkum}(1993)}]{Corkum1993}%
  \BibitemOpen
  \bibfield  {author} {\bibinfo {author} {\bibfnamefont {P.~B.}\ \bibnamefont
  {Corkum}},\ }\href {https://link.aps.org/doi/10.1103/PhysRevLett.71.1994}
  {\bibfield  {journal} {\bibinfo  {journal} {Physical Review Letters}\
  }\textbf {\bibinfo {volume} {71}},\ \bibinfo {pages} {1994} (\bibinfo {year}
  {1993})}\BibitemShut {NoStop}%
\bibitem [{\citenamefont {Lewenstein}\ \emph {et~al.}(1994)\citenamefont
  {Lewenstein}, \citenamefont {Balcou}, \citenamefont {Ivanov}, \citenamefont
  {L'Huillier},\ and\ \citenamefont {Corkum}}]{Lewenstein1994}%
  \BibitemOpen
  \bibfield  {author} {\bibinfo {author} {\bibfnamefont {M.}~\bibnamefont
  {Lewenstein}}, \bibinfo {author} {\bibfnamefont {P.}~\bibnamefont {Balcou}},
  \bibinfo {author} {\bibfnamefont {M.~Y.}\ \bibnamefont {Ivanov}}, \bibinfo
  {author} {\bibfnamefont {A.}~\bibnamefont {L'Huillier}}, \ and\ \bibinfo
  {author} {\bibfnamefont {P.~B.}\ \bibnamefont {Corkum}},\ }\href
  {https://link.aps.org/doi/10.1103/PhysRevA.49.2117} {\bibfield  {journal}
  {\bibinfo  {journal} {Physical Review A}\ }\textbf {\bibinfo {volume} {49}},\
  \bibinfo {pages} {2117} (\bibinfo {year} {1994})}\BibitemShut {NoStop}%
\bibitem [{\citenamefont {Brabec}\ and\ \citenamefont
  {Krausz}(2000)}]{Brabec2000}%
  \BibitemOpen
  \bibfield  {author} {\bibinfo {author} {\bibfnamefont {T.}~\bibnamefont
  {Brabec}}\ and\ \bibinfo {author} {\bibfnamefont {F.}~\bibnamefont
  {Krausz}},\ }\href {\doibase 10.1103/RevModPhys.72.545} {\bibfield  {journal}
  {\bibinfo  {journal} {Reviews of Modern Physics}\ }\textbf {\bibinfo {volume}
  {72}},\ \bibinfo {pages} {545} (\bibinfo {year} {2000})}\BibitemShut
  {NoStop}%
\bibitem [{\citenamefont {Corkum}\ and\ \citenamefont
  {Krausz}(2007)}]{Corkum2007}%
  \BibitemOpen
  \bibfield  {author} {\bibinfo {author} {\bibfnamefont {P.~B.}\ \bibnamefont
  {Corkum}}\ and\ \bibinfo {author} {\bibfnamefont {F.}~\bibnamefont
  {Krausz}},\ }\href {\doibase doi:10.1038/nphys620} {\bibfield  {journal}
  {\bibinfo  {journal} {Nature Physics}\ }\textbf {\bibinfo {volume} {3}},\
  \bibinfo {pages} {381} (\bibinfo {year} {2007})}\BibitemShut {NoStop}%
\bibitem [{\citenamefont {Krausz}\ and\ \citenamefont
  {Ivanov}(2009)}]{Krausz2009}%
  \BibitemOpen
  \bibfield  {author} {\bibinfo {author} {\bibfnamefont {F.}~\bibnamefont
  {Krausz}}\ and\ \bibinfo {author} {\bibfnamefont {M.}~\bibnamefont
  {Ivanov}},\ }\href {\doibase 10.1103/RevModPhys.81.163} {\bibfield  {journal}
  {\bibinfo  {journal} {Reviews of Modern Physics}\ }\textbf {\bibinfo {volume}
  {81}},\ \bibinfo {pages} {163} (\bibinfo {year} {2009})},\ \Eprint
  {http://arxiv.org/abs/1102.1291} {arXiv:1102.1291} \BibitemShut {NoStop}%
\bibitem [{\citenamefont {Li}\ \emph {et~al.}(2020)\citenamefont {Li},
  \citenamefont {Lu}, \citenamefont {Chew}, \citenamefont {Han}, \citenamefont
  {Li}, \citenamefont {Wu}, \citenamefont {Wang}, \citenamefont {Ghimire},\
  and\ \citenamefont {Chang}}]{Li2020}%
  \BibitemOpen
  \bibfield  {author} {\bibinfo {author} {\bibfnamefont {J.}~\bibnamefont
  {Li}}, \bibinfo {author} {\bibfnamefont {J.}~\bibnamefont {Lu}}, \bibinfo
  {author} {\bibfnamefont {A.}~\bibnamefont {Chew}}, \bibinfo {author}
  {\bibfnamefont {S.}~\bibnamefont {Han}}, \bibinfo {author} {\bibfnamefont
  {J.}~\bibnamefont {Li}}, \bibinfo {author} {\bibfnamefont {Y.}~\bibnamefont
  {Wu}}, \bibinfo {author} {\bibfnamefont {H.}~\bibnamefont {Wang}}, \bibinfo
  {author} {\bibfnamefont {S.}~\bibnamefont {Ghimire}}, \ and\ \bibinfo
  {author} {\bibfnamefont {Z.}~\bibnamefont {Chang}},\ }\href {\doibase
  10.1038/s41467-020-16480-6} {\bibfield  {journal} {\bibinfo  {journal}
  {Nature Communications}\ }\textbf {\bibinfo {volume} {11}},\ \bibinfo {pages}
  {2748} (\bibinfo {year} {2020})}\BibitemShut {NoStop}%
\bibitem [{\citenamefont {Itatani}\ \emph {et~al.}(2004)\citenamefont
  {Itatani}, \citenamefont {Levesque}, \citenamefont {Zeidler}, \citenamefont
  {Niikura}, \citenamefont {P{\'{e}}pin}, \citenamefont {Kieffer},
  \citenamefont {Corkum},\ and\ \citenamefont {Villeneuve}}]{Itatani2004}%
  \BibitemOpen
  \bibfield  {author} {\bibinfo {author} {\bibfnamefont {J.}~\bibnamefont
  {Itatani}}, \bibinfo {author} {\bibfnamefont {J.}~\bibnamefont {Levesque}},
  \bibinfo {author} {\bibfnamefont {D.}~\bibnamefont {Zeidler}}, \bibinfo
  {author} {\bibfnamefont {H.}~\bibnamefont {Niikura}}, \bibinfo {author}
  {\bibfnamefont {H.}~\bibnamefont {P{\'{e}}pin}}, \bibinfo {author}
  {\bibfnamefont {J.~C.}\ \bibnamefont {Kieffer}}, \bibinfo {author}
  {\bibfnamefont {P.~B.}\ \bibnamefont {Corkum}}, \ and\ \bibinfo {author}
  {\bibfnamefont {D.~M.}\ \bibnamefont {Villeneuve}},\ }\href {\doibase
  10.1038/nature03183} {\bibfield  {journal} {\bibinfo  {journal} {Nature}\
  }\textbf {\bibinfo {volume} {432}},\ \bibinfo {pages} {867} (\bibinfo {year}
  {2004})}\BibitemShut {NoStop}%
\bibitem [{\citenamefont {L{\'{e}}pine}\ \emph {et~al.}(2014)\citenamefont
  {L{\'{e}}pine}, \citenamefont {Ivanov},\ and\ \citenamefont
  {Vrakking}}]{Lepine2014}%
  \BibitemOpen
  \bibfield  {author} {\bibinfo {author} {\bibfnamefont {F.}~\bibnamefont
  {L{\'{e}}pine}}, \bibinfo {author} {\bibfnamefont {M.~Y.}\ \bibnamefont
  {Ivanov}}, \ and\ \bibinfo {author} {\bibfnamefont {M.~J.}\ \bibnamefont
  {Vrakking}},\ }\href {\doibase 10.1038/nphoton.2014.25} {\bibfield  {journal}
  {\bibinfo  {journal} {Nature Photonics}\ }\textbf {\bibinfo {volume} {8}},\
  \bibinfo {pages} {195} (\bibinfo {year} {2014})}\BibitemShut {NoStop}%
\bibitem [{\citenamefont {Ghimire}\ \emph {et~al.}(2011)\citenamefont
  {Ghimire}, \citenamefont {DiChiara}, \citenamefont {Sistrunk}, \citenamefont
  {Agostini}, \citenamefont {DiMauro},\ and\ \citenamefont
  {Reis}}]{Ghimire2011}%
  \BibitemOpen
  \bibfield  {author} {\bibinfo {author} {\bibfnamefont {S.}~\bibnamefont
  {Ghimire}}, \bibinfo {author} {\bibfnamefont {A.~D.}\ \bibnamefont
  {DiChiara}}, \bibinfo {author} {\bibfnamefont {E.}~\bibnamefont {Sistrunk}},
  \bibinfo {author} {\bibfnamefont {P.}~\bibnamefont {Agostini}}, \bibinfo
  {author} {\bibfnamefont {L.~F.}\ \bibnamefont {DiMauro}}, \ and\ \bibinfo
  {author} {\bibfnamefont {D.~A.}\ \bibnamefont {Reis}},\ }\href {\doibase
  10.1038/nphys1847} {\bibfield  {journal} {\bibinfo  {journal} {Nature
  Physics}\ }\textbf {\bibinfo {volume} {7}},\ \bibinfo {pages} {138} (\bibinfo
  {year} {2011})}\BibitemShut {NoStop}%
\bibitem [{\citenamefont {Ghimire}\ and\ \citenamefont
  {Reis}(2018)}]{Ghimire2018}%
  \BibitemOpen
  \bibfield  {author} {\bibinfo {author} {\bibfnamefont {S.}~\bibnamefont
  {Ghimire}}\ and\ \bibinfo {author} {\bibfnamefont {D.~A.}\ \bibnamefont
  {Reis}},\ }\href {\doibase 10.1038/s41567-018-0315-5} {\bibfield  {journal}
  {\bibinfo  {journal} {Nature Physics}\ } (\bibinfo {year} {2018}),\
  10.1038/s41567-018-0315-5}\BibitemShut {NoStop}%
\bibitem [{\citenamefont {Schubert}\ \emph {et~al.}(2014)\citenamefont
  {Schubert}, \citenamefont {Hohenleutner}, \citenamefont {Langer},
  \citenamefont {Urbanek}, \citenamefont {Lange}, \citenamefont {Huttner},
  \citenamefont {Golde}, \citenamefont {Meier}, \citenamefont {Kira},
  \citenamefont {Koch},\ and\ \citenamefont {Huber}}]{Schubert2014}%
  \BibitemOpen
  \bibfield  {author} {\bibinfo {author} {\bibfnamefont {O.}~\bibnamefont
  {Schubert}}, \bibinfo {author} {\bibfnamefont {M.}~\bibnamefont
  {Hohenleutner}}, \bibinfo {author} {\bibfnamefont {F.}~\bibnamefont
  {Langer}}, \bibinfo {author} {\bibfnamefont {B.}~\bibnamefont {Urbanek}},
  \bibinfo {author} {\bibfnamefont {C.}~\bibnamefont {Lange}}, \bibinfo
  {author} {\bibfnamefont {U.}~\bibnamefont {Huttner}}, \bibinfo {author}
  {\bibfnamefont {D.}~\bibnamefont {Golde}}, \bibinfo {author} {\bibfnamefont
  {T.}~\bibnamefont {Meier}}, \bibinfo {author} {\bibfnamefont
  {M.}~\bibnamefont {Kira}}, \bibinfo {author} {\bibfnamefont {S.~W.}\
  \bibnamefont {Koch}}, \ and\ \bibinfo {author} {\bibfnamefont
  {R.}~\bibnamefont {Huber}},\ }\href {\doibase 10.1038/nphoton.2013.349}
  {\bibfield  {journal} {\bibinfo  {journal} {Nature Photonics}\ }\textbf
  {\bibinfo {volume} {8}},\ \bibinfo {pages} {119} (\bibinfo {year}
  {2014})}\BibitemShut {NoStop}%
\bibitem [{\citenamefont {Luu}\ \emph {et~al.}(2015)\citenamefont {Luu},
  \citenamefont {Garg}, \citenamefont {{Yu. Kruchinin}}, \citenamefont
  {Moulet}, \citenamefont {Hassan},\ and\ \citenamefont
  {Goulielmakis}}]{Luu2015}%
  \BibitemOpen
  \bibfield  {author} {\bibinfo {author} {\bibfnamefont {T.~T.}\ \bibnamefont
  {Luu}}, \bibinfo {author} {\bibfnamefont {M.}~\bibnamefont {Garg}}, \bibinfo
  {author} {\bibfnamefont {S.}~\bibnamefont {{Yu. Kruchinin}}}, \bibinfo
  {author} {\bibfnamefont {A.}~\bibnamefont {Moulet}}, \bibinfo {author}
  {\bibfnamefont {M.~T.}\ \bibnamefont {Hassan}}, \ and\ \bibinfo {author}
  {\bibfnamefont {E.}~\bibnamefont {Goulielmakis}},\ }\href {\doibase
  10.1038/nature14456} {\bibfield  {journal} {\bibinfo  {journal} {Nature}\
  }\textbf {\bibinfo {volume} {521}},\ \bibinfo {pages} {498} (\bibinfo {year}
  {2015})},\ \Eprint {http://arxiv.org/abs/1604.03768} {arXiv:1604.03768}
  \BibitemShut {NoStop}%
\bibitem [{\citenamefont {Vampa}\ \emph {et~al.}(2015)\citenamefont {Vampa},
  \citenamefont {Hammond}, \citenamefont {Thir{\'{e}}}, \citenamefont
  {Schmidt}, \citenamefont {L{\'{e}}gar{\'{e}}}, \citenamefont {McDonald},
  \citenamefont {Brabec}, \citenamefont {Klug},\ and\ \citenamefont
  {Corkum}}]{Vampa2015}%
  \BibitemOpen
  \bibfield  {author} {\bibinfo {author} {\bibfnamefont {G.}~\bibnamefont
  {Vampa}}, \bibinfo {author} {\bibfnamefont {T.~J.}\ \bibnamefont {Hammond}},
  \bibinfo {author} {\bibfnamefont {N.}~\bibnamefont {Thir{\'{e}}}}, \bibinfo
  {author} {\bibfnamefont {B.~E.}\ \bibnamefont {Schmidt}}, \bibinfo {author}
  {\bibfnamefont {F.}~\bibnamefont {L{\'{e}}gar{\'{e}}}}, \bibinfo {author}
  {\bibfnamefont {C.~R.}\ \bibnamefont {McDonald}}, \bibinfo {author}
  {\bibfnamefont {T.}~\bibnamefont {Brabec}}, \bibinfo {author} {\bibfnamefont
  {D.~D.}\ \bibnamefont {Klug}}, \ and\ \bibinfo {author} {\bibfnamefont
  {P.~B.}\ \bibnamefont {Corkum}},\ }\href {\doibase
  10.1103/PhysRevLett.115.193603} {\bibfield  {journal} {\bibinfo  {journal}
  {Physical Review Letters}\ }\textbf {\bibinfo {volume} {115}},\ \bibinfo
  {pages} {193603} (\bibinfo {year} {2015})}\BibitemShut {NoStop}%
\bibitem [{\citenamefont {Hohenleutner}\ \emph {et~al.}(2015)\citenamefont
  {Hohenleutner}, \citenamefont {Langer}, \citenamefont {Schubert},
  \citenamefont {Knorr}, \citenamefont {Huttner}, \citenamefont {Koch},
  \citenamefont {Kira},\ and\ \citenamefont {Huber}}]{Hohenleutner2015}%
  \BibitemOpen
  \bibfield  {author} {\bibinfo {author} {\bibfnamefont {M.}~\bibnamefont
  {Hohenleutner}}, \bibinfo {author} {\bibfnamefont {F.}~\bibnamefont
  {Langer}}, \bibinfo {author} {\bibfnamefont {O.}~\bibnamefont {Schubert}},
  \bibinfo {author} {\bibfnamefont {M.}~\bibnamefont {Knorr}}, \bibinfo
  {author} {\bibfnamefont {U.}~\bibnamefont {Huttner}}, \bibinfo {author}
  {\bibfnamefont {S.~W.}\ \bibnamefont {Koch}}, \bibinfo {author}
  {\bibfnamefont {M.}~\bibnamefont {Kira}}, \ and\ \bibinfo {author}
  {\bibfnamefont {R.}~\bibnamefont {Huber}},\ }\href {\doibase
  10.1038/nature14652} {\bibfield  {journal} {\bibinfo  {journal} {Nature}\
  }\textbf {\bibinfo {volume} {523}},\ \bibinfo {pages} {572} (\bibinfo {year}
  {2015})},\ \Eprint {http://arxiv.org/abs/1604.03768} {arXiv:1604.03768}
  \BibitemShut {NoStop}%
\bibitem [{\citenamefont {McDonald}\ \emph {et~al.}(2015)\citenamefont
  {McDonald}, \citenamefont {Vampa}, \citenamefont {Corkum},\ and\
  \citenamefont {Brabec}}]{McDonald2015}%
  \BibitemOpen
  \bibfield  {author} {\bibinfo {author} {\bibfnamefont {C.~R.}\ \bibnamefont
  {McDonald}}, \bibinfo {author} {\bibfnamefont {G.}~\bibnamefont {Vampa}},
  \bibinfo {author} {\bibfnamefont {P.~B.}\ \bibnamefont {Corkum}}, \ and\
  \bibinfo {author} {\bibfnamefont {T.}~\bibnamefont {Brabec}},\ }\href
  {\doibase 10.1103/PhysRevA.92.033845} {\bibfield  {journal} {\bibinfo
  {journal} {Physical Review A - Atomic, Molecular, and Optical Physics}\
  }\textbf {\bibinfo {volume} {92}},\ \bibinfo {pages} {033845} (\bibinfo
  {year} {2015})}\BibitemShut {NoStop}%
\bibitem [{\citenamefont {You}\ \emph {et~al.}(2017)\citenamefont {You},
  \citenamefont {Wu}, \citenamefont {Yin}, \citenamefont {Chew}, \citenamefont
  {Ren}, \citenamefont {Gholam-Mirzaei}, \citenamefont {Browne}, \citenamefont
  {Chini}, \citenamefont {Chang}, \citenamefont {Schafer}, \citenamefont
  {Gaarde},\ and\ \citenamefont {Ghimire}}]{You2017b}%
  \BibitemOpen
  \bibfield  {author} {\bibinfo {author} {\bibfnamefont {Y.~S.}\ \bibnamefont
  {You}}, \bibinfo {author} {\bibfnamefont {M.}~\bibnamefont {Wu}}, \bibinfo
  {author} {\bibfnamefont {Y.}~\bibnamefont {Yin}}, \bibinfo {author}
  {\bibfnamefont {A.}~\bibnamefont {Chew}}, \bibinfo {author} {\bibfnamefont
  {X.}~\bibnamefont {Ren}}, \bibinfo {author} {\bibfnamefont {S.}~\bibnamefont
  {Gholam-Mirzaei}}, \bibinfo {author} {\bibfnamefont {D.~A.}\ \bibnamefont
  {Browne}}, \bibinfo {author} {\bibfnamefont {M.}~\bibnamefont {Chini}},
  \bibinfo {author} {\bibfnamefont {Z.}~\bibnamefont {Chang}}, \bibinfo
  {author} {\bibfnamefont {K.~J.}\ \bibnamefont {Schafer}}, \bibinfo {author}
  {\bibfnamefont {M.~B.}\ \bibnamefont {Gaarde}}, \ and\ \bibinfo {author}
  {\bibfnamefont {S.}~\bibnamefont {Ghimire}},\ }\href {\doibase
  10.1364/OL.42.001816} {\bibfield  {journal} {\bibinfo  {journal} {Optics
  Letters}\ }\textbf {\bibinfo {volume} {42}},\ \bibinfo {pages} {1816}
  (\bibinfo {year} {2017})}\BibitemShut {NoStop}%
\bibitem [{\citenamefont {Liu}\ \emph {et~al.}(2017)\citenamefont {Liu},
  \citenamefont {Li}, \citenamefont {You}, \citenamefont {Ghimire},
  \citenamefont {Heinz},\ and\ \citenamefont {Reis}}]{Liu2017a}%
  \BibitemOpen
  \bibfield  {author} {\bibinfo {author} {\bibfnamefont {H.}~\bibnamefont
  {Liu}}, \bibinfo {author} {\bibfnamefont {Y.}~\bibnamefont {Li}}, \bibinfo
  {author} {\bibfnamefont {Y.~S.}\ \bibnamefont {You}}, \bibinfo {author}
  {\bibfnamefont {S.}~\bibnamefont {Ghimire}}, \bibinfo {author} {\bibfnamefont
  {T.~F.}\ \bibnamefont {Heinz}}, \ and\ \bibinfo {author} {\bibfnamefont
  {D.~A.}\ \bibnamefont {Reis}},\ }\href {\doibase 10.1038/nphys3946}
  {\bibfield  {journal} {\bibinfo  {journal} {Nature Physics}\ }\textbf
  {\bibinfo {volume} {13}},\ \bibinfo {pages} {262} (\bibinfo {year}
  {2017})}\BibitemShut {NoStop}%
\bibitem [{\citenamefont {Luu}\ and\ \citenamefont
  {W{\"{o}}rner}(2018)}]{Luu2018}%
  \BibitemOpen
  \bibfield  {author} {\bibinfo {author} {\bibfnamefont {T.~T.}\ \bibnamefont
  {Luu}}\ and\ \bibinfo {author} {\bibfnamefont {H.~J.}\ \bibnamefont
  {W{\"{o}}rner}},\ }\href {\doibase 10.1038/s41467-018-03397-4} {\bibfield
  {journal} {\bibinfo  {journal} {Nature Communications}\ }\textbf {\bibinfo
  {volume} {9}},\ \bibinfo {pages} {916} (\bibinfo {year} {2018})}\BibitemShut
  {NoStop}%
\bibitem [{\citenamefont {Hsieh}\ \emph {et~al.}(2008)\citenamefont {Hsieh},
  \citenamefont {Qian}, \citenamefont {Wray}, \citenamefont {Xia},
  \citenamefont {Hor}, \citenamefont {Cava},\ and\ \citenamefont
  {Hasan}}]{Hsieh2008}%
  \BibitemOpen
  \bibfield  {author} {\bibinfo {author} {\bibfnamefont {D.}~\bibnamefont
  {Hsieh}}, \bibinfo {author} {\bibfnamefont {D.}~\bibnamefont {Qian}},
  \bibinfo {author} {\bibfnamefont {L.}~\bibnamefont {Wray}}, \bibinfo {author}
  {\bibfnamefont {Y.}~\bibnamefont {Xia}}, \bibinfo {author} {\bibfnamefont
  {Y.~S.}\ \bibnamefont {Hor}}, \bibinfo {author} {\bibfnamefont {R.~J.}\
  \bibnamefont {Cava}}, \ and\ \bibinfo {author} {\bibfnamefont {M.~Z.}\
  \bibnamefont {Hasan}},\ }\href {\doibase 10.1038/nature06843} {\bibfield
  {journal} {\bibinfo  {journal} {Nature}\ }\textbf {\bibinfo {volume} {452}},\
  \bibinfo {pages} {970} (\bibinfo {year} {2008})},\ \Eprint
  {http://arxiv.org/abs/0910.2420} {arXiv:0910.2420} \BibitemShut {NoStop}%
\bibitem [{\citenamefont {Xia}\ \emph {et~al.}(2009)\citenamefont {Xia},
  \citenamefont {Qian}, \citenamefont {Hsieh}, \citenamefont {Wray},
  \citenamefont {Pal}, \citenamefont {Lin}, \citenamefont {Bansil},
  \citenamefont {Grauer}, \citenamefont {Hor}, \citenamefont {Cava},\ and\
  \citenamefont {Hasan}}]{Xia2009}%
  \BibitemOpen
  \bibfield  {author} {\bibinfo {author} {\bibfnamefont {Y.}~\bibnamefont
  {Xia}}, \bibinfo {author} {\bibfnamefont {D.}~\bibnamefont {Qian}}, \bibinfo
  {author} {\bibfnamefont {D.}~\bibnamefont {Hsieh}}, \bibinfo {author}
  {\bibfnamefont {L.}~\bibnamefont {Wray}}, \bibinfo {author} {\bibfnamefont
  {A.}~\bibnamefont {Pal}}, \bibinfo {author} {\bibfnamefont {H.}~\bibnamefont
  {Lin}}, \bibinfo {author} {\bibfnamefont {A.}~\bibnamefont {Bansil}},
  \bibinfo {author} {\bibfnamefont {D.}~\bibnamefont {Grauer}}, \bibinfo
  {author} {\bibfnamefont {Y.~S.}\ \bibnamefont {Hor}}, \bibinfo {author}
  {\bibfnamefont {R.~J.}\ \bibnamefont {Cava}}, \ and\ \bibinfo {author}
  {\bibfnamefont {M.~Z.}\ \bibnamefont {Hasan}},\ }\href {\doibase
  10.1038/nphys1274} {\bibfield  {journal} {\bibinfo  {journal} {Nature
  Physics}\ }\textbf {\bibinfo {volume} {5}},\ \bibinfo {pages} {398} (\bibinfo
  {year} {2009})},\ \Eprint {http://arxiv.org/abs/0908.3513} {arXiv:0908.3513}
  \BibitemShut {NoStop}%
\bibitem [{\citenamefont {Hsieh}\ \emph
  {et~al.}(2009{\natexlab{a}})\citenamefont {Hsieh}, \citenamefont {Xia},
  \citenamefont {Qian}, \citenamefont {Wray}, \citenamefont {Dil},
  \citenamefont {Meier}, \citenamefont {Osterwalder}, \citenamefont {Patthey},
  \citenamefont {Checkelsky}, \citenamefont {Ong}, \citenamefont {Fedorov},
  \citenamefont {Lin}, \citenamefont {Bansil}, \citenamefont {Grauer},
  \citenamefont {Hor}, \citenamefont {Cava},\ and\ \citenamefont
  {Hasan}}]{Hsieh2009b}%
  \BibitemOpen
  \bibfield  {author} {\bibinfo {author} {\bibfnamefont {D.}~\bibnamefont
  {Hsieh}}, \bibinfo {author} {\bibfnamefont {Y.}~\bibnamefont {Xia}}, \bibinfo
  {author} {\bibfnamefont {D.}~\bibnamefont {Qian}}, \bibinfo {author}
  {\bibfnamefont {L.}~\bibnamefont {Wray}}, \bibinfo {author} {\bibfnamefont
  {J.~H.}\ \bibnamefont {Dil}}, \bibinfo {author} {\bibfnamefont
  {F.}~\bibnamefont {Meier}}, \bibinfo {author} {\bibfnamefont
  {J.}~\bibnamefont {Osterwalder}}, \bibinfo {author} {\bibfnamefont
  {L.}~\bibnamefont {Patthey}}, \bibinfo {author} {\bibfnamefont {J.~G.}\
  \bibnamefont {Checkelsky}}, \bibinfo {author} {\bibfnamefont {N.~P.}\
  \bibnamefont {Ong}}, \bibinfo {author} {\bibfnamefont {A.~V.}\ \bibnamefont
  {Fedorov}}, \bibinfo {author} {\bibfnamefont {H.}~\bibnamefont {Lin}},
  \bibinfo {author} {\bibfnamefont {A.}~\bibnamefont {Bansil}}, \bibinfo
  {author} {\bibfnamefont {D.}~\bibnamefont {Grauer}}, \bibinfo {author}
  {\bibfnamefont {Y.~S.}\ \bibnamefont {Hor}}, \bibinfo {author} {\bibfnamefont
  {R.~J.}\ \bibnamefont {Cava}}, \ and\ \bibinfo {author} {\bibfnamefont
  {M.~Z.}\ \bibnamefont {Hasan}},\ }\href {\doibase 10.1038/nature08234}
  {\bibfield  {journal} {\bibinfo  {journal} {Nature}\ }\textbf {\bibinfo
  {volume} {460}},\ \bibinfo {pages} {1101} (\bibinfo {year}
  {2009}{\natexlab{a}})},\ \Eprint {http://arxiv.org/abs/arXiv:0908.0564}
  {arXiv:arXiv:0908.0564} \BibitemShut {NoStop}%
\bibitem [{\citenamefont {Chen}\ \emph {et~al.}(2009)\citenamefont {Chen},
  \citenamefont {Analytis}, \citenamefont {Chu}, \citenamefont {Liu},
  \citenamefont {Mo}, \citenamefont {Qi}, \citenamefont {Zhang}, \citenamefont
  {Lu}, \citenamefont {Dai}, \citenamefont {Fang}, \citenamefont {Zhang},
  \citenamefont {Fisher}, \citenamefont {Hussain},\ and\ \citenamefont
  {Shen}}]{Chen2009}%
  \BibitemOpen
  \bibfield  {author} {\bibinfo {author} {\bibfnamefont {Y.~L.}\ \bibnamefont
  {Chen}}, \bibinfo {author} {\bibfnamefont {J.~G.}\ \bibnamefont {Analytis}},
  \bibinfo {author} {\bibfnamefont {J.-H.}\ \bibnamefont {Chu}}, \bibinfo
  {author} {\bibfnamefont {Z.~K.}\ \bibnamefont {Liu}}, \bibinfo {author}
  {\bibfnamefont {S.-K.}\ \bibnamefont {Mo}}, \bibinfo {author} {\bibfnamefont
  {X.~L.}\ \bibnamefont {Qi}}, \bibinfo {author} {\bibfnamefont {H.~J.}\
  \bibnamefont {Zhang}}, \bibinfo {author} {\bibfnamefont {D.~H.}\ \bibnamefont
  {Lu}}, \bibinfo {author} {\bibfnamefont {X.}~\bibnamefont {Dai}}, \bibinfo
  {author} {\bibfnamefont {Z.}~\bibnamefont {Fang}}, \bibinfo {author}
  {\bibfnamefont {S.~C.}\ \bibnamefont {Zhang}}, \bibinfo {author}
  {\bibfnamefont {I.~R.}\ \bibnamefont {Fisher}}, \bibinfo {author}
  {\bibfnamefont {Z.}~\bibnamefont {Hussain}}, \ and\ \bibinfo {author}
  {\bibfnamefont {Z.-X.}\ \bibnamefont {Shen}},\ }\href {\doibase
  10.1126/science.1173034} {\bibfield  {journal} {\bibinfo  {journal}
  {Science}\ }\textbf {\bibinfo {volume} {325}},\ \bibinfo {pages} {178}
  (\bibinfo {year} {2009})},\ \Eprint {http://arxiv.org/abs/arXiv:0904.1829v1}
  {arXiv:arXiv:0904.1829v1} \BibitemShut {NoStop}%
\bibitem [{\citenamefont {Moore}(2010)}]{Moore2010}%
  \BibitemOpen
  \bibfield  {author} {\bibinfo {author} {\bibfnamefont {J.~E.}\ \bibnamefont
  {Moore}},\ }\href {\doibase 10.1038/nature08916} {\bibfield  {journal}
  {\bibinfo  {journal} {Nature}\ }\textbf {\bibinfo {volume} {464}},\ \bibinfo
  {pages} {194} (\bibinfo {year} {2010})},\ \Eprint
  {http://arxiv.org/abs/arXiv:1011.5462v1} {arXiv:arXiv:1011.5462v1}
  \BibitemShut {NoStop}%
\bibitem [{\citenamefont {Hasan}\ and\ \citenamefont
  {Kane}(2010)}]{Hasan2010a}%
  \BibitemOpen
  \bibfield  {author} {\bibinfo {author} {\bibfnamefont {M.~Z.}\ \bibnamefont
  {Hasan}}\ and\ \bibinfo {author} {\bibfnamefont {C.~L.}\ \bibnamefont
  {Kane}},\ }\href {\doibase 10.1103/RevModPhys.82.3045} {\bibfield  {journal}
  {\bibinfo  {journal} {Reviews of Modern Physics}\ }\textbf {\bibinfo {volume}
  {82}},\ \bibinfo {pages} {3045} (\bibinfo {year} {2010})},\ \Eprint
  {http://arxiv.org/abs/1002.3895} {arXiv:1002.3895} \BibitemShut {NoStop}%
\bibitem [{\citenamefont {Qi}\ and\ \citenamefont {Zhang}(2011)}]{Qi2011}%
  \BibitemOpen
  \bibfield  {author} {\bibinfo {author} {\bibfnamefont {X.~L.}\ \bibnamefont
  {Qi}}\ and\ \bibinfo {author} {\bibfnamefont {S.~C.}\ \bibnamefont {Zhang}},\
  }\href {\doibase 10.1103/RevModPhys.83.1057} {\bibfield  {journal} {\bibinfo
  {journal} {Reviews of Modern Physics}\ }\textbf {\bibinfo {volume} {83}},\
  \bibinfo {pages} {1057} (\bibinfo {year} {2011})},\ \Eprint
  {http://arxiv.org/abs/1008.2026} {arXiv:1008.2026} \BibitemShut {NoStop}%
\bibitem [{\citenamefont {Zhang}\ \emph {et~al.}(2009)\citenamefont {Zhang},
  \citenamefont {Liu}, \citenamefont {Qi}, \citenamefont {Dai}, \citenamefont
  {Fang},\ and\ \citenamefont {Zhang}}]{Zhang2009}%
  \BibitemOpen
  \bibfield  {author} {\bibinfo {author} {\bibfnamefont {H.}~\bibnamefont
  {Zhang}}, \bibinfo {author} {\bibfnamefont {C.-X.}\ \bibnamefont {Liu}},
  \bibinfo {author} {\bibfnamefont {X.-L.}\ \bibnamefont {Qi}}, \bibinfo
  {author} {\bibfnamefont {X.}~\bibnamefont {Dai}}, \bibinfo {author}
  {\bibfnamefont {Z.}~\bibnamefont {Fang}}, \ and\ \bibinfo {author}
  {\bibfnamefont {S.-C.}\ \bibnamefont {Zhang}},\ }\href {\doibase
  10.1038/nphys1270} {\bibfield  {journal} {\bibinfo  {journal} {Nature
  Physics}\ }\textbf {\bibinfo {volume} {5}},\ \bibinfo {pages} {438} (\bibinfo
  {year} {2009})},\ \Eprint {http://arxiv.org/abs/1405.2036} {arXiv:1405.2036}
  \BibitemShut {NoStop}%
\bibitem [{\citenamefont {Roushan}\ \emph {et~al.}(2009)\citenamefont
  {Roushan}, \citenamefont {Seo}, \citenamefont {Parker}, \citenamefont {Hor},
  \citenamefont {Hsieh}, \citenamefont {Qian}, \citenamefont {Richardella},
  \citenamefont {Hasan}, \citenamefont {Cava},\ and\ \citenamefont
  {Yazdani}}]{Roushan2009}%
  \BibitemOpen
  \bibfield  {author} {\bibinfo {author} {\bibfnamefont {P.}~\bibnamefont
  {Roushan}}, \bibinfo {author} {\bibfnamefont {J.}~\bibnamefont {Seo}},
  \bibinfo {author} {\bibfnamefont {C.~V.}\ \bibnamefont {Parker}}, \bibinfo
  {author} {\bibfnamefont {Y.~S.}\ \bibnamefont {Hor}}, \bibinfo {author}
  {\bibfnamefont {D.}~\bibnamefont {Hsieh}}, \bibinfo {author} {\bibfnamefont
  {D.}~\bibnamefont {Qian}}, \bibinfo {author} {\bibfnamefont {A.}~\bibnamefont
  {Richardella}}, \bibinfo {author} {\bibfnamefont {M.~Z.}\ \bibnamefont
  {Hasan}}, \bibinfo {author} {\bibfnamefont {R.~J.}\ \bibnamefont {Cava}}, \
  and\ \bibinfo {author} {\bibfnamefont {A.}~\bibnamefont {Yazdani}},\ }\href
  {\doibase 10.1038/nature08308} {\bibfield  {journal} {\bibinfo  {journal}
  {Nature}\ }\textbf {\bibinfo {volume} {460}},\ \bibinfo {pages} {1106}
  (\bibinfo {year} {2009})},\ \Eprint {http://arxiv.org/abs/0908.1247}
  {arXiv:0908.1247} \BibitemShut {NoStop}%
\bibitem [{\citenamefont {Hsieh}\ \emph
  {et~al.}(2009{\natexlab{b}})\citenamefont {Hsieh}, \citenamefont {Xia},
  \citenamefont {Wray}, \citenamefont {Qian}, \citenamefont {Pal},
  \citenamefont {Dil}, \citenamefont {Osterwalder}, \citenamefont {Meier},
  \citenamefont {Bihlmayer}, \citenamefont {Kane}, \citenamefont {Hor},
  \citenamefont {Cava},\ and\ \citenamefont {Hasan}}]{Hsieh2009}%
  \BibitemOpen
  \bibfield  {author} {\bibinfo {author} {\bibfnamefont {D.}~\bibnamefont
  {Hsieh}}, \bibinfo {author} {\bibfnamefont {Y.}~\bibnamefont {Xia}}, \bibinfo
  {author} {\bibfnamefont {L.}~\bibnamefont {Wray}}, \bibinfo {author}
  {\bibfnamefont {D.}~\bibnamefont {Qian}}, \bibinfo {author} {\bibfnamefont
  {A.}~\bibnamefont {Pal}}, \bibinfo {author} {\bibfnamefont {J.~H.}\
  \bibnamefont {Dil}}, \bibinfo {author} {\bibfnamefont {J.}~\bibnamefont
  {Osterwalder}}, \bibinfo {author} {\bibfnamefont {F.}~\bibnamefont {Meier}},
  \bibinfo {author} {\bibfnamefont {G.}~\bibnamefont {Bihlmayer}}, \bibinfo
  {author} {\bibfnamefont {C.~L.}\ \bibnamefont {Kane}}, \bibinfo {author}
  {\bibfnamefont {Y.~S.}\ \bibnamefont {Hor}}, \bibinfo {author} {\bibfnamefont
  {R.~J.}\ \bibnamefont {Cava}}, \ and\ \bibinfo {author} {\bibfnamefont
  {M.~Z.}\ \bibnamefont {Hasan}},\ }\href {\doibase 10.1126/science.1167733}
  {\bibfield  {journal} {\bibinfo  {journal} {Science}\ }\textbf {\bibinfo
  {volume} {323}},\ \bibinfo {pages} {915} (\bibinfo {year}
  {2009}{\natexlab{b}})}\BibitemShut {NoStop}%
\bibitem [{\citenamefont {McIver}\ \emph {et~al.}(2012)\citenamefont {McIver},
  \citenamefont {Hsieh}, \citenamefont {Steinberg}, \citenamefont
  {Jarillo-Herrero},\ and\ \citenamefont {Gedik}}]{McIver2012}%
  \BibitemOpen
  \bibfield  {author} {\bibinfo {author} {\bibfnamefont {J.~W.}\ \bibnamefont
  {McIver}}, \bibinfo {author} {\bibfnamefont {D.}~\bibnamefont {Hsieh}},
  \bibinfo {author} {\bibfnamefont {H.}~\bibnamefont {Steinberg}}, \bibinfo
  {author} {\bibfnamefont {P.}~\bibnamefont {Jarillo-Herrero}}, \ and\ \bibinfo
  {author} {\bibfnamefont {N.}~\bibnamefont {Gedik}},\ }\href {\doibase
  10.1038/nnano.2011.214} {\bibfield  {journal} {\bibinfo  {journal} {Nature
  Nanotechnology}\ }\textbf {\bibinfo {volume} {7}},\ \bibinfo {pages} {96}
  (\bibinfo {year} {2012})},\ \Eprint {http://arxiv.org/abs/1111.3694}
  {arXiv:1111.3694} \BibitemShut {NoStop}%
\bibitem [{\citenamefont {{Oliaei Motlagh}}\ \emph {et~al.}(2017)\citenamefont
  {{Oliaei Motlagh}}, \citenamefont {Apalkov},\ and\ \citenamefont
  {Stockman}}]{OliaeiMotlagh2017}%
  \BibitemOpen
  \bibfield  {author} {\bibinfo {author} {\bibfnamefont {S.~A.}\ \bibnamefont
  {{Oliaei Motlagh}}}, \bibinfo {author} {\bibfnamefont {V.}~\bibnamefont
  {Apalkov}}, \ and\ \bibinfo {author} {\bibfnamefont {M.~I.}\ \bibnamefont
  {Stockman}},\ }\href {\doibase 10.1103/PhysRevB.95.085438} {\bibfield
  {journal} {\bibinfo  {journal} {Physical Review B}\ }\textbf {\bibinfo
  {volume} {95}},\ \bibinfo {pages} {085438} (\bibinfo {year}
  {2017})}\BibitemShut {NoStop}%
\bibitem [{\citenamefont {{Oliaei Motlagh}}\ \emph {et~al.}(2018)\citenamefont
  {{Oliaei Motlagh}}, \citenamefont {Wu}, \citenamefont {Apalkov},\ and\
  \citenamefont {Stockman}}]{OliaeiMotlagh2018}%
  \BibitemOpen
  \bibfield  {author} {\bibinfo {author} {\bibfnamefont {S.~A.}\ \bibnamefont
  {{Oliaei Motlagh}}}, \bibinfo {author} {\bibfnamefont {J.~S.}\ \bibnamefont
  {Wu}}, \bibinfo {author} {\bibfnamefont {V.}~\bibnamefont {Apalkov}}, \ and\
  \bibinfo {author} {\bibfnamefont {M.~I.}\ \bibnamefont {Stockman}},\ }\href
  {\doibase 10.1103/PhysRevB.98.125410} {\bibfield  {journal} {\bibinfo
  {journal} {Physical Review B}\ }\textbf {\bibinfo {volume} {98}},\ \bibinfo
  {pages} {125410} (\bibinfo {year} {2018})}\BibitemShut {NoStop}%
\bibitem [{\citenamefont {H{\"{u}}bener}\ \emph {et~al.}(2017)\citenamefont
  {H{\"{u}}bener}, \citenamefont {Sentef}, \citenamefont {{De Giovannini}},
  \citenamefont {Kemper},\ and\ \citenamefont {Rubio}}]{Hubener2017}%
  \BibitemOpen
  \bibfield  {author} {\bibinfo {author} {\bibfnamefont {H.}~\bibnamefont
  {H{\"{u}}bener}}, \bibinfo {author} {\bibfnamefont {M.~A.}\ \bibnamefont
  {Sentef}}, \bibinfo {author} {\bibfnamefont {U.}~\bibnamefont {{De
  Giovannini}}}, \bibinfo {author} {\bibfnamefont {A.~F.}\ \bibnamefont
  {Kemper}}, \ and\ \bibinfo {author} {\bibfnamefont {A.}~\bibnamefont
  {Rubio}},\ }\href {\doibase 10.1038/ncomms13940} {\bibfield  {journal}
  {\bibinfo  {journal} {Nature Communications}\ }\textbf {\bibinfo {volume}
  {8}},\ \bibinfo {pages} {1} (\bibinfo {year} {2017})},\ \Eprint
  {http://arxiv.org/abs/1604.03399} {arXiv:1604.03399} \BibitemShut {NoStop}%
\bibitem [{\citenamefont {Morimoto}\ and\ \citenamefont
  {Nagaosa}(2016)}]{Morimoto2016a}%
  \BibitemOpen
  \bibfield  {author} {\bibinfo {author} {\bibfnamefont {T.}~\bibnamefont
  {Morimoto}}\ and\ \bibinfo {author} {\bibfnamefont {N.}~\bibnamefont
  {Nagaosa}},\ }\href {\doibase 10.1126/sciadv.1501524} {\bibfield  {journal}
  {\bibinfo  {journal} {Science Advances}\ }\textbf {\bibinfo {volume} {2}}
  (\bibinfo {year} {2016}),\ 10.1126/sciadv.1501524},\ \Eprint
  {http://arxiv.org/abs/1510.08112} {arXiv:1510.08112} \BibitemShut {NoStop}%
\bibitem [{\citenamefont {Shin}\ \emph {et~al.}(2019)\citenamefont {Shin},
  \citenamefont {Sato}, \citenamefont {H{\"{u}}bener}, \citenamefont {{De
  Giovannini}}, \citenamefont {Kim}, \citenamefont {Park},\ and\ \citenamefont
  {Rubio}}]{Shin2019}%
  \BibitemOpen
  \bibfield  {author} {\bibinfo {author} {\bibfnamefont {D.}~\bibnamefont
  {Shin}}, \bibinfo {author} {\bibfnamefont {S.~A.}\ \bibnamefont {Sato}},
  \bibinfo {author} {\bibfnamefont {H.}~\bibnamefont {H{\"{u}}bener}}, \bibinfo
  {author} {\bibfnamefont {U.}~\bibnamefont {{De Giovannini}}}, \bibinfo
  {author} {\bibfnamefont {J.}~\bibnamefont {Kim}}, \bibinfo {author}
  {\bibfnamefont {N.}~\bibnamefont {Park}}, \ and\ \bibinfo {author}
  {\bibfnamefont {A.}~\bibnamefont {Rubio}},\ }\href {\doibase
  10.1073/pnas.1816904116} {\bibfield  {journal} {\bibinfo  {journal}
  {Proceedings of the National Academy of Sciences of the United States of
  America}\ }\textbf {\bibinfo {volume} {116}},\ \bibinfo {pages} {4135}
  (\bibinfo {year} {2019})}\BibitemShut {NoStop}%
\bibitem [{\citenamefont {Silva}\ \emph {et~al.}(2019)\citenamefont {Silva},
  \citenamefont {Amorim}, \citenamefont {Smirnova}, \citenamefont {Ivanov},
  \citenamefont {Jim{\'{e}}nez-Gal{\'{a}}n}, \citenamefont {Amorim},
  \citenamefont {Smirnova},\ and\ \citenamefont {Ivanov}}]{Silva2019}%
  \BibitemOpen
  \bibfield  {author} {\bibinfo {author} {\bibfnamefont {R.~E.~F.}\
  \bibnamefont {Silva}}, \bibinfo {author} {\bibfnamefont {B.}~\bibnamefont
  {Amorim}}, \bibinfo {author} {\bibfnamefont {O.}~\bibnamefont {Smirnova}},
  \bibinfo {author} {\bibfnamefont {M.}~\bibnamefont {Ivanov}}, \bibinfo
  {author} {\bibfnamefont {{\'{A}}.}~\bibnamefont {Jim{\'{e}}nez-Gal{\'{a}}n}},
  \bibinfo {author} {\bibfnamefont {B.}~\bibnamefont {Amorim}}, \bibinfo
  {author} {\bibfnamefont {O.}~\bibnamefont {Smirnova}}, \ and\ \bibinfo
  {author} {\bibfnamefont {M.}~\bibnamefont {Ivanov}},\ }\href {\doibase
  10.1038/s41566-019-0516-1} {\bibfield  {journal} {\bibinfo  {journal} {Nature
  Photonics}\ } (\bibinfo {year} {2019}),\
  10.1038/s41566-019-0516-1}\BibitemShut {NoStop}%
\bibitem [{\citenamefont {Chac{\'{o}}n}\ \emph {et~al.}(2018)\citenamefont
  {Chac{\'{o}}n}, \citenamefont {Zhu}, \citenamefont {Kelly}, \citenamefont
  {Dauphin}, \citenamefont {Pisanty}, \citenamefont {Pic{\'{o}}n},
  \citenamefont {Ticknor}, \citenamefont {Ciappina}, \citenamefont {Saxena},\
  and\ \citenamefont {Lewenstein}}]{Chacon2018}%
  \BibitemOpen
  \bibfield  {author} {\bibinfo {author} {\bibfnamefont {A.}~\bibnamefont
  {Chac{\'{o}}n}}, \bibinfo {author} {\bibfnamefont {W.}~\bibnamefont {Zhu}},
  \bibinfo {author} {\bibfnamefont {S.~P.}\ \bibnamefont {Kelly}}, \bibinfo
  {author} {\bibfnamefont {A.}~\bibnamefont {Dauphin}}, \bibinfo {author}
  {\bibfnamefont {E.}~\bibnamefont {Pisanty}}, \bibinfo {author} {\bibfnamefont
  {A.}~\bibnamefont {Pic{\'{o}}n}}, \bibinfo {author} {\bibfnamefont
  {C.}~\bibnamefont {Ticknor}}, \bibinfo {author} {\bibfnamefont {M.~F.}\
  \bibnamefont {Ciappina}}, \bibinfo {author} {\bibfnamefont {A.}~\bibnamefont
  {Saxena}}, \ and\ \bibinfo {author} {\bibfnamefont {M.}~\bibnamefont
  {Lewenstein}},\ }\href {http://arxiv.org/abs/1807.01616} {\bibfield
  {journal} {\bibinfo  {journal} {Arxiv Preprint}\ ,\ \bibinfo {pages}
  {arXiv:1807.01616v1}} (\bibinfo {year} {2018})},\ \Eprint
  {http://arxiv.org/abs/1807.01616} {arXiv:1807.01616} \BibitemShut {NoStop}%
\bibitem [{\citenamefont {Bauer}\ and\ \citenamefont
  {Hansen}(2018)}]{Bauer2018}%
  \BibitemOpen
  \bibfield  {author} {\bibinfo {author} {\bibfnamefont {D.}~\bibnamefont
  {Bauer}}\ and\ \bibinfo {author} {\bibfnamefont {K.~K.}\ \bibnamefont
  {Hansen}},\ }\href {\doibase 10.1103/PhysRevLett.120.177401} {\bibfield
  {journal} {\bibinfo  {journal} {Physical Review Letters}\ }\textbf {\bibinfo
  {volume} {120}},\ \bibinfo {pages} {177401} (\bibinfo {year} {2018})},\
  \Eprint {http://arxiv.org/abs/1711.05783} {arXiv:1711.05783} \BibitemShut
  {NoStop}%
\bibitem [{\citenamefont {Dr{\"{u}}eke}\ and\ \citenamefont
  {Bauer}(2019)}]{Drueeke2019}%
  \BibitemOpen
  \bibfield  {author} {\bibinfo {author} {\bibfnamefont {H.}~\bibnamefont
  {Dr{\"{u}}eke}}\ and\ \bibinfo {author} {\bibfnamefont {D.}~\bibnamefont
  {Bauer}},\ }\href {\doibase 10.1103/PhysRevA.99.053402} {\bibfield  {journal}
  {\bibinfo  {journal} {Physical Review A}\ }\textbf {\bibinfo {volume} {99}},\
  \bibinfo {pages} {053402} (\bibinfo {year} {2019})},\ \Eprint
  {http://arxiv.org/abs/1901.01437} {arXiv:1901.01437} \BibitemShut {NoStop}%
\bibitem [{\citenamefont {J{\"{u}}r{\ss}}\ and\ \citenamefont
  {Bauer}(2019)}]{Juerss2019}%
  \BibitemOpen
  \bibfield  {author} {\bibinfo {author} {\bibfnamefont {C.}~\bibnamefont
  {J{\"{u}}r{\ss}}}\ and\ \bibinfo {author} {\bibfnamefont {D.}~\bibnamefont
  {Bauer}},\ }\href {\doibase 10.1103/PhysRevB.99.195428} {\bibfield  {journal}
  {\bibinfo  {journal} {Physical Review B}\ }\textbf {\bibinfo {volume} {99}},\
  \bibinfo {pages} {195428} (\bibinfo {year} {2019})},\ \Eprint
  {http://arxiv.org/abs/1902.04120} {arXiv:1902.04120} \BibitemShut {NoStop}%
\bibitem [{\citenamefont {Mao}\ \emph {et~al.}(2011)\citenamefont {Mao},
  \citenamefont {Yamakage},\ and\ \citenamefont {Kuramoto}}]{Mao2011}%
  \BibitemOpen
  \bibfield  {author} {\bibinfo {author} {\bibfnamefont {S.}~\bibnamefont
  {Mao}}, \bibinfo {author} {\bibfnamefont {A.}~\bibnamefont {Yamakage}}, \
  and\ \bibinfo {author} {\bibfnamefont {Y.}~\bibnamefont {Kuramoto}},\ }\href
  {\doibase 10.1103/PhysRevB.84.115413} {\bibfield  {journal} {\bibinfo
  {journal} {Physical Review B - Condensed Matter and Materials Physics}\
  }\textbf {\bibinfo {volume} {84}},\ \bibinfo {pages} {115413} (\bibinfo
  {year} {2011})}\BibitemShut {NoStop}%
\bibitem [{\citenamefont {Liu}\ \emph {et~al.}(2018)\citenamefont {Liu},
  \citenamefont {Zheng}, \citenamefont {Zeng},\ and\ \citenamefont
  {Li}}]{Liu2018a}%
  \BibitemOpen
  \bibfield  {author} {\bibinfo {author} {\bibfnamefont {C.}~\bibnamefont
  {Liu}}, \bibinfo {author} {\bibfnamefont {Y.}~\bibnamefont {Zheng}}, \bibinfo
  {author} {\bibfnamefont {Z.}~\bibnamefont {Zeng}}, \ and\ \bibinfo {author}
  {\bibfnamefont {R.}~\bibnamefont {Li}},\ }\href {\doibase
  10.1103/PhysRevA.97.063412} {\bibfield  {journal} {\bibinfo  {journal}
  {Physical Review A}\ }\textbf {\bibinfo {volume} {97}},\ \bibinfo {pages}
  {063412} (\bibinfo {year} {2018})}\BibitemShut {NoStop}%
\bibitem [{\citenamefont {Fu}(2009)}]{Fu2009a}%
  \BibitemOpen
  \bibfield  {author} {\bibinfo {author} {\bibfnamefont {L.}~\bibnamefont
  {Fu}},\ }\href {\doibase 10.1103/PhysRevLett.103.266801} {\bibfield
  {journal} {\bibinfo  {journal} {Physical Review Letters}\ }\textbf {\bibinfo
  {volume} {103}},\ \bibinfo {pages} {266801} (\bibinfo {year}
  {2009})}\BibitemShut {NoStop}%
\bibitem [{\citenamefont {Liu}\ \emph {et~al.}(2010)\citenamefont {Liu},
  \citenamefont {Qi}, \citenamefont {Zhang}, \citenamefont {Dai}, \citenamefont
  {Fang},\ and\ \citenamefont {Zhang}}]{Liu2010}%
  \BibitemOpen
  \bibfield  {author} {\bibinfo {author} {\bibfnamefont {C.~X.}\ \bibnamefont
  {Liu}}, \bibinfo {author} {\bibfnamefont {X.~L.}\ \bibnamefont {Qi}},
  \bibinfo {author} {\bibfnamefont {H.}~\bibnamefont {Zhang}}, \bibinfo
  {author} {\bibfnamefont {X.}~\bibnamefont {Dai}}, \bibinfo {author}
  {\bibfnamefont {Z.}~\bibnamefont {Fang}}, \ and\ \bibinfo {author}
  {\bibfnamefont {S.~C.}\ \bibnamefont {Zhang}},\ }\href {\doibase
  10.1103/PhysRevB.82.045122} {\bibfield  {journal} {\bibinfo  {journal}
  {Physical Review B - Condensed Matter and Materials Physics}\ }\textbf
  {\bibinfo {volume} {82}},\ \bibinfo {pages} {045122} (\bibinfo {year}
  {2010})}\BibitemShut {NoStop}%
\bibitem [{\citenamefont {Wang}\ \emph {et~al.}(2011)\citenamefont {Wang},
  \citenamefont {Hsieh}, \citenamefont {Pilon}, \citenamefont {Fu},
  \citenamefont {Gardner}, \citenamefont {Lee},\ and\ \citenamefont
  {Gedik}}]{Wang2011}%
  \BibitemOpen
  \bibfield  {author} {\bibinfo {author} {\bibfnamefont {Y.~H.}\ \bibnamefont
  {Wang}}, \bibinfo {author} {\bibfnamefont {D.}~\bibnamefont {Hsieh}},
  \bibinfo {author} {\bibfnamefont {D.}~\bibnamefont {Pilon}}, \bibinfo
  {author} {\bibfnamefont {L.}~\bibnamefont {Fu}}, \bibinfo {author}
  {\bibfnamefont {D.~R.}\ \bibnamefont {Gardner}}, \bibinfo {author}
  {\bibfnamefont {Y.~S.}\ \bibnamefont {Lee}}, \ and\ \bibinfo {author}
  {\bibfnamefont {N.}~\bibnamefont {Gedik}},\ }\href {\doibase
  10.1103/PhysRevLett.107.207602} {\bibfield  {journal} {\bibinfo  {journal}
  {Physical Review Letters}\ }\textbf {\bibinfo {volume} {107}},\ \bibinfo
  {pages} {207602} (\bibinfo {year} {2011})}\BibitemShut {NoStop}%
\bibitem [{\citenamefont {Kira}\ and\ \citenamefont {Koch}(2011)}]{Kira2011}%
  \BibitemOpen
  \bibfield  {author} {\bibinfo {author} {\bibfnamefont {M.}~\bibnamefont
  {Kira}}\ and\ \bibinfo {author} {\bibfnamefont {S.~W.}\ \bibnamefont
  {Koch}},\ }\href {\doibase DOI: 10.1017/CBO9781139016926} {\emph {\bibinfo
  {title} {{Semiconductor Quantum Optics}}}}\ (\bibinfo  {publisher} {Cambridge
  University Press},\ \bibinfo {address} {Cambridge},\ \bibinfo {year}
  {2011})\BibitemShut {NoStop}%
\bibitem [{\citenamefont {Luu}\ and\ \citenamefont
  {W{\"{o}}rner}(2016)}]{Luu2016}%
  \BibitemOpen
  \bibfield  {author} {\bibinfo {author} {\bibfnamefont {T.~T.}\ \bibnamefont
  {Luu}}\ and\ \bibinfo {author} {\bibfnamefont {H.~J.}\ \bibnamefont
  {W{\"{o}}rner}},\ }\href {\doibase 10.1103/PhysRevB.94.115164} {\bibfield
  {journal} {\bibinfo  {journal} {Physical Review B}\ }\textbf {\bibinfo
  {volume} {94}},\ \bibinfo {pages} {115164} (\bibinfo {year}
  {2016})}\BibitemShut {NoStop}%
\bibitem [{\citenamefont {Li}\ \emph {et~al.}(2019)\citenamefont {Li},
  \citenamefont {Zhang}, \citenamefont {Fu}, \citenamefont {Feng},
  \citenamefont {Hu},\ and\ \citenamefont {Du}}]{Li2019}%
  \BibitemOpen
  \bibfield  {author} {\bibinfo {author} {\bibfnamefont {J.}~\bibnamefont
  {Li}}, \bibinfo {author} {\bibfnamefont {X.}~\bibnamefont {Zhang}}, \bibinfo
  {author} {\bibfnamefont {S.}~\bibnamefont {Fu}}, \bibinfo {author}
  {\bibfnamefont {Y.}~\bibnamefont {Feng}}, \bibinfo {author} {\bibfnamefont
  {B.}~\bibnamefont {Hu}}, \ and\ \bibinfo {author} {\bibfnamefont
  {H.}~\bibnamefont {Du}},\ }\href {\doibase 10.1103/PhysRevA.100.043404}
  {\bibfield  {journal} {\bibinfo  {journal} {Phys. Rev. A}\ }\textbf {\bibinfo
  {volume} {100}},\ \bibinfo {pages} {43404} (\bibinfo {year}
  {2019})}\BibitemShut {NoStop}%
\bibitem [{\citenamefont {Yue}\ and\ \citenamefont
  {Gaarde}(2020{\natexlab{a}})}]{Yue2020}%
  \BibitemOpen
  \bibfield  {author} {\bibinfo {author} {\bibfnamefont {L.}~\bibnamefont
  {Yue}}\ and\ \bibinfo {author} {\bibfnamefont {M.~B.}\ \bibnamefont
  {Gaarde}},\ }\href {\doibase 10.1103/PhysRevA.101.053411} {\bibfield
  {journal} {\bibinfo  {journal} {Phys. Rev. A}\ }\textbf {\bibinfo {volume}
  {101}},\ \bibinfo {pages} {53411} (\bibinfo {year}
  {2020}{\natexlab{a}})}\BibitemShut {NoStop}%
\bibitem [{\citenamefont {Vampa}\ \emph {et~al.}(2014)\citenamefont {Vampa},
  \citenamefont {McDonald}, \citenamefont {Orlando}, \citenamefont {Klug},
  \citenamefont {Corkum},\ and\ \citenamefont {Brabec}}]{Vampa2014}%
  \BibitemOpen
  \bibfield  {author} {\bibinfo {author} {\bibfnamefont {G.}~\bibnamefont
  {Vampa}}, \bibinfo {author} {\bibfnamefont {C.~R.}\ \bibnamefont {McDonald}},
  \bibinfo {author} {\bibfnamefont {G.}~\bibnamefont {Orlando}}, \bibinfo
  {author} {\bibfnamefont {D.~D.}\ \bibnamefont {Klug}}, \bibinfo {author}
  {\bibfnamefont {P.~B.}\ \bibnamefont {Corkum}}, \ and\ \bibinfo {author}
  {\bibfnamefont {T.}~\bibnamefont {Brabec}},\ }\href {\doibase
  10.1103/PhysRevLett.113.073901} {\bibfield  {journal} {\bibinfo  {journal}
  {Physical Review Letters}\ }\textbf {\bibinfo {volume} {113}},\ \bibinfo
  {pages} {073901} (\bibinfo {year} {2014})}\BibitemShut {NoStop}%
\bibitem [{\citenamefont {Xiao}\ \emph {et~al.}(2010)\citenamefont {Xiao},
  \citenamefont {Chang},\ and\ \citenamefont {Niu}}]{Xiao2010}%
  \BibitemOpen
  \bibfield  {author} {\bibinfo {author} {\bibfnamefont {D.}~\bibnamefont
  {Xiao}}, \bibinfo {author} {\bibfnamefont {M.~C.}\ \bibnamefont {Chang}}, \
  and\ \bibinfo {author} {\bibfnamefont {Q.}~\bibnamefont {Niu}},\ }\href
  {\doibase 10.1103/RevModPhys.82.1959} {\bibfield  {journal} {\bibinfo
  {journal} {Reviews of Modern Physics}\ }\textbf {\bibinfo {volume} {82}},\
  \bibinfo {pages} {1959} (\bibinfo {year} {2010})},\ \Eprint
  {http://arxiv.org/abs/0907.2021} {arXiv:0907.2021} \BibitemShut {NoStop}%
\bibitem [{\citenamefont {Yang}\ and\ \citenamefont {Liu}(2014)}]{Yang2014}%
  \BibitemOpen
  \bibfield  {author} {\bibinfo {author} {\bibfnamefont {F.}~\bibnamefont
  {Yang}}\ and\ \bibinfo {author} {\bibfnamefont {R.~B.}\ \bibnamefont {Liu}},\
  }\href {\doibase 10.1103/PhysRevB.90.245205} {\bibfield  {journal} {\bibinfo
  {journal} {Physical Review B}\ }\textbf {\bibinfo {volume} {90}},\ \bibinfo
  {pages} {245205} (\bibinfo {year} {2014})}\BibitemShut {NoStop}%
\bibitem [{\citenamefont {Shindou}\ and\ \citenamefont
  {Imura}(2005)}]{Shindou2005}%
  \BibitemOpen
  \bibfield  {author} {\bibinfo {author} {\bibfnamefont {R.}~\bibnamefont
  {Shindou}}\ and\ \bibinfo {author} {\bibfnamefont {K.~I.}\ \bibnamefont
  {Imura}},\ }\href {\doibase 10.1016/j.nuclphysb.2005.05.019} {\bibfield
  {journal} {\bibinfo  {journal} {Nuclear Physics B}\ }\textbf {\bibinfo
  {volume} {720}},\ \bibinfo {pages} {399} (\bibinfo {year}
  {2005})}\BibitemShut {NoStop}%
\bibitem [{\citenamefont {Gradhand}\ \emph {et~al.}(2012)\citenamefont
  {Gradhand}, \citenamefont {Fedorov}, \citenamefont {Pientka}, \citenamefont
  {Zahn}, \citenamefont {Mertig}, \citenamefont {Gy{\"{o}}rffy}, \citenamefont
  {Gradhand}, \citenamefont {Fedorov}, \citenamefont {Pientka}, \citenamefont
  {Zahn}, \citenamefont {Mertig},\ and\ \citenamefont
  {Gy{\"{o}}rffy}}]{Gradhand2012}%
  \BibitemOpen
  \bibfield  {author} {\bibinfo {author} {\bibfnamefont {M.}~\bibnamefont
  {Gradhand}}, \bibinfo {author} {\bibfnamefont {V.}~\bibnamefont {Fedorov}},
  \bibinfo {author} {\bibfnamefont {F.}~\bibnamefont {Pientka}}, \bibinfo
  {author} {\bibfnamefont {P.}~\bibnamefont {Zahn}}, \bibinfo {author}
  {\bibfnamefont {I.}~\bibnamefont {Mertig}}, \bibinfo {author} {\bibfnamefont
  {L.}~\bibnamefont {Gy{\"{o}}rffy}}, \bibinfo {author} {\bibfnamefont
  {M.}~\bibnamefont {Gradhand}}, \bibinfo {author} {\bibfnamefont {D.~V.}\
  \bibnamefont {Fedorov}}, \bibinfo {author} {\bibfnamefont {F.}~\bibnamefont
  {Pientka}}, \bibinfo {author} {\bibfnamefont {P.}~\bibnamefont {Zahn}},
  \bibinfo {author} {\bibfnamefont {I.}~\bibnamefont {Mertig}}, \ and\ \bibinfo
  {author} {\bibfnamefont {B.~L.}\ \bibnamefont {Gy{\"{o}}rffy}},\ }\href
  {\doibase 10.1088/0953-8984/24/21/213202} {\bibfield  {journal} {\bibinfo
  {journal} {Journal of Physics Condensed Matter}\ }\textbf {\bibinfo {volume}
  {24}} (\bibinfo {year} {2012}),\ 10.1088/0953-8984/24/21/213202}\BibitemShut
  {NoStop}%
\bibitem [{\citenamefont {Tang}\ and\ \citenamefont {Rabin}(1971)}]{Tang1971}%
  \BibitemOpen
  \bibfield  {author} {\bibinfo {author} {\bibfnamefont {C.~L.}\ \bibnamefont
  {Tang}}\ and\ \bibinfo {author} {\bibfnamefont {H.}~\bibnamefont {Rabin}},\
  }\href {\doibase 10.1103/PhysRevB.3.4025} {\bibfield  {journal} {\bibinfo
  {journal} {Physical Review B}\ }\textbf {\bibinfo {volume} {3}},\ \bibinfo
  {pages} {4025} (\bibinfo {year} {1971})}\BibitemShut {NoStop}%
\bibitem [{\citenamefont {Saito}\ \emph {et~al.}(2017)\citenamefont {Saito},
  \citenamefont {Xia}, \citenamefont {Lu}, \citenamefont {Kanai}, \citenamefont
  {Itatani},\ and\ \citenamefont {Ishii}}]{Saito2017}%
  \BibitemOpen
  \bibfield  {author} {\bibinfo {author} {\bibfnamefont {N.}~\bibnamefont
  {Saito}}, \bibinfo {author} {\bibfnamefont {P.}~\bibnamefont {Xia}}, \bibinfo
  {author} {\bibfnamefont {F.}~\bibnamefont {Lu}}, \bibinfo {author}
  {\bibfnamefont {T.}~\bibnamefont {Kanai}}, \bibinfo {author} {\bibfnamefont
  {J.}~\bibnamefont {Itatani}}, \ and\ \bibinfo {author} {\bibfnamefont
  {N.}~\bibnamefont {Ishii}},\ }\href {\doibase 10.1364/OPTICA.4.001333}
  {\bibfield  {journal} {\bibinfo  {journal} {Optica}\ }\textbf {\bibinfo
  {volume} {4}},\ \bibinfo {pages} {1333} (\bibinfo {year} {2017})}\BibitemShut
  {NoStop}%
\bibitem [{\citenamefont {Yue}\ and\ \citenamefont
  {Gaarde}(2020{\natexlab{b}})}]{Yue2020a}%
  \BibitemOpen
  \bibfield  {author} {\bibinfo {author} {\bibfnamefont {L.}~\bibnamefont
  {Yue}}\ and\ \bibinfo {author} {\bibfnamefont {M.~B.}\ \bibnamefont
  {Gaarde}},\ }\href {\doibase 10.1103/PhysRevLett.124.153204} {\bibfield
  {journal} {\bibinfo  {journal} {Physical Review Letters}\ }\textbf {\bibinfo
  {volume} {124}},\ \bibinfo {pages} {1} (\bibinfo {year}
  {2020}{\natexlab{b}})},\ \Eprint {http://arxiv.org/abs/2001.04626}
  {arXiv:2001.04626} \BibitemShut {NoStop}%
\bibitem [{\citenamefont {Yoshikawa}\ \emph {et~al.}(2017)\citenamefont
  {Yoshikawa}, \citenamefont {Tamaya},\ and\ \citenamefont
  {Tanaka}}]{Yoshikawa2017}%
  \BibitemOpen
  \bibfield  {author} {\bibinfo {author} {\bibfnamefont {N.}~\bibnamefont
  {Yoshikawa}}, \bibinfo {author} {\bibfnamefont {T.}~\bibnamefont {Tamaya}}, \
  and\ \bibinfo {author} {\bibfnamefont {K.}~\bibnamefont {Tanaka}},\ }\href
  {\doibase 10.1126/science.aam8861} {\bibfield  {journal} {\bibinfo  {journal}
  {Science}\ }\textbf {\bibinfo {volume} {356}},\ \bibinfo {pages} {736}
  (\bibinfo {year} {2017})}\BibitemShut {NoStop}%
\bibitem [{\citenamefont {Shan}\ \emph {et~al.}(2010)\citenamefont {Shan},
  \citenamefont {Lu},\ and\ \citenamefont {Shen}}]{Shan2010}%
  \BibitemOpen
  \bibfield  {author} {\bibinfo {author} {\bibfnamefont {W.~Y.}\ \bibnamefont
  {Shan}}, \bibinfo {author} {\bibfnamefont {H.~Z.}\ \bibnamefont {Lu}}, \ and\
  \bibinfo {author} {\bibfnamefont {S.~Q.}\ \bibnamefont {Shen}},\ }\href
  {\doibase 10.1088/1367-2630/12/4/043048} {\bibfield  {journal} {\bibinfo
  {journal} {New Journal of Physics}\ }\textbf {\bibinfo {volume} {12}},\
  \bibinfo {pages} {043048} (\bibinfo {year} {2010})},\ \Eprint
  {http://arxiv.org/abs/arXiv:0911.3706} {arXiv:arXiv:0911.3706} \BibitemShut
  {NoStop}%
\bibitem [{\citenamefont {{Shun-Qing Shen}}(2012)}]{Shun-QingShen2012}%
  \BibitemOpen
  \bibfield  {author} {\bibinfo {author} {\bibnamefont {{Shun-Qing Shen}}},\
  }\href@noop {} {\emph {\bibinfo {title} {{Topological Insulators: Dirac
  Equations in Condensed Matters}}}},\ \bibinfo {edition} {1st}\ ed.,\ edited
  by\ \bibinfo {editor} {\bibfnamefont {H.~S.}\ \bibnamefont {{M. Cardona, P.
  Fulde, K. von Kitzling, R. Merlin, H.-J. Queisser}}}\ (\bibinfo  {publisher}
  {Springer-Verlag Berlin Heidelberg},\ \bibinfo {year} {2012})\ p.\ \bibinfo
  {pages} {232}\BibitemShut {NoStop}%
\bibitem [{\citenamefont {Neufeld}\ \emph {et~al.}(2019)\citenamefont
  {Neufeld}, \citenamefont {Podolsky},\ and\ \citenamefont
  {Cohen}}]{Neufeld2019}%
  \BibitemOpen
  \bibfield  {author} {\bibinfo {author} {\bibfnamefont {O.}~\bibnamefont
  {Neufeld}}, \bibinfo {author} {\bibfnamefont {D.}~\bibnamefont {Podolsky}}, \
  and\ \bibinfo {author} {\bibfnamefont {O.}~\bibnamefont {Cohen}},\ }\href
  {\doibase 10.1038/s41467-018-07935-y} {\bibfield  {journal} {\bibinfo
  {journal} {Nature Communications}\ }\textbf {\bibinfo {volume} {10}},\
  \bibinfo {pages} {405} (\bibinfo {year} {2019})}\BibitemShut {NoStop}%
\bibitem [{\citenamefont {Jiang}\ \emph {et~al.}(2019)\citenamefont {Jiang},
  \citenamefont {Gholam-Mirzaei}, \citenamefont {Crites}, \citenamefont
  {Beetar}, \citenamefont {Singh}, \citenamefont {Lu}, \citenamefont {Chini},\
  and\ \citenamefont {Lin}}]{Jiang2019}%
  \BibitemOpen
  \bibfield  {author} {\bibinfo {author} {\bibfnamefont {S.}~\bibnamefont
  {Jiang}}, \bibinfo {author} {\bibfnamefont {S.}~\bibnamefont
  {Gholam-Mirzaei}}, \bibinfo {author} {\bibfnamefont {E.}~\bibnamefont
  {Crites}}, \bibinfo {author} {\bibfnamefont {J.~E.}\ \bibnamefont {Beetar}},
  \bibinfo {author} {\bibfnamefont {M.}~\bibnamefont {Singh}}, \bibinfo
  {author} {\bibfnamefont {R.}~\bibnamefont {Lu}}, \bibinfo {author}
  {\bibfnamefont {M.}~\bibnamefont {Chini}}, \ and\ \bibinfo {author}
  {\bibfnamefont {C.~D.}\ \bibnamefont {Lin}},\ }\href {\doibase
  10.1088/1361-6455/ab470d} {\bibfield  {journal} {\bibinfo  {journal} {Journal
  of Physics B: Atomic, Molecular and Optical Physics}\ }\textbf {\bibinfo
  {volume} {52}},\ \bibinfo {pages} {225601} (\bibinfo {year}
  {2019})}\BibitemShut {NoStop}%
\bibitem [{\citenamefont {Langer}\ \emph {et~al.}(2016)\citenamefont {Langer},
  \citenamefont {Hohenleutner}, \citenamefont {Schmid}, \citenamefont
  {Poellmann}, \citenamefont {Nagler}, \citenamefont {Korn}, \citenamefont
  {Sch{\"{u}}ller}, \citenamefont {Sherwin}, \citenamefont {Huttner},
  \citenamefont {Steiner}, \citenamefont {Koch}, \citenamefont {Kira},\ and\
  \citenamefont {Huber}}]{Langer2016}%
  \BibitemOpen
  \bibfield  {author} {\bibinfo {author} {\bibfnamefont {F.}~\bibnamefont
  {Langer}}, \bibinfo {author} {\bibfnamefont {M.}~\bibnamefont
  {Hohenleutner}}, \bibinfo {author} {\bibfnamefont {C.~P.}\ \bibnamefont
  {Schmid}}, \bibinfo {author} {\bibfnamefont {C.}~\bibnamefont {Poellmann}},
  \bibinfo {author} {\bibfnamefont {P.}~\bibnamefont {Nagler}}, \bibinfo
  {author} {\bibfnamefont {T.}~\bibnamefont {Korn}}, \bibinfo {author}
  {\bibfnamefont {C.}~\bibnamefont {Sch{\"{u}}ller}}, \bibinfo {author}
  {\bibfnamefont {M.~S.}\ \bibnamefont {Sherwin}}, \bibinfo {author}
  {\bibfnamefont {U.}~\bibnamefont {Huttner}}, \bibinfo {author} {\bibfnamefont
  {J.~T.}\ \bibnamefont {Steiner}}, \bibinfo {author} {\bibfnamefont {S.~W.}\
  \bibnamefont {Koch}}, \bibinfo {author} {\bibfnamefont {M.}~\bibnamefont
  {Kira}}, \ and\ \bibinfo {author} {\bibfnamefont {R.}~\bibnamefont {Huber}},\
  }\href {\doibase 10.1038/nature17958} {\bibfield  {journal} {\bibinfo
  {journal} {Nature}\ }\textbf {\bibinfo {volume} {533}},\ \bibinfo {pages}
  {225} (\bibinfo {year} {2016})}\BibitemShut {NoStop}%
\end{thebibliography}%

\end{document}